\documentclass[twocolumn,tighten]{aastex63}
\usepackage{amsmath}
\usepackage{threeparttable}
\usepackage{float}
\usepackage{graphicx}
\usepackage{CJK}

\newcommand{\package}[1]{\tt{#1}}

\newcommand{\elz}{$E-L_{\rm{z}}$}
\newcommand{\elzs}{$E-L_{\rm{z}}\ $}

\newcommand{\afe}{[$\alpha$/Fe]}
\newcommand{\feh}{[Fe/H]}

\shorttitle{The H3 Survey: Halo Substructure}
\shortauthors{Naidu et al.}

\begin{document}
\begin{CJK*}{UTF8}{gbsn}

\title{Evidence from the H3 Survey that the Stellar Halo is Entirely Comprised of Substructure}

\correspondingauthor{Rohan P. Naidu}
\email{rohan.naidu@cfa.harvard.edu}
\author[0000-0003-3997-5705]{Rohan P. Naidu}
\affiliation{Center for Astrophysics $|$ Harvard \& Smithsonian, 60 Garden Street, Cambridge, MA 02138, USA}
\author[0000-0002-1590-8551]{Charlie Conroy}
\affiliation{Center for Astrophysics $|$ Harvard \& Smithsonian, 60 Garden Street, Cambridge, MA 02138, USA}
\author[0000-0002-7846-9787]{Ana Bonaca}
\affiliation{Center for Astrophysics $|$ Harvard \& Smithsonian, 60 Garden Street, Cambridge, MA 02138, USA}
\author[0000-0002-9280-7594]{Benjamin D. Johnson}
\affiliation{Center for Astrophysics $|$ Harvard \& Smithsonian, 60 Garden Street, Cambridge, MA 02138, USA}
\author[0000-0001-5082-9536]{Yuan-Sen Ting (丁源森)}
\altaffiliation{Hubble Fellow}
\affiliation{Institute for Advanced Study, Princeton, NJ 08540, USA}
\affiliation{Department of Astrophysical Sciences, Princeton University, Princeton, NJ 08544, USA}
\affiliation{Observatories of the Carnegie Institution of Washington, 813 Santa Barbara Street, Pasadena, CA 91101, USA}
\affiliation{Research School of Astronomy and Astrophysics, Mount Stromlo Observatory, Cotter Road, Weston Creek, ACT 2611, Canberra, Australia}
\author[0000-0003-2352-3202]{Nelson Caldwell}
\affiliation{Center for Astrophysics $|$ Harvard \& Smithsonian, 60 Garden Street, Cambridge, MA 02138, USA}
\author[0000-0002-5177-727X]{Dennis Zaritsky}
\affiliation{Steward Observatory, University of Arizona, 933 North Cherry Avenue, Tucson, AZ 85721-0065, USA}
\author[0000-0002-1617-8917]{Phillip A. Cargile}
\affiliation{Center for Astrophysics $|$ Harvard \& Smithsonian, 60 Garden Street, Cambridge, MA 02138, USA}

\begin{abstract} 
In the $\Lambda$CDM paradigm the Galactic stellar halo is predicted to harbor the accreted debris of smaller systems. To identify these systems, the H3 Spectroscopic Survey, combined with \textit{Gaia}, is gathering 6D phase-space and chemical information in the distant Galaxy. Here we present a comprehensive inventory of structure within 50 kpc from the Galactic center using a sample of 5684 giants at $|b|>40^{\circ}$ and $|Z|>2$ kpc. We identify known structures including the high-$\alpha$ disk, the in-situ halo (disk stars heated to eccentric orbits), Sagittarius (Sgr), \textit{Gaia}-Sausage-Enceladus (GSE), the Helmi Streams, Sequoia, and Thamnos. Additionally, we identify the following new structures: (i) Aleph ([Fe/H]$=-0.5$), a low eccentricity structure that rises a surprising 10 kpc off the plane, (ii, iii) Arjuna ([Fe/H]$=-1.2$)  and I'itoi ([Fe/H]$<-2$), which comprise the high-energy retrograde halo along with Sequoia, and (iv) Wukong ([Fe/H]$=-1.6$), a prograde phase-space overdensity chemically distinct from GSE. For each structure we provide [Fe/H], [$\alpha$/Fe], and orbital parameters. Stars born within the Galaxy are a major component at $|Z|\sim$2 kpc ($\approx$60$\%$), but their relative fraction declines sharply to $\lesssim$5$\%$ past 15 kpc. Beyond 15 kpc, $>$80$\%$ of the halo is built by two massive ($M_{\star}\sim10^{8}-10^{9}M_{\odot}$) accreted dwarfs: GSE ([Fe/H]$=-1.2$) within 25 kpc, and Sgr ([Fe/H]$=-1.0$) beyond 25 kpc. This explains the relatively high overall metallicity of the halo ([Fe/H]$\approx-1.2$). We attribute $\gtrsim$95$\%$ of the sample to one of the listed structures, pointing to a halo built entirely from accreted dwarfs and heating of the disk.
\end{abstract}


\section{Introduction}
\label{sec:introduction}

The Milky Way's stellar halo comprises only $\sim1\%$ of its stellar mass \citep[e.g.,][]{Deason19, Mackereth20}, and yet it is an object of intense interest because it acts as a time capsule, preserving memory of the Galaxy's assembly history with high fidelity. As early as \citet{Woolley57} it was realized that ``the time of relaxation of stellar motions in this part of the galaxy is at least $10^{12}$ years, whereas the stars themselves have not existed in their present form for much more than $10^{10}$ years". In detail, halo stars belonging to the same structure, even when they are scattered across the sky, retain similar coordinates in their integrals of motion (e.g., energy, angular momenta, actions). Further, stars belonging to the same structure share similar chemical abundance patterns \citep[e.g.,][]{Freeman02, Venn04, Lee15}. This expected clustering of halo stars in both integrals of motion and chemistry opens the door to ``reconstruct the galactic past" \citep{Eggen62}.

Thanks to large stellar spectroscopic surveys, e.g., RAVE \citep{RAVE}, SEGUE \citep{SEGUE}, LAMOST \citep{LAMOST}, GALAH \citep{GALAH}, APOGEE \citep{APOGEE}, and the \textit{Gaia} mission \citep{gaia}, integrals of motion and chemical abundances have become available for millions of stars in the solar neighborhood. Strikingly, more than half of the [Fe/H]$<-1$ stars in the local halo\footnote{By ``local halo" we mean the portion of the kinematic halo within a few kpc from the Sun that is typically selected using 3D Galactocentric velocity (e.g., $|V-V_{\rm{LSR}}|>210\ \rm{km\ s^{-1}}$, \citealt[][]{Helmi18}) with a view to avoid the disk.} appear to originate from a single system, the accreted \textit{Gaia}-Sausage-Enceladus (GSE) dwarf galaxy \citep[e.g.,][]{Belokurov18,Koppelman18,Myeong18,Haywood18, Helmi18,Mackereth19}. However, the most substantive component ($\gtrsim50\%$) of the local halo, the ``in-situ halo"/``Splash", likely arose from the heating of the primordial high-$\alpha$ disk by early mergers \citep[e.g.,][]{Bonaca17,Bonaca20,Haywood18,DiMatteo19, Belokurov20}.

While the local halo has provided these vital insights into the Galaxy's assembly, a complete census of accretion events requires going beyond the solar neighborhood. A number of simulations show that debris from minor mergers, higher-mass but recently accreted galaxies, and galaxies accreted along particular inclinations (e.g., at high angular momentum) are under-represented in the local halo \citep[e.g.,][]{bj05_1,Amorisco17, Fattahi20, Pfeffer20}. As a consequence, studies that rely on local high-energy orbits to deduce the nature of the distant halo are biased against these populations. The disrupting Sagittarius dwarf galaxy \citep[e.g.,][]{Ibata94, Majewski03} is a prime example of a massive structure that is completely absent from the local halo. 

Fully characterizing the global extent of structures discovered in local samples also demands pushing farther into the halo. Debris from low-mass structures like Thamnos ($M_{\star}<5\times10^{6}M_{\odot}$, \citealt{Koppelman19}), which is barely discernible in local samples, might be more apparent at larger distances due to ``apocenter pile-up" \citep[][]{Deason18} or due to higher contrast once the density of GSE and disk-like stars falls off. Studying massive accreted structures like GSE and the Helmi Streams \citep{Helmi99} -- e.g., the presence/absence of metallicity gradients, robust stellar masses from star counts, if they even are a single contiguous structure -- will also become more tractable with samples spanning their full extent.

Ranging beyond the local halo is also necessary to settle long-standing debates about the origin and nature of the halo. Is the halo largely formed in-situ or ex-situ (e.g., \citealt[][]{Eggen62} vs. \citealt[][]{Searle78})? To what radius does the recently discovered in-situ component of the halo dominate the halo mass function? Some simulations \citep[e.g.,][]{Monachesi19} show disk stars, heated by mergers, comprising $\sim20\%$ of the halo even beyond 50 kpc. Consequently, the extent and relative fraction of the in-situ halo should provide an independent constraint on the Galaxy's accretion history \citep[e.g.,][]{Zolotov09, Purcell10}. More generally, the fraction of in-situ halo stars (not only the heated disk, but also stars formed from stripped gas from satellites or through cosmological accretion) varies widely across simulations, ranging from negligible to comparable to the accreted mass, and could act as a sensitive constraint on sub-grid physics like star-formation and feedback prescriptions \citep[e.g.,][]{Cooper15, Pillepich15,Fattahi20}. 

Intertwined questions about the ex-situ component persist. Is it built from a handful of massive galaxies ($M_{\star}\sim10^{8}-10^{9} M_{\odot}$), or a multitude of metal-poor ultra-faints ($M_{\star}\lesssim10^{5} M_{\odot}$) \citep[e.g.,][]{bj05_2,Frebel10,Deason15, Deason16, dsouza18}? How does the metallicity of the halo change as a function of radius? Does the halo transition into a metal-poor ([Fe/H]$\sim-2.2$), spherical structure beyond 20 kpc as predicted by local energetic orbits in the popular ``dual halo" scenario \citep[e.g.,][]{Carollo07,Carollo10, Beers12}? Are different accreted galaxies responsible for this shift? Or could this be due to a smooth component from dissolved ancient globular clusters \citep[e.g.,][]{Martell11,Carretta16,Koch19}? Is the traditional conception of the distant halo as a metal-poor structure a selection artifact, arising from color cuts designed to avoid the disk, and from metallicity-biased standard candles \citep[e.g.,][]{Conroy19b}? 

Studying the stellar halo also enables new forms of near-field cosmology. For instance, accreted debris from $M_{\star} = 10^{6}-10^{7} M_{\odot}$ galaxies gives us access on a star-by-star level to high-redshift galaxies whose evolution was frozen at the time of infall. This provides a complementary view on issues of the distant universe -- for instance, the evolution of the ISM \citep[e.g.,][]{Steidel16,Bian20}, the interplay between reionization and low-mass galaxies \citep[e.g.,][]{Barkana99,Naidu20}, the shape of high-$z$ star-formation histories \citep[e.g.,][]{Carnall19,Leja19} -- at a resolution and mass-limit even beyond the reach of upcoming ELTs (Extremely Large Telescopes) and the \textit{James Webb Space Telescope} \citep[e.g.,][]{Boylan-Kolchin15,Boylan-Kolchin16,Weisz14}.

Previous efforts to directly probe the distant halo have had to overcome the challenge of targeting rare, distant stars without the benefit of \textit{Gaia} parallaxes to filter out nearby contaminants. One common solution has been to use color cuts that implicitly or explicitly select for low-metallicities to avoid the disk \citep[e.g.,][]{Chiba00,Carollo07,Ivezic08,Sesar11,Xue15,Zuo17}.  Another common choice is to rely on rare, standardizable candles like RR Lyrae and blue horizontal branch stars (BHBs) that are inherently metal-poor and more abundant in older populations \citep[e.g.,][]{Deason11,Kafle12,Janesh16,Cohen17,Iorio19}. Studies based on these tracers have collectively shown the distant Galaxy to display a high degree of substructure, which has been interpreted as support for an accretion origin of the halo \citep[e.g.,][]{Bell08,Starkenburg09,Xue11,Schlaufman12,Deason18}. In order to make further progress, and ask more detailed questions -- which accreted structure dominates at what radius? how far does the in-situ halo extend? what is the mass function of accreted material? -- we require a homogeneously selected, metallicity-unbiased sample with full 6D phase-space coordinates, chemical information, and an easily interpretable selection function.

The H3 (``Hectochelle in the Halo at High Resolution") Survey \citep{Conroy19} is fulfilling this need. H3 is a spectroscopic survey of 200,000 stars in high-latitude fields designed to study the distant Galaxy. A defining feature of H3 is a simple, \textit{Gaia}-based selection function (parallaxes implying $d_{\rm{helio}}>2$ kpc) that, critically, is unbiased in metallicity. With this survey we aim to search for new structure in the distant halo, trace known structures out to their apocenters, clarify long-standing debates about the nature of the halo, and explore promising avenues for near-field cosmology. 

\begin{figure*}
\centering
\includegraphics[width=\linewidth]{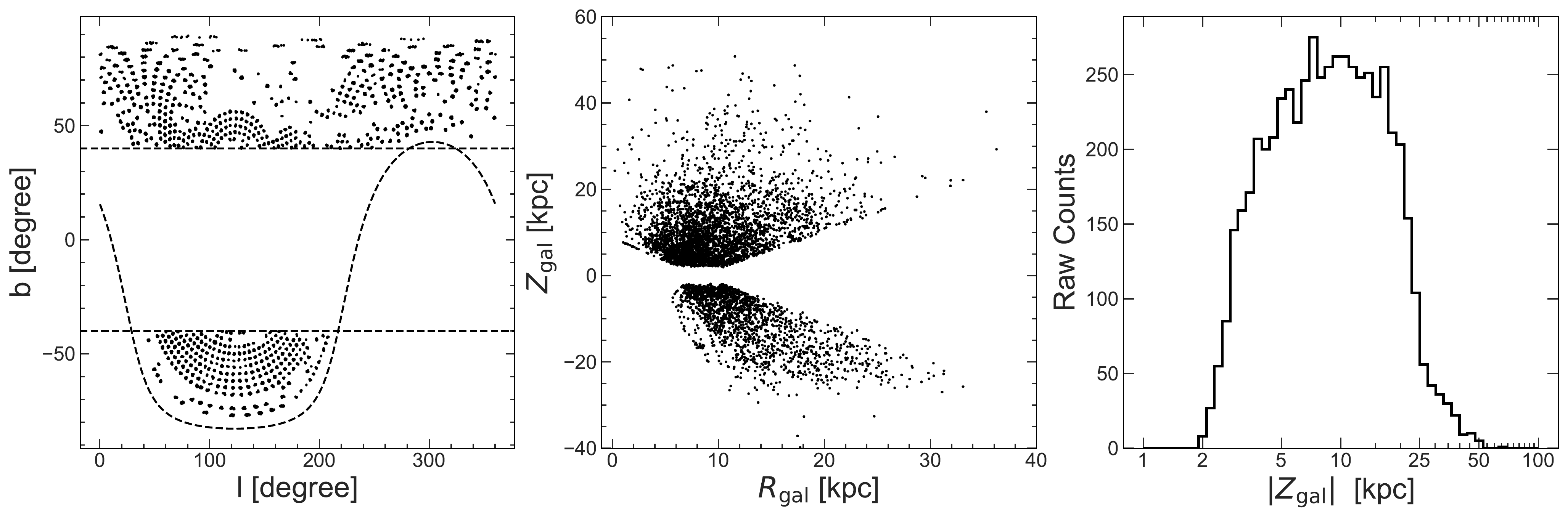}
\caption{Overview of the H3 Survey. \textbf{Left:} Current footprint in Galactic coordinates. Dashed lines demarcate $|b|=40^{\circ}$ and Dec.$=-20^{\circ}$. The survey will eventually cover $|b|>30^{\circ}$. A majority of fields ($\approx65\%$) are in the northern Galactic sky due to the location of the survey telescope ($+32^{\circ}$, AZ, USA). \textbf{Center:} Spatial extent of the sample used in this work in cylindrical, Galactocentric coordinates. \textbf{Right:} Distribution of distance from the plane. $99.9\%$ of the sample lies at an elevation of $>2$ kpc, with a median elevation of $\approx9$ kpc.}
\label{fig:data-1}
\end{figure*}

\begin{figure}
\centering
\includegraphics[width=0.9\linewidth]{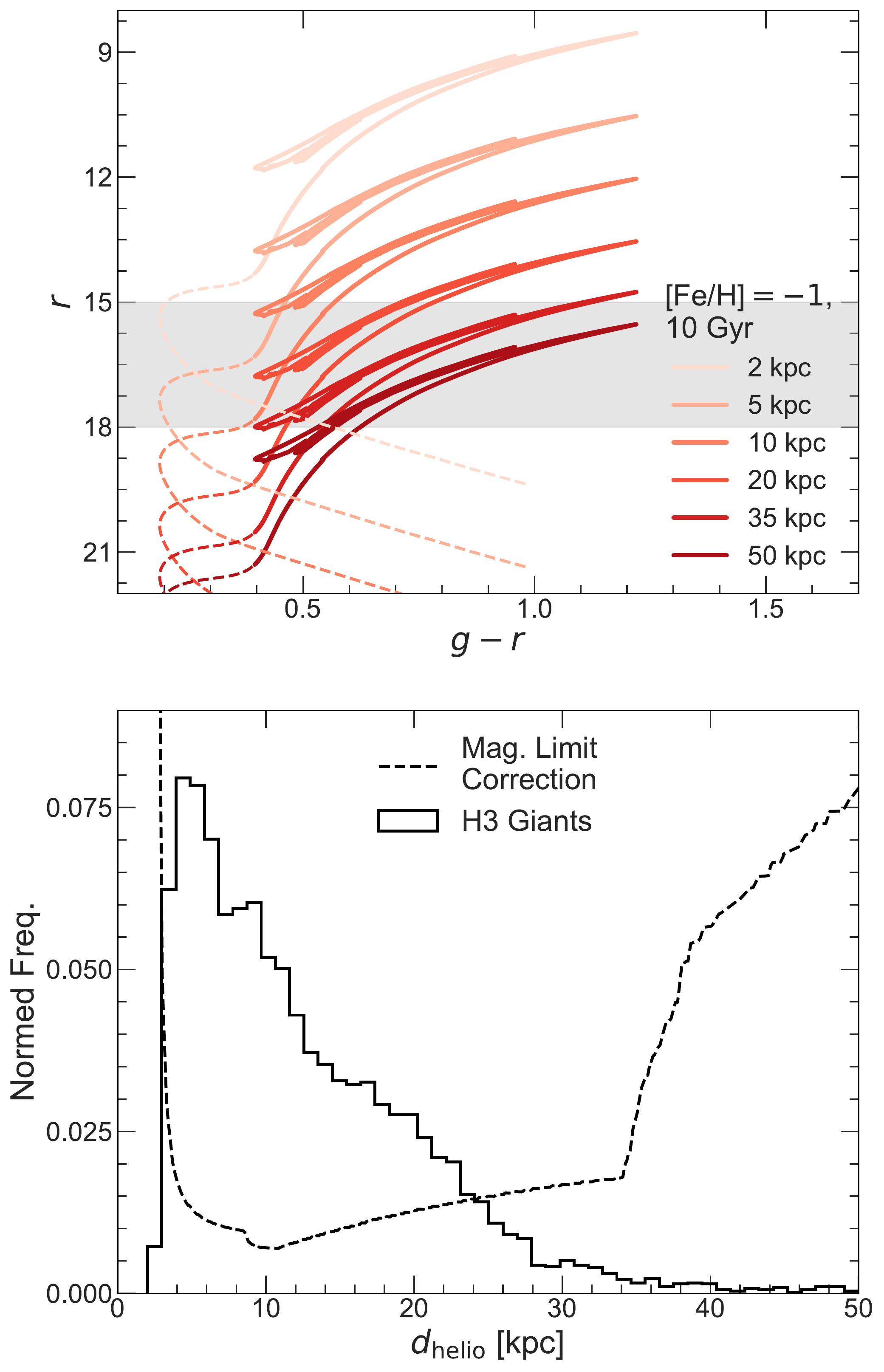}
\caption{Correcting for the magnitude selection in the H3 selection function. \textbf{Top:} An [Fe/H]$=-1$, 10 Gyr isochrone at different distances, with giants ($\log{g}<3.5$) highlighted with solid lines. At $d_{\rm{helio}}\sim4-35$ kpc the silver band representing the survey magnitude limit ($15<r<18$) almost completely contains the sections of the red giant branch which have a high number density. Therefore, the stars at these distances require little correction for the magnitude limit. \textbf{Bottom:} Correction weights as a function of heliocentric distance (dashed line) overplotted on the distance distribution for stars in this work. The weights are derived using isochrones and assuming a \citet{Kroupa01} IMF (see \S\ref{subsec:magcorr}), and are remarkably flat for the bulk of the sample. The rise at 35 kpc coincides with the red clump moving out of our magnitude range.}
\label{fig:magcorr}
\end{figure}

\begin{figure*}
\centering
\includegraphics[width=\linewidth]{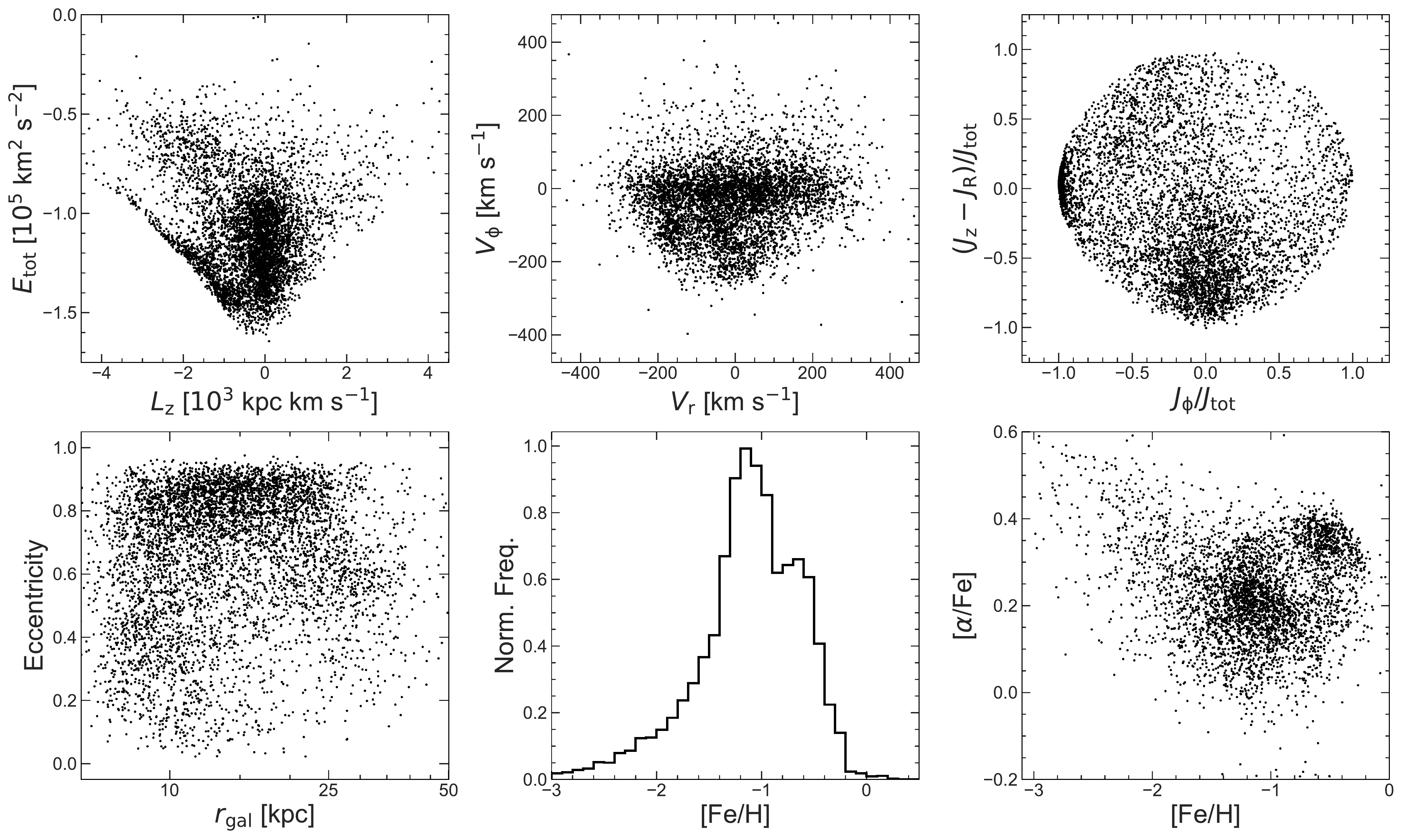}
\caption{Overview of sample in phase-space and chemistry. 
\textbf{Top left:} Total energy ($E_{\rm{tot}}$) versus the $z$-component of angular momentum ($L_{\rm{z}}$). 
Stars belonging to the same physical structure are expected to cluster in integrals of motion such as energy and angular momentum ($L_{\rm{z}}$, in particular, in an axisymmetric potential). \textbf{Top center:} Azimuthal velocity ($V_{\rm{\phi}}$) versus radial velocity ($V_{\rm{r}}$). Disk-like populations appear at negative $V_{\rm{\phi}}$ around our assumed $V_{\rm{\phi,\odot}}=-245.6\ \rm{km}\ \rm{s^{-1}}$. \textbf{Top right:} Summary of actions. Structures with strong vertical action ($J_{\rm{z}}$) occupy the top half of the diagram while those with a strong radial action ($J_{\rm{R}}$) occupy the bottom half. Prograde stars fall in the left hemisphere and retrograde stars in the right. \textbf{Bottom left:} Eccentricity vs. Galactocentric distance. In this space stars from the same accreted object have similar eccentricities because they are on similar orbits and show density breaks around their apocenters. \textbf{Bottom center:} Metallicity distribution function (MDF). The bins are 0.1 dex in size, corresponding to the typical uncertainty in [Fe/H] ($<0.1$ dex). \textbf{Bottom right:} $\alpha$-abundance ([$\alpha$/Fe]) vs iron abundance ([Fe/H]). We only show SNR$>5$ stars in this and all such subsequent panels to improve clarity. Distinct stellar populations are expected to follow chemical evolutionary tracks corresponding to their star-formation history and mass.}
\label{fig:data}
\end{figure*}

In this work we present a census of substructure, previously known and unknown, out to 50 kpc and link the results to the questions outlined in this section. In \S\ref{sec:data_methods} we provide details of H3 pertinent to this study (\S\ref{subsec:H3}), outline how we compute dynamical quantities (\S\ref{subsec:pot}), and correct for the survey selection function (\S\ref{subsec:magcorr}). In \S\ref{subsec:results_overview} we present an overview of our sample in integrals of motion and chemistry, revealing a high degree of substructure. \S\ref{subsec:inventory} forms the bulk of the paper -- here we identify and define individual structures, and remark on their chemodynamical properties. \S\ref{subsec:structuresummary} provides a synopsis of all the structures identified. In \S\ref{sec:discussion} we discuss the implications of the inventory -- we chart the relative fractions of structures with distance (\S\ref{subsec:relfrac}), interpret what this means for the origin of the halo (\S\ref{subsec:origin}), evaluate the net rotation of the halo (\S\ref{subsec:protretro}), dissect the halo in chemical space (\S\ref{subsec:chemistry}), and discuss caveats  (\S\ref{subsec:caveats}). A  summary of our results is provided in \S\ref{sec:summary}.

To describe central values of distributions we generally report the median, along with 16th and 84th percentiles. We use $\langle x\rangle$ to denote the mean of the quantity $x$, and report the corresponding error on the mean as 16th and 84th percentiles estimated via bootstrapping. We use $r_{\rm{gal}}$ to denote 3D Galactocentric distance, $R_{\rm{gal}}$ to denote axial distance in Galactocentric cylindrical coordinates, $Z_{\rm{gal}}$ to denote distance from the plane, and $d_{\rm{helio}}$ to refer to 3D heliocentric distance. We use $V_{r}$,  $V_{\rm{\phi}}$,  $V_{\rm{\theta}}$ to represent velocities in a right-handed spherical coordinate system with origin at the Galactic center. That is, prograde stars have negative $V_{\rm{\phi}}$ and $L_{\rm{z}}$.  In the context of photometric magnitudes``$r$"  refers to the Pan-STARRS $r$-band \citep{psack1,psack2} that is used in the H3 selection function. Magnitudes are in the AB system \citep{Oke83}. When converting between redshifts and ages we use a cosmology with $\Omega_M=0.3, \Omega_\Lambda=0.7, H_0=70$ km s$^{-1}$ Mpc$^{-1}$, i.e., $h=0.7$. Unless mentioned otherwise, total orbital energy ($E_{\rm{tot}}$) is always reported in units of $10^{5}\ \rm{km^{2}\ s^{-2}}$ and angular momenta ($L_{\rm{x}}$, $L_{\rm{y}}$, $L_{\rm{z}}$) are reported in units of $10^{3}\ \rm{kpc}\ \rm{km\ s^{-1}}$.

\begin{figure*}
\centering
\includegraphics[width=.95\linewidth]{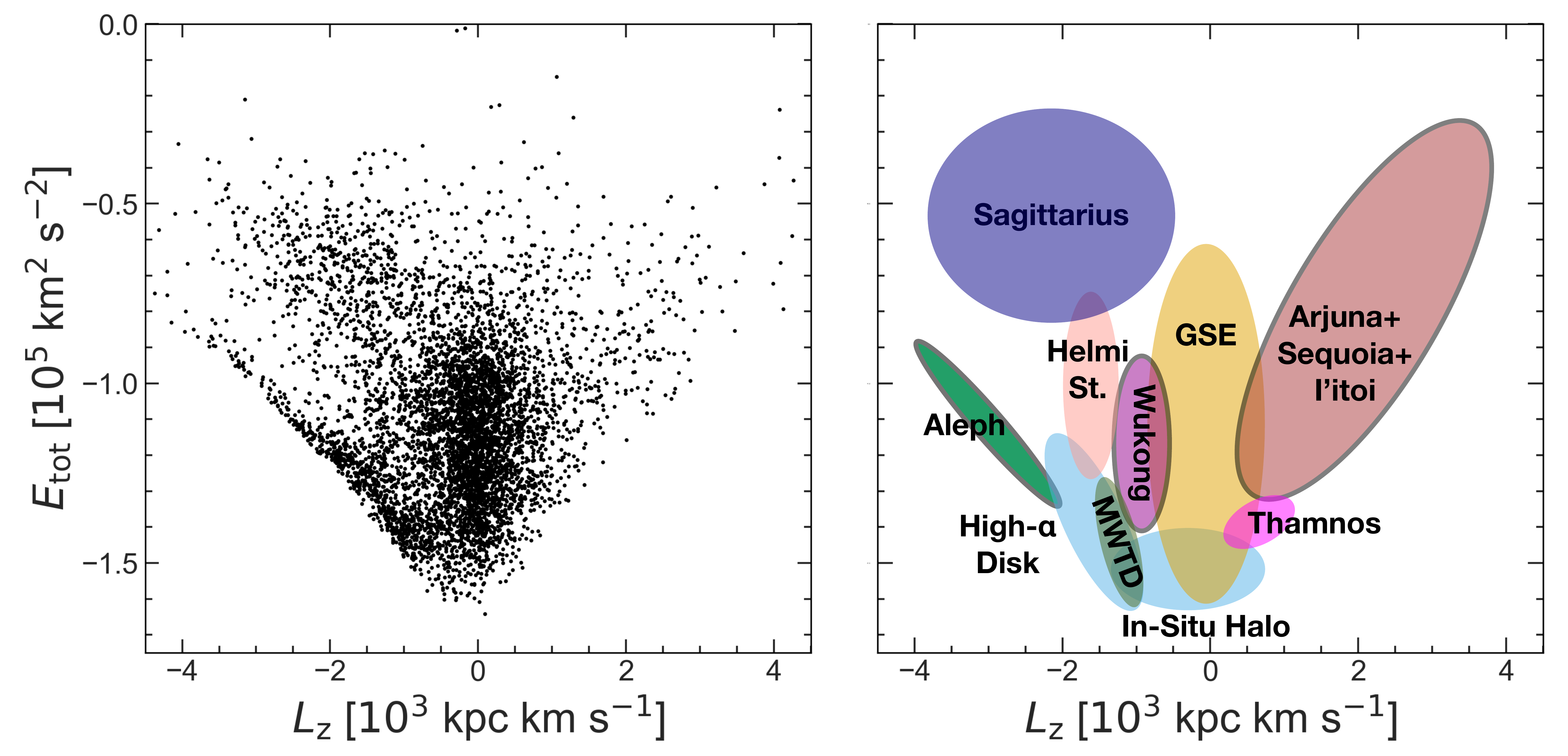}
\caption{An overview of structure in \elz. \textbf{Left} panel same as Figure \ref{fig:data}. In the \textbf{right} panel we provide a schematic of the various structures we will identify in this work (structures highlighted with solid boundaries, except for Sequoia, are new) -- Sagittarius (``Sgr", \S\ref{subsec:sgr}, Fig. \ref{fig:sgr}), Aleph (\S\ref{subsec:aleph}, Fig. \ref{fig:aleph}), the high-$\alpha$ disk and in-situ halo (\S\ref{subsec:insitu}, Fig. \ref{fig:insitu}), \textit{Gaia}-Sausage-Enceladus (``GSE", \S\ref{subsec:ge}, Fig. \ref{fig:ge}), the Helmi Streams (``Helmi St.", \S\ref{subsec:hs}, Fig. \ref{fig:helmi}), Thamnos (\S\ref{subsec:thamnos}, Fig. \ref{fig:thamnos}), Arjuna, Sequoia, and I'itoi (\S\ref{subsec:retrograde}, Fig. \ref{fig:retrograde}), Wukong (\S\ref{subsec:wukong}, Fig. \ref{fig:wukong}), and the metal-weak thick disk (MWTD, \S\ref{subsec:mwtd}, Fig. \ref{fig:MWTD}). There is significant overlap among these structures in chemodynamical space, so in defining and discussing them sequentially it is impossible to avoid referring to objects that are yet to be introduced. We provide this schematic to build a common frame of reference and so readers may notice these structures in figures to come.}
\label{fig:elz_opener}
\end{figure*}

\section{Data and Methods}
\label{sec:data_methods}
\subsection{The H3 Survey}
\label{subsec:H3}
H3 \citep{Conroy19} is the first spectroscopic survey to leverage \textit{Gaia} parallaxes, $\pi$, in its selection of targets.  The selection function of the primary sample is composed of the following conditions: (i) $15<r<18$, (ii) $\pi-2\sigma_{\pi}<0.5$, implying $d_{\rm{helio}}>2$ kpc, (iii) $|b|>30^{\circ}$, to avoid the disk, and (iv) Dec.$>-20^{\circ}$, observable from the MMT located in Arizona, USA. This simple selection function ensures a view of the halo that is free from metallicity biases due to color cuts or metal-poor stellar tracers (e.g., BHBs, RR Lyrae). While H3 will eventually cover $|b|>30^{\circ}$ and the survey selection function requires \textit{Gaia} parallaxes consistent with $d_{\rm{helio}}>2$ kpc, the data presented in this paper is at $|b|>40^{\circ}$, and also limited to $d_{\rm{helio}}>3$ kpc (for reasons outlined in \S\ref{subsec:magcorr}).

Complementing the primary selection, we target a small number ($\approx6\%$ of the final sample used in this work) of color-selected K giants ($\approx5\%$, cuts from \citealt{Conroy18}), BHBs ($\approx1\%$, cuts from \citealt{Deason14}), and RR Lyrae (7 in number, sourced from \citealt{Sesar17rrl}). We take care to appropriately weight these specially targeted stars while accounting for the selection function in \S\ref{subsec:magcorr}. Inspection of stellar parameters of the BHBs reveals that while their distances and radial velocities are robust, their abundances are not reliable, so they are omitted from plots featuring [Fe/H] and [$\alpha$/Fe]. 

The key outputs from the survey are radial velocities precise to $\lesssim$1 km $\rm{s^{-1}}$, [Fe/H] and [$\alpha$/Fe] abundances precise to $\lesssim$0.1 dex, and spectrophotometric distances precise to $\lesssim$10$\%$  (see \citealt{Cargile19} for details on the stellar parameter pipeline). Combined with \textit{Gaia} proper motions (SNR$>$3 for $>$90$\%$ of the sample), H3 thus provides the full 6D phase-space and 2D chemical-space for the sample stars. The survey is ongoing -- $\approx$125,000 targets have been observed as of March 2020 and they form the basis of this work. 

In this paper we focus on an SNR$>3$ sub-sample whose stellar parameters are deemed robust (``\texttt{flag}$=0$" in \texttt{v2.4} of the survey catalogs, but also allowing for BHBs and RR Lyrae). We work only with the primary parallax-selected and secondary color-selected K giant/BHB/RR Lyrae samples described earlier (\texttt{xfit\_rank}$=$ 1 or 2), leaving out the fainter and higher parallax filler targets. We restrict our sample to the 6799 giants ($\log{g}<3.5$) to ensure a relatively uniform view of the halo. The dwarfs, while numerous, are complete only out to $d_{\rm{helio}}\sim10$ kpc, and would require significant selection function corrections (see \S\ref{subsec:magcorr} for details) to be interpreted on the same footing as the giants used in this study.  Visual inspection of the spectra and corner-plots of the stellar parameters suggest metallicities below $-3$ are less reliable at SNR$\approx3$ so we remove the 23 stars that would have otherwise made it into our sample. We further limit this sample as per the considerations in \S\ref{subsec:pot} and \S\ref{subsec:magcorr}.

\begin{figure*}
\centering
\includegraphics[width=\linewidth]{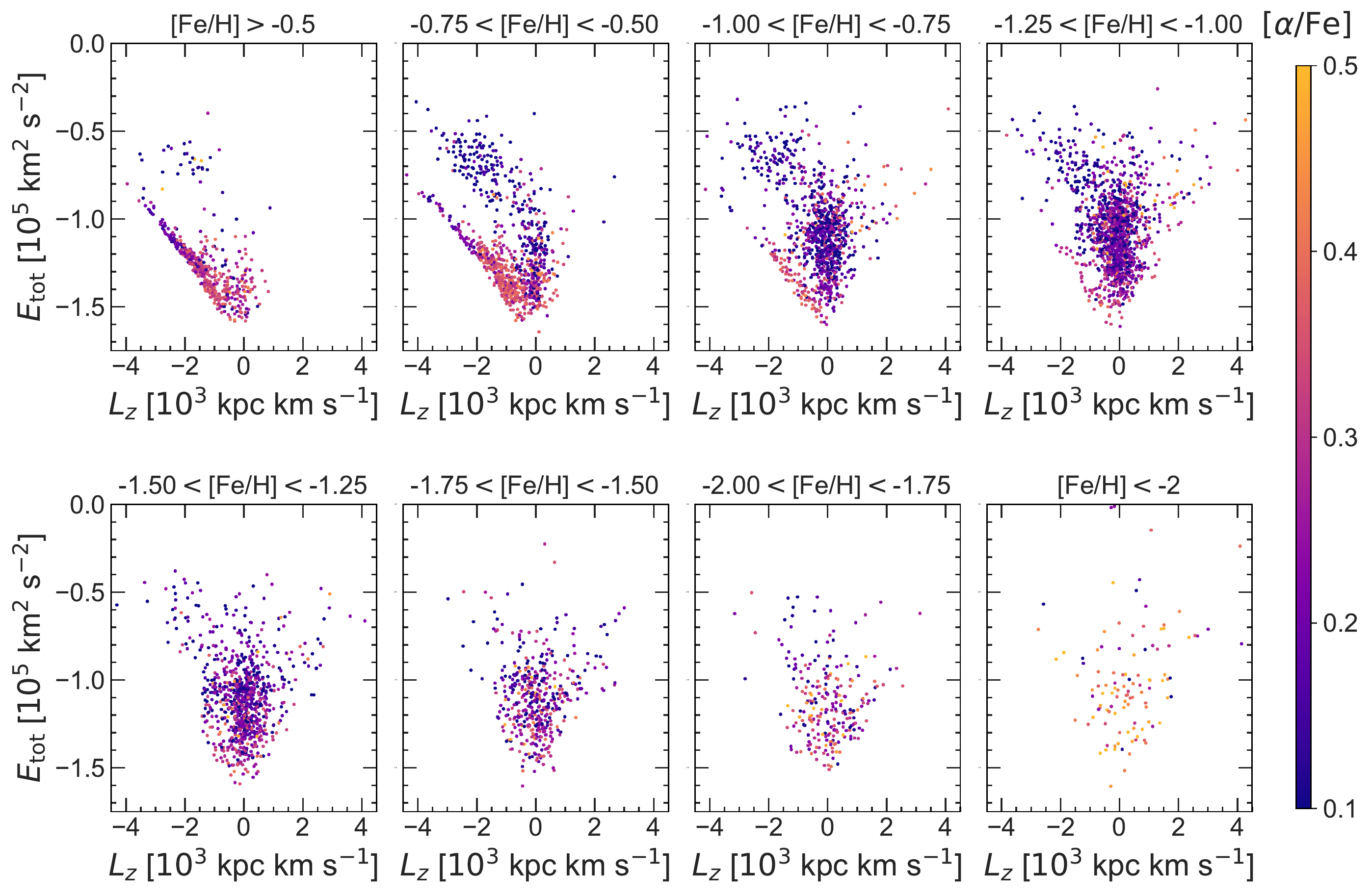}
\caption{\elzs binned by [Fe/H], color-coded by [$\alpha$/Fe], ordered by decreasing [Fe/H]. The most metal-rich bin is largely comprised of stars on disk-like orbits with negative $L_{\rm{z}}$, that extend smoothly to eccentric, $L_{\rm{z}}\sim0$ orbits. The most prograde stars ($L_{\rm{z}}\lesssim-2$) define an $\alpha$-poor sequence confined to the first two panels. Two populations, one centered at $L_{\rm{z}}\sim0$ and another at $E_{\rm{tot}}\sim-0.75$ emerge in the second panel and comprise the bulk of the stars in the remaining panels. High-energy retrograde stars appear at [Fe/H]$<-0.75$. The very metal-poor bins at [Fe/H]$<-1.75$ are sparsely populated, but still clumpy. The most metal-poor bins are not biased to particularly high energies (i.e., larger distances).}
\label{fig:feh_slices}
\end{figure*}

\begin{figure*}
\centering
\includegraphics[width=\linewidth]{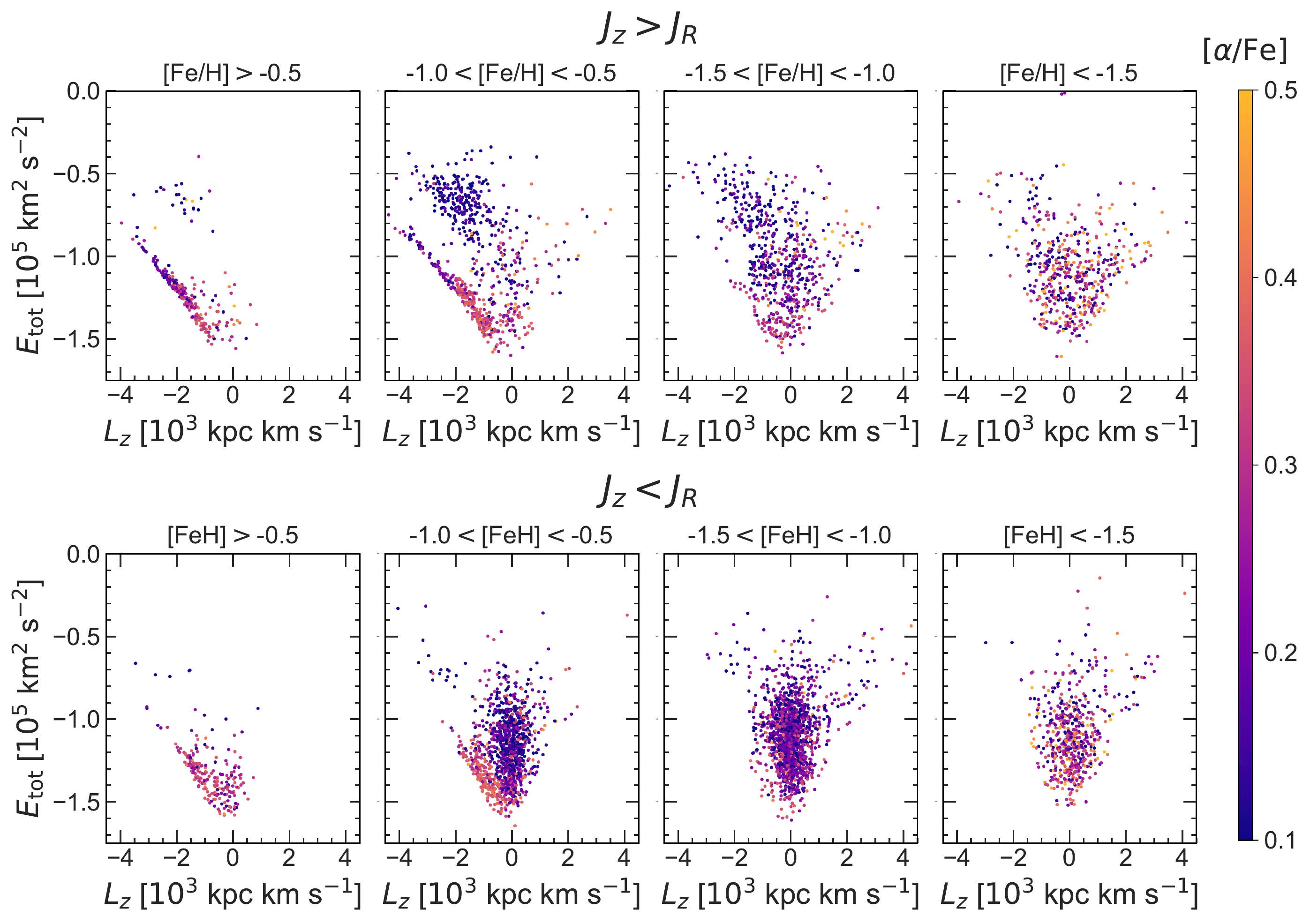}
\caption{\elzs binned by [Fe/H] and actions, color-coded by [$\alpha$/Fe]. \textbf{Top:} Stars with $J_{\rm{z}}>J_{\rm{R}}$, which rise to high elevations off the plane. Some populations (the highly prograde $\alpha$-poor sequence, the stars at $E_{\rm{tot}}\sim-0.75$) are completely confined to $J_{\rm{z}}>J_{\rm{R}}$ orbits, while others (e.g., the high-energy retrograde stars) show no such preference and are equally distributed between $J_{\rm{z}}>J_{\rm{R}}$ and $J_{\rm{z}}<J_{\rm{R}}$. \textbf{Bottom:} Stars with $J_{\rm{z}}<J_{\rm{R}}$ that are on radial or eccentric orbits. The most metal-rich bins show $\alpha$-rich, disk-like stars, that extend to eccentric orbits at $L_{\rm{z}}\approx0$. At lower metallicity a dense cloud of $\alpha$-poor stars appears at $L_{\rm z}\sim0$, with retrograde structure appearing at [Fe/H]$<-1$.} 
\label{fig:feh_orbit_slices}
\end{figure*}

\begin{figure*}
\centering
\includegraphics[width=1.0\linewidth]{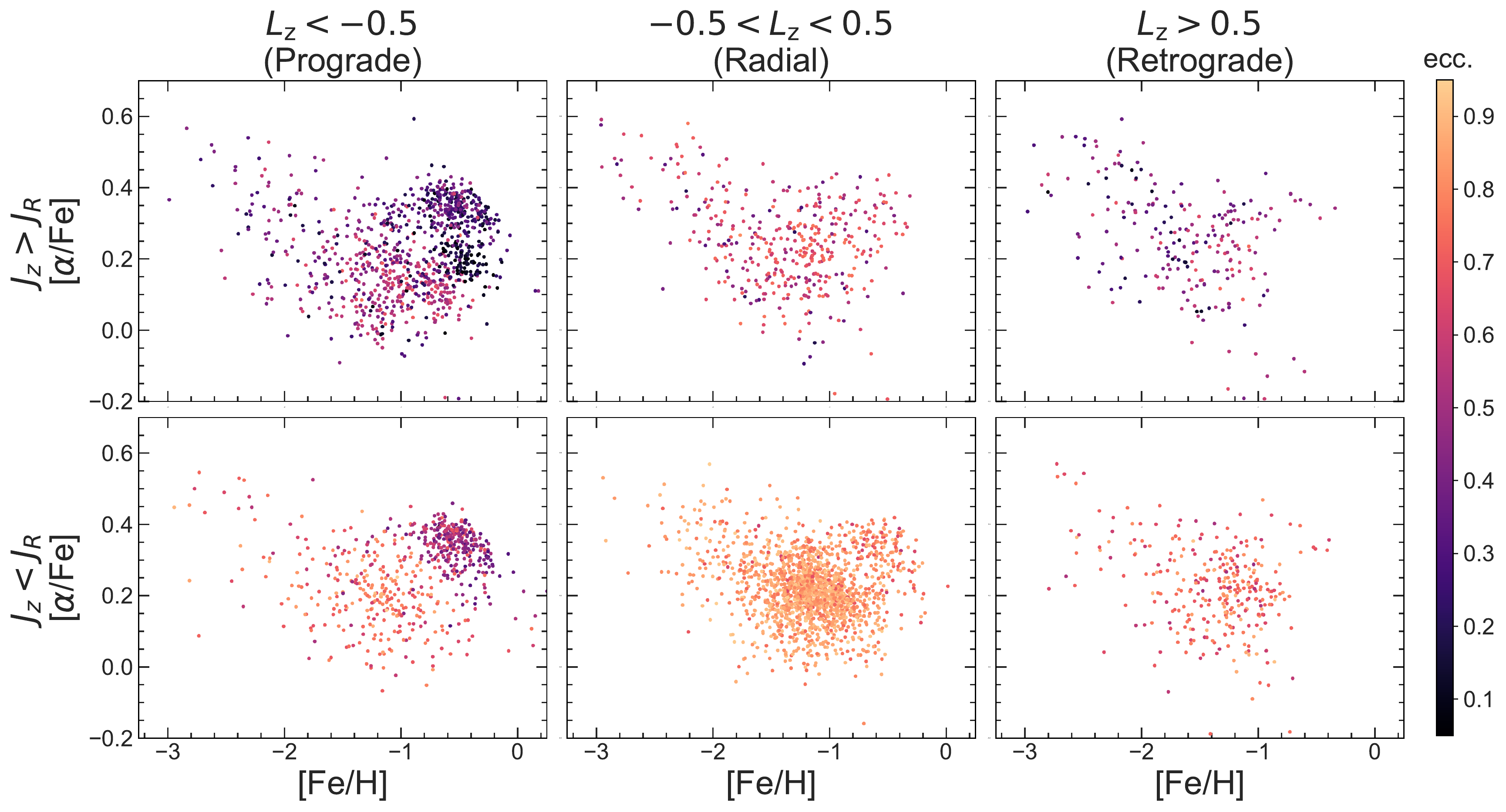}
\caption{[$\alpha$/Fe] vs [Fe/H] color-coded by eccentricity and binned by orbit type -- $L_{\rm{z}}/[10^{3}\rm{\ kpc\ km\ s^{-1}}]<-0.5$ (prograde, left), $-0.5<L_{\rm{z}}/[10^{3}\rm{\ kpc\ km\ s^{-1}}]<0.5$ (radial, center), $L_{\rm{z}}/[10^{3}\rm{\ kpc\ km\ s^{-1}}]>0.5$ (retrograde, right), $J_{\rm{z}}>J_{\rm{R}}$ (top), $J_{\rm{z}}<J_{\rm{R}}$ (bottom). The morphology of chemical space varies strongly with orbit type. We highlight some prominent groups: (i) a metal-rich ([Fe/H]$>-1$), $\alpha$-rich ([$\alpha$/Fe]$>0.25$) population spread across the first two columns (the high-$\alpha$ disk and in-situ halo), (ii) a highly circular ($e<0.2$), metal-rich ([Fe/H]$>-0.75$) population completely confined to the top-left panel (Aleph), (iii) a more eccentric ($e\sim0.5$), relatively $\alpha$-poor population adjacent to Aleph in the top-left panel (Sgr), (iv) a highly eccentric ($e>0.7$), well populated chemical sequence in the bottom-center panel (GSE), (v) an [Fe/H]$\sim-1.2$ retrograde population in the bottom-right panel (Arjuna).}
\label{fig:confusogram_feh}
\end{figure*}

\begin{figure*}
\centering
\includegraphics[width=\linewidth]{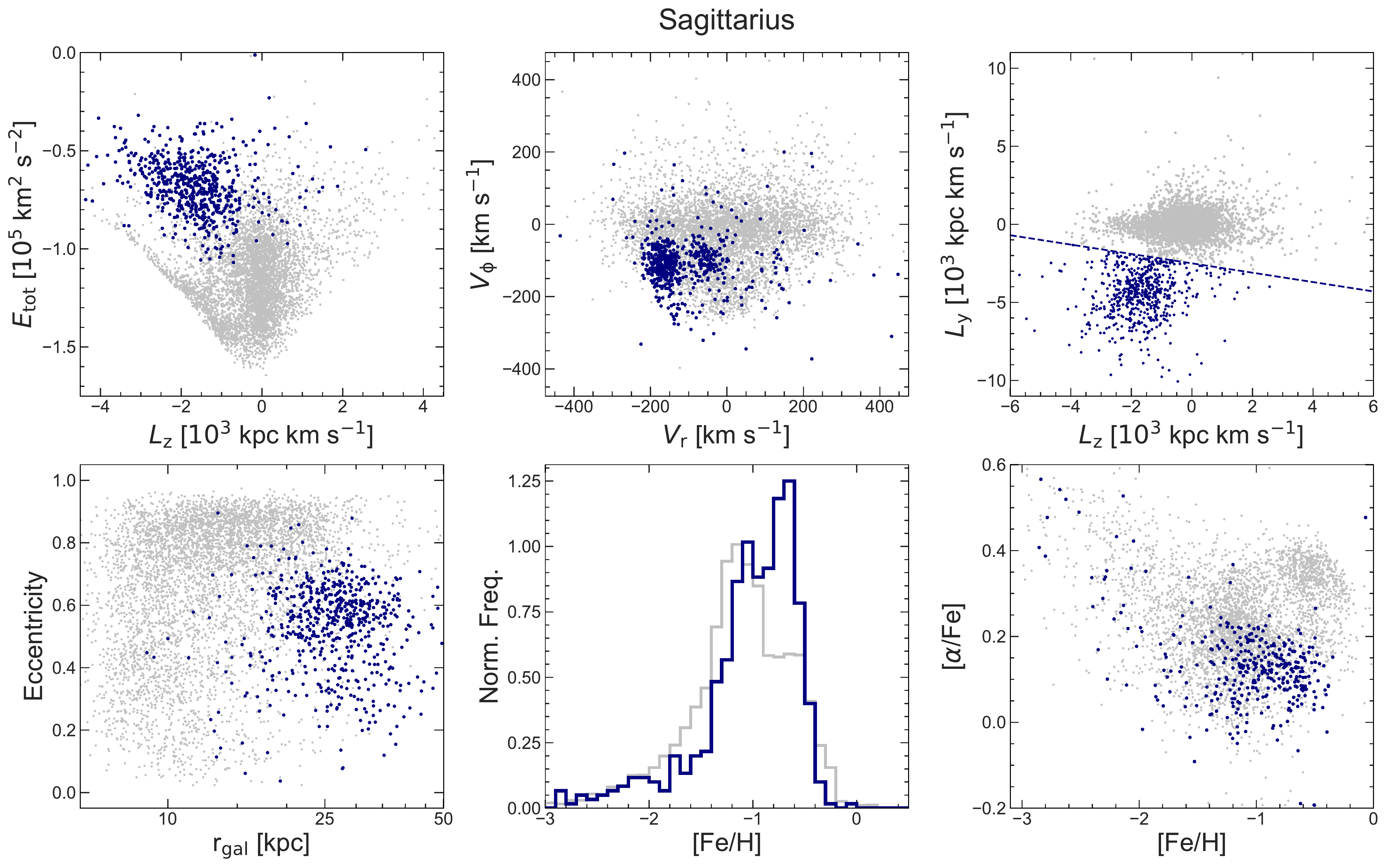}
\caption{Sagittarius in chemodynamical space. Sgr stars are colored navy blue, and the rest of the sample is shown in gray. Panels are as in Figure \ref{fig:data}, except for the top-right, which shows the selection plane of $L_{\rm{z}}$-$L_{\rm{y}}$. Because of its relatively recent accretion, Sgr is highly coherent in phase-space (first four panels). It forms a striking sequence in $L_{\rm{z}}-L_{\rm{y}}$ extending to very negative $L_{\rm{y}}$ (top-right panel), allowing us to make a clean selection using $L_{\rm{y}}$. The leading and trailing arms are visible in $V_{\rm{r}}-V_{\rm{\phi}}$ (top-center) at $V_{\rm{r}}\approx-50, -175$ km s$^{-1}$. The MDF is multi-peaked, with an extended tail to lower metallicity. In [Fe/H] vs [$\alpha$/Fe] Sgr is $\alpha$-poor compared to the halo overall, as expected for a galaxy accreted relatively recently that has had more time for enrichment via Type Ia supernovae.}
\label{fig:sgr}
\end{figure*}

\subsection{Computing Phase-Space Quantities}
\label{subsec:pot}

We adopt the Galactocentric frame implemented in \texttt{Astropy v4.0} \citep{astropy1, astropy2} which has the following parameters: $R_{0}=8.122$ kpc \citep{Gravity19}, $[V_{\rm{R,\odot}}, V_{\rm{\phi,\odot}}, V_{\rm{Z,\odot}}]=[-12.9, 245.6, 7.78]$ km s$^{-1}$ \citep{Drimmel18}, $Z_{\rm{\odot}}=20.8$ pc \citep{BennettBovy19}. This frame is right-handed, i.e., prograde (retrograde) orbits have $L_{\rm{z}}<0$ ($L_{\rm{z}}>0$). 

Potential-related quantities (actions, eccentricities, energies) are computed using \texttt{gala v1.1} \citep{gala1, gala2} with its default \texttt{MilkyWayPotential}. This potential, based on \citet{Bovy15}, is composed of a \citet{Hernquist90} nucleus ($m=1.7\times10^{9}\ \rm{M}_{\rm{\odot}}$, $a=1$ kpc) and bulge ($m=5\times10^{9}\ \rm{M}_{\rm{\odot}}$, $a=1$ kpc), a \citet{MiyamotoNagai75} disk ($m=6.8\times10^{10}\ \rm{M}_{\rm{\odot}}$, $a=3$ kpc, $b=0.28$ kpc), and a spherical \citet{Navarro97} dark matter halo ($m=5.4\times10^{11}\ \rm{M}_{\rm{\odot}}$, $a=15.62$ kpc), where $m$, $a$, $b$ are the characteristic mass and scale radii of these models respectively. The mass enclosed within 200 kpc is $9.9\times10^{11}$ M$_{\rm{\odot}}$ consistent with recent estimates \citep[e.g.,][and references therein]{Zaritsky20}. We also show the final summary plots in the \citet{McMillan17} potential in Appendix \ref{appendix:pot}, both to ease comparison with studies that use this potential \citep[e.g.][]{Myeong19, Koppelman19}, and to demonstrate that the features described in this paper are not specific to our choice of the Galactic potential. 

Orbits are computed using the \citet{DormandPrince} explicit integration scheme, which belongs to the Range-Kutta family of ordinary differential equation solvers, with time-steps of 1 Myr and a total integration time of 25 Gyrs. Eccentricities are computed from these orbits as $e=\frac{r_{\rm{apo}}-r_{\rm{peri}}}{r_{\rm{apo}}+r_{\rm{peri}}}$ where $r_{\rm{apo}}$ and $r_{\rm{peri}}$ are the orbital apocenter and pericenter respectively. Actions ($J_{\mathrm{R}}, J_{\mathrm{\phi}}, J_{\mathrm{z}}, J_{\rm{tot}}=\sqrt{J_{\rm{\phi}}^{2}+J_{\rm{z}}^{2}+J_{\rm{R}}^{2}}$) are estimated from the computed orbits as per the torus-mapping method described in \citet{Sanders16}. We test the robustness of the computed actions by checking that (a) $J_{\mathrm{\phi}}$ converges to $L_{\mathrm{z}}$ within $5\%$, as expected in an axisymmetric potential like the one adopted here, and (b) actions calculated using $75\%$ of the orbit and $100\%$ of the orbit differ by no more than $1\%$. In the few cases ($\approx5\%$) where these conditions are not met (typically long-period orbits for stars at $>20$ kpc), we recompute the actions by extending the 25 Gyr integration period by $2\times$, up to 200 Gyrs (this is a choice made purely for numerical stability to collect a statistical number of orbital periods for long-period orbits). After this, only a small number of bound stars ($\approx50$), mostly with $L_{\mathrm{z}}\approx0$ fail our tests (as expected for very eccentric orbits with $J_{\rm{R}}\sim J_{\rm{tot}}$, Figure 3 of \citealt[][]{Sanders16}) and we exclude them from any analysis involving actions.

The error-budget on phase-space quantities is dominated by uncertainties in spectrophotometric distances and \textit{Gaia} PMs. For an illustration of how measurement errors distort substructure in \elzs we point readers to Appendix \ref{appendix:errorvec}. Since \elzs is a key diagram in the analysis to follow, we limit the sample to stars that satisfy (i) $(|E_{\rm{tot}}|/\sigma_{E_{\rm{tot}}}>3) \land (|L_{\rm{z}}|/\sigma_{L_{\rm{z}}}>3)$, or  (ii) $(\sigma_{E_{\rm{tot}}}<0.1 \times 10^{5} \rm{km}^{2}) \land (\sigma_{L_{\rm{z}}}<0.5\times 10^{3} \rm{kpc\  km\ s^{-1}})$, where ``$\land$" stands for the Boolean ``and" operator. Condition (i) is a relative error cut, and condition (ii) ensures we do not discriminate against low $|L_{\rm{z}}|$ stars that comprise a large fraction of the halo. 1024 stars, a majority of which have uncertain \textit{Gaia} PMs (SNR$<3$) are excised due to these cuts. The excised stars lie at larger distances, including some of our most distant giants, and $\approx300$ of them judging by their PMs are likely members of the Sagittarius stream. The excised stars have an MDF similar to the sample used in this work, except they have fewer metal-rich disk stars (since the excised stars lie at larger distances). We expect improvement in the PM SNR for these stars from future data releases of the \textit{Gaia} mission. This leaves us with a current sample of 5752 giants.

\subsection{Correcting for the H3 Selection Function}
\label{subsec:magcorr}

Every spectroscopic survey has a selection function that can be thought of as the conditional probability  $p\ (obs.|\ \theta)$ that a star with parameters $\theta$ ($l$, $b$, $d_{\rm{helio}}$, age, [Fe/H], [$\alpha$/Fe]...) will be observed by the survey \citep[for a comprehensive overview see][]{Everall20}. In general, the survey selection function represents a biased view of the underlying population. For instance, in a purely magnitude-limited survey, $p\ (obs.)$ is higher for nearby stars, so if one compares the fractions of two accreted structures with different mean heliocentric distances, raw star counts would provide a biased picture. In order to obtain an unbiased view, one must correct for the selection function by using weights that are  proportional to $p\ (obs.|\ \theta)^{-1}$. In what follows, we outline how these weights are computed for our sample.

The H3 selection function can be decomposed into three independent components: (i) where we point the telescope ($|b|>30^{\circ}$, Dec$>-20^{\circ}$), i.e., the ``window selection", (ii) the sample definition ($15<r<18$,  $\pi-2\sigma_{\pi}<0.5$), i.e., the ``magnitude selection" and (iii) the fraction of stars from the input sample that end up with spectra, which leads to a ``targeting selection".

\textit{Window Selection:} H3 is limited to $|b|>30^{\circ}$ by design and to Dec.$>-20^{\circ}$ by geography. Any structures that are anisotropically distributed on the sky will require some correction for the survey window function.  However, correction for the window is difficult as it requires a model for the underlying anisotropy.  In this work we limit ourselves to demonstrating the existence of various substructures, and commenting on their relative contribution to the high-latitude Galaxy sampled within our survey fields.

\textit{Magnitude Selection:} The H3 selection function imposes a magnitude cut ($15<r<18$), which introduces a bias against distant and intrinsically less-luminous sources. We also limit the sample in this work to SNR$>3$, which further discriminates against fainter sources. Further, by restricting the sample only to giants ($\log{g}<3.5$) we are excluding bright, nearby dwarfs that satisfy the magnitude and parallax selections. These effects are illustrated in the top-panel of Figure \ref{fig:magcorr} for an example MIST \texttt{v2.0} isochrone (10 Gyr, [Fe/H]$=-1$, [$\alpha$/Fe]$=0$, \citealt[][]{Choi16}, Dotter et al. in prep).

To correct for this, we sample from an isochrone matched to each star's derived parameters (age, [Fe/H], [$\alpha$/Fe], $A_{\rm{V}}$) using a \citet{Kroupa01} IMF and calculate $f_{\rm{mag}}\ (d_{\rm{helio}})$, the fraction of $\log{g}<3.5$ stars at the star's distance that fall at $15<r<18$. We shrink the magnitude range according to the SNR$>$3 cut-off for each field, which varies with observing conditions. For the subsample of color-selected rare stars (K giants and BHBs, $\approx6\%$ of the sample), instead of calculating $f_{\rm{mag}}\ (d_{\rm{helio}})$ using $15<r<18$ we use the appropriate color cuts and magnitude limits.

The correction weights ($1/f_{\rm{mag}}$) for the example isochrone are plotted as a function of $d_{\rm{helio}}$ in the bottom panel of Figure \ref{fig:magcorr}. The curve has a ``U" shape with a steep rise below 4 kpc and above 35 kpc. The number density of stars predicted by the IMF is high close to the main-sequence turnoff and falls off precipitously as one goes up the red giant branch. At $d_{\rm{helio}}=4-35$ kpc the sections of the red giant branch with the highest number density as well as the red clump are almost entirely contained within $15<r<18$, and so at these distances the correction factor is fairly flat. Importantly, this distance range is where the bulk of our sample ($>90\%$) lies. This means the H3 giants (even without any applied corrections) provide a relatively unbiased view of the halo at these distances.

\textit{Targeting Selection:}
Of all the stars that satisfy our selection function, we assign fibers to $\approx200$ per field. In fields closer to the dense galactic plane, the fraction of stars that are assigned a fiber is lower compared to higher latitudes -- that is, at low $|b|$ the stars in our sample represent a larger underlying population. 
Another targeting bias arises from the higher fiber assignment rank we award to the small number ($<$1 per field on average) of rare, color-selected BHBs and K giants that we complement our main parallax-selected sample with. The higher rank means that fibers are assigned to all possible BHBs and K giants in a field before the other sources. As a result, the median fiber assignment probability for these stars is slightly higher than the main parallax-selected sample ($\approx85\%$ vs $\approx65\%$). Correcting for both these effects is  straightforward. For a given field we compute $f_{\rm{target}} \rm{(rank)}$, the fraction of stars of a given rank that ended up with SNR$>$3 spectra out of all the stars of that rank that satisfied our selection function. For stars of the same rank, $f_{\rm{target}}$ is completely independent of stellar properties and hence can be simply multiplied with $f_{\rm{mag}}$ to produce the total weight.

We note that our approach here is conceptually similar to previous work \citep[e.g.,][]{Bovy14, Stonkute16, Das16, Vickers18, Everall20}. We excise all 68 stars at $d_{\rm{helio}}<3$ kpc from our sample due to their very high weights (see Fig. \ref{fig:magcorr}), and because we are interested in the distant Galaxy, leaving us with a final sample of 5684 stars. In the summary plots where we interpret the relative fractions of various substructures (Figs. \ref{fig:summary1}, \ref{fig:accretedinsitu}, \ref{fig:FeHaFe}), and in Table \ref{table:summary} we employ weights equal to $(f_{\rm{mag}}f_{\rm{target}})^{-1}$. In all other figures we display raw counts. The distinction between raw and weighted quantities is made explicit throughout the text.

\section{Results}
\label{sec:results}

\subsection{Overview of the High-Latitude Galaxy}
\label{subsec:results_overview}

We begin with a general overview of the data in chemistry and integrals of motion (\elz, actions, eccentricity) to motivate the selection criteria for various structures in the sections to come. Figure \ref{fig:data} introduces the projections of phase-space and chemistry we will use frequently. Figure \ref{fig:elz_opener} provides an \elzs ``map" that identifies structure that will be presented in subsequent sections. We do this so readers can see the entire landscape at once, which will be helpful as we discuss individual structures in depth. Figures \ref{fig:feh_slices}, \ref{fig:feh_orbit_slices}, \ref{fig:confusogram_feh} present a high-level overview of features in \elz, actions and chemistry.

In Figure \ref{fig:data}, the top-left panel shows \elz, which we use as our primary workspace. It has long been recognized that groups of stars accreted together display coherence in their energies, and in the $z$-component of their angular momentum, even when they are thoroughly dispersed in configuration space \citep[e.g.,][]{Helmi00,Brown05,Gomez10, Gomez13,Simpson19}. In the second panel we display $V_{\rm{r}}$ vs. $V_{\rm{\phi}}$ -- in this space stars on disk-like orbits intuitively occupy the region around the assumed rotation velocity of the Sun. This is also the space in which GSE was discovered by \citet{Belokurov18} as an overdensity of stars around $V_{\rm{\phi}}=0$ that is also prominent in our data. In the third panel we depict a summary of actions in the form of $(J_{\rm{z}}-J_{\rm{R}})/J_{\rm{tot}}$ vs. $J_{\phi}/J_{\rm{tot}}$ following \citet{BT,Vasiliev19}. Generally, stars on very radial or eccentric orbits are confined to the bottom half of this diagram while stars on polar or circular ($J_{\rm{R}}=0$) orbits occupy the top half of this diagram. Circular, in-plane disk orbits have $J_{\rm{z}}=J_{\rm{R}}=0$, and $J_{\rm{\phi}}/J_{\rm{tot}}=-1$. Purely planar ($J_{\rm{z}}=0$) orbits, that would also occupy the bottom half of this diagram, are under-represented in our $|Z_{\rm{gal}}|>2$ kpc sample. A similar diagram was used by \citet{Myeong19} to discover the retrograde, accreted Sequoia structure, and is a useful way to isolate GSE since it is largely confined to $J_{\rm{z}}<J_{\rm{R}}$ orbits. Eccentricities (bottom-left panel of Figure \ref{fig:data}) are similarly useful, in that GSE is almost completely confined to $e>0.7$. Plotting eccentricity vs $r_{\rm{gal}}$ also shows abrupt changes in the density of stars around the pericenter/apocenter of various structures.

While actions are useful, we favor \elzs while defining selections in part because this space is simpler to understand. Further, a large body of local halo studies has primarily deployed energy, eccentricities, and angular momenta, and we seek to draw direct connections and build on it \citep[e.g.,][]{Helmi17,Belokurov18,Koppelman19}. A high degree of overlap of multiple accreted structures is expected in \elzs and other projections of phase-space \citep[e.g.,][]{bj05_3,Jean-Baptiste17,Pfeffer20}, which we resolve when possible using chemistry (bottom-center and bottom-right panels of Figure \ref{fig:data}). Stars belonging to the same structure are expected to show coherent MDFs and distinct chemical evolutionary tracks in the [Fe/H] vs. [$\alpha$/Fe] plane that are a function of their mass, star-formation history and formation redshift (discussed further in \S\ref{subsec:chemistry}).

In Figure \ref{fig:feh_slices} we show \elzs in bins of metallicity and color-coded by [$\alpha$/Fe]. The most metal-rich stars define two sequences: one at higher energy, lower [$\alpha$/Fe] (Aleph), and the other at lower energy, higher [$\alpha$/Fe] (the high-$\alpha$ disk) that extends to $L_{\rm{z}}\sim0$ orbits (the in-situ halo). At $-0.75<$[Fe/H]$<-0.5$ two structures appear, one centered at $L_{\rm{z}}\sim0$ (GSE), and the other at high energy, $L_{\rm{z}}\sim-2$, $E_{\rm{tot}}\sim-0.75$ (Sgr). The density of the $L_{\rm{z}}\sim0$ population (GSE) peaks in the $-1.25<$[Fe/H]$<-1$ panel. High-energy retrograde stars only begin to appear at [Fe/H]$<-0.75$, and are almost entirely absent from the higher metallicity bins. Several smaller clumps appear in various [Fe/H] intervals. From this figure it is already clear that a very small fraction of the halo within 50 kpc is metal poor ([Fe/H]$\leq-1.75$). 

Figure \ref{fig:feh_orbit_slices} is similar to Figure \ref{fig:feh_slices}, but here we separate the stars by actions.  Stars with $J_{\rm{z}}>J_{\rm{R}}$, on polar or circular orbits are limited to the top row and stars with $J_{\rm{z}}<J_{\rm{R}}$, on radial or eccentric orbits are limited to the bottom row. The most prograde structure ($L_{\rm{z}}<-2$) is $\alpha$-poor and confined to the first two panels (Aleph). The high-energy population at $L_{\rm{z}}\sim-2$, $E_{\rm{tot}}\sim-0.75$ is also confined to the top-row (Sgr). A prominent structure centered at $L_{\rm{z}}\sim0$ appears largely at $J_{\rm{z}}<J_{\rm{R}}$ and dominates the bottom row (GSE). The other disk-like prograde population that extends to $L_{\rm{z}}\sim0$ is spread across the top and bottom rows (high-$\alpha$ disk and in-situ halo), as are the high-energy retrograde halo stars. 

Figure \ref{fig:confusogram_feh} depicts [$\alpha$/Fe] vs [Fe/H], binned by actions and angular momenta, and color-coded by eccentricity. This figure is particularly rich in structure, and underscores the power of combining chemistry with dynamics. We highlight a few prominent populations apparent in this figure. There is a highly circular (shaded black), metal-rich ([Fe/H]$<-0.5$), $J_{\rm{z}}>J_{\rm{R}}$ population completely contained within the top-left panel (Aleph). Adjacent to it, at lower [Fe/H] and [$\alpha$/Fe], appears an agglomeration of more eccentric stars that is also confined purely to the top-left panel (Sgr). The most $\alpha$-rich population is dispersed across the first two columns, and has orbits ranging from highly eccentric to circular, and extends from prograde to radial $L_{\rm{z}}$ (high-$\alpha$ disk + in-situ halo). Among the $J_{\rm{z}}<J_{\rm{R}}$ orbits we see a well-populated sequence largely contained within the bottom-center panel (GSE). 

Through these figures we have demonstrated the distant halo to be highly structured in chemodynamical space, with various populations appearing preferentially in certain regions of metallicity and orbital space.  We now proceed to define and characterize these individual structures in detail.

\subsection{Substructure Inventory}
\label{subsec:inventory}

In what follows, we provide a detailed inventory of the $|b|>40^{\circ}, d_{\rm{helio}}>3$ kpc Milky Way, one component at a time. We provide relevant background on each component, justify our selection, and comment on any noteworthy features. We support this discussion with a corresponding 6-panel figure for each component that situates it in chemodynamical space. Each 6-panel figure follows the layout introduced in Figure \ref{fig:data}, with the top-right panel changing across figures to highlight a particular projection of chemodynamical space most relevant for the structure under discussion. We emphasize that the primary goal of this work is a high-level inventory. This means we focus on cleanly selecting various components rather than on a thorough characterization and analysis of their nature, which we defer to forthcoming work.

We first outline our overall strategy. We begin by selecting the most coherent, well-defined structures in chemodynamical space -- Sagittarius, Aleph, the high-$\alpha$ disk $\&$ and the in-situ halo. Having accounted for the eccentric stars of the in-situ halo and Sgr, we assign the remaining highly eccentric ($e>0.7$) stars to GSE. Next, we isolate other known halo structures in the literature (the Helmi Streams, Thamnos, Sequoia). While investigating Sequoia in the high-energy retrograde halo we identify a relatively metal-poor (I'itoi) and metal-rich population (Arjuna) in the same \elzs region. After subtracting out all these structures, we turn to a remaining prograde \elzs overdensity (Wukong). We also highlight a metal-poor, $\alpha$-rich, rotationally supported population that we identify as the metal-weak thick disk. The remaining stars are labeled unclassified debris.

While selecting a structure we exclude all the previously defined structures. This ensures that new structures (Arjuna, I'itoi, Wukong) have minimal overlap with previously identified ones. We often rely on chemistry in our selections due to the high degree of overlap expected for accreted structures in integrals of motion \citep[e.g.,][]{bj05_3, Jean-Baptiste17,Pfeffer20}. For instance, the low eccentricity tail of GSE ($e<0.7$) is a major contaminant in purely phase-space selections of lower-mass objects, but we are able to exclude it by appealing to chemistry. As much as possible, we incorporate insights from the existing literature in our selections -- for instance, for the Helmi Streams and Thamnos we use literature definitions as our starting point, and for Sequoia we are guided by previous studies of its chemistry.

Instead of using clustering algorithms \citep[e.g.,][]{Yuan18, Yuan20,Mackereth19,Koppelman19}, we take an artisanal approach, making simple, physically motivated, easily reproducible selections.  Through extensive experimentation we have found that clustering algorithms (e.g., DBSCAN, HDBSCAN, k-means) either fracture the space into too many clusters, or assign the entire sample to Sagittarius, GSE, the high-$\alpha$ disk, and Aleph (i.e., the structures apparent by eye in Figures \ref{fig:feh_slices}, \ref{fig:feh_orbit_slices}, \ref{fig:confusogram_feh}). In the case of a high degree of fracturing, we then had to consider one at a time the nature of each mini-cluster, akin to \citet[][]{Yuan20} whose algorithm applied to [Fe/H]$<-1.8$ stars in LAMOST yielded 57 distinct groups, almost all of which they coalesced back into GSE and Sequoia. A downside of our approach compared to clustering methods is the deterministic assignment of every star as belonging to one structure or another instead of assigning membership probabilities and marginalizing over them (also discussed in \S\ref{subsec:caveats}).

\subsubsection{Sagittarius}
\label{subsec:sgr}

The stream of debris associated with the Sagittarius (Sgr) dwarf spheroidal galaxy \citep{Ibata94} provides the clearest demonstration of the hierarchical build-up of the stellar halo. In recent years Sgr debris has been traced out to $\sim100$ kpc, showing surprising features \citep[e.g.,][]{Belokurov14, Hernitschek17,Sesar17,Li19}, and inspiring a new generation of numerical models \citep[e.g.,][]{Dierickx17,Fardal19,Laporte18} that builds on earlier work \citep[e.g.,][]{Johnston95, LM05,LM10}. In tandem, its chemistry is beginning to be resolved in ever greater detail by large spectroscopic efforts \citep[e.g.,][]{Alfaro-Cuello19, Li19, Hayes20} that are building on earlier efforts \citep[e.g.,][]{Bellazzini06, Chou07, Moncao07, Carlin12,Gibbons17}.

Before \textit{Gaia}, studies made the best of incomplete phase-space data to select Sgr stream stars, typically relying on heuristics such as distance from the orbital plane \citep[e.g.,][]{Newberg03, Belokurov14, Lancaster19}. However, with full phase-space information, clean selections that fully exploit the highly coherent Sgr features are now possible \citep[e.g.,][]{Li19,Yang19,Hayes20}. Sgr is a high-energy, prograde overdensity in $E-L_{\mathrm{z}}$, and owing to its polar orbit, its angular momentum is concentrated in $L_{\rm{y}}$. Capitalizing on this, we define Sgr stars as those which satisfy:

\begin{equation}
\label{eq:sgr}
\begin{aligned}
L_{\rm{y}} < -0.3L_{\rm{z}}-2.5\times10^{3}\ \rm{kpc\ km\ s^{-1}}.
\end{aligned}
\end{equation}

\noindent
We verify that this simple criterion selects $>99.5\%$ of the star particles in a version of the \citet{LM10} model that is matched to the current H3 footprint with a $10\%$ distance uncertainty (see \citealt[][]{Johnson20} for more detailed comparisons with models). The selected 612 stars are shown in Figure \ref{fig:sgr} -- Sgr comprises the  majority of our distant stars. Additionally, we identified 63 stars that do not satisfy Eq. \ref{eq:sgr} but have PMs highly aligned with stars selected by Eq. \ref{eq:sgr}. Closer inspection revealed that the distances to these stars are incorrect, due to confusion between the red clump and red giant branch \citep[][]{Masseron17,Mackereth17}. When the distances to these stars are doubled, they satisfy Eq. \ref{eq:sgr}. In our pipeline this confusion arises for low-$[\alpha\rm{/Fe}]$ stars ($<0.1$) where Sgr is the dominant structure. These stars have a similar magnitude distribution to those selected by Eq. \ref{eq:sgr}, so when computing relative fractions (e.g., in Table \ref{table:summary} and Figure \ref{fig:summary2}) we adjust our Sgr numbers upwards by $10\%$ (i.e., 63/612) but do not show these stars in projections of phase-space.

A detailed characterization of Sgr in H3 is forthcoming \citep{Johnson20}. Here we only remark on prominent features in the MDF, which displays two peaks in [Fe/H] separated by $\approx 0.4$ dex. This is consistent with the picture in \citet[][their Fig. 7]{Hayes20}, who find different mean metallicities in the leading and trailing arms and a similarly multi-peaked MDF. There may also be a link to the complex star-formation history and distinct chemical populations recently shown to exist in the core of the Sgr dwarf \citep{Alfaro-Cuello19}. Stars in the metal poor ([Fe/H]$\lesssim-2$) tail are highly aligned with Sgr in angular momenta as well as in \textit{Gaia} proper motions (that are independent of the measured distances and radial velocities) and will be a point of focus of \citet[][]{Johnson20}.

\subsubsection{Aleph}
\label{subsec:aleph}

\begin{figure*}
\centering
\includegraphics[width=\linewidth]{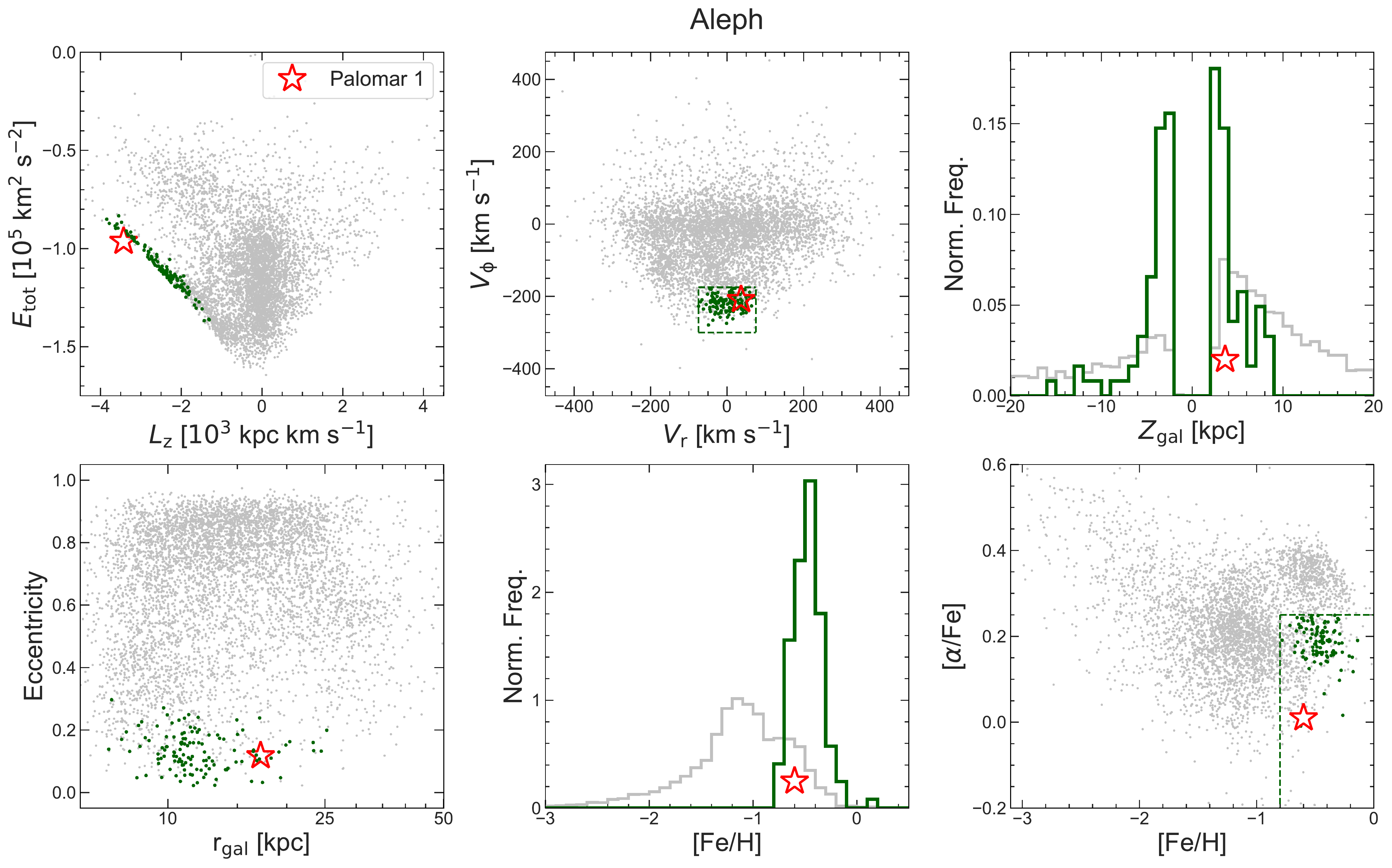}
\caption{Aleph (dark green) in chemodynamical space. Panels are as in Figure \ref{fig:data}, except for the top-right, which shows the distribution of $Z_{\rm gal}$.  Dashed green lines indicate the selection planes ($V_{\rm{r}}-V_{\rm{\phi}}$, and [$\alpha$/Fe] vs. [Fe/H]). The globular cluster Palomar 1 is represented in red. Aleph is rapidly rotating, circular ($e<0.3$), metal-rich, and relatively $\alpha$-poor. It is at higher energy than the high-$\alpha$ disk and is clearly distinct in chemistry (see also top-left panel of Figure \ref{fig:confusogram_feh}). Aleph extends up to $\approx 10$ kpc off the plane. Palomar 1 and Aleph share many chemodynamical properties in common, suggesting a possible association.}
\label{fig:aleph}
\end{figure*}

\begin{figure*}
\centering
\includegraphics[width=\linewidth]{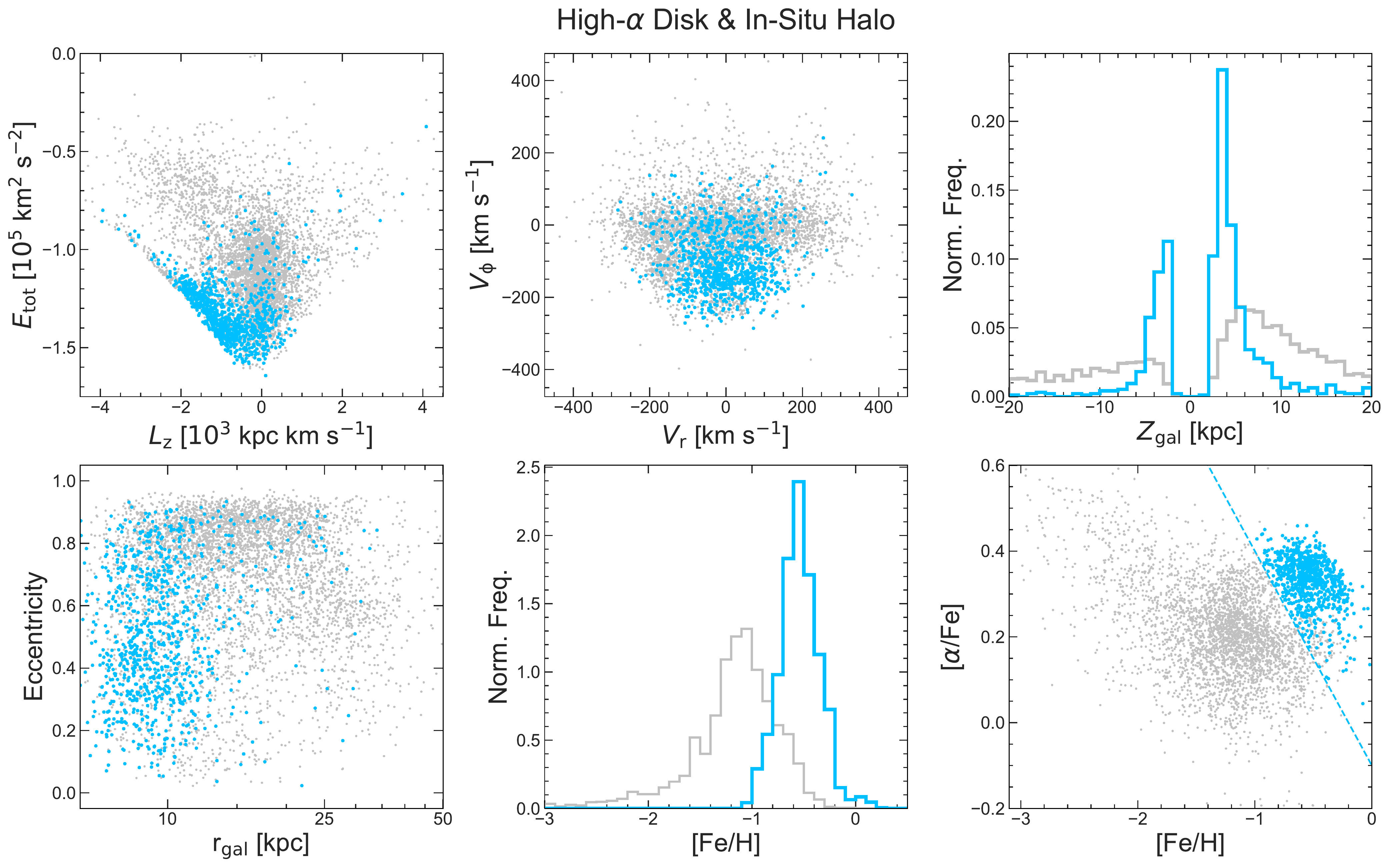}
\caption{High-$\alpha$ disk and in-situ halo (light blue) in chemodynamical space. Panels are as in Figure \ref{fig:data}, except for the top-right, which shows the distribution of $Z_{\rm gal}$.  [Fe/H] vs [$\alpha$/Fe] is the selection plane. Stars with chemistry resembling the high-$\alpha$ disk span the full range of eccentricities. In \elzs the rotationally-supported stars form the diffuse, inclined sequence while the eccentric stars have $L_{\rm{z}}\sim0$. In $V_{\rm{r}}-V_{\rm{\phi}}$ the low eccentricity stars lie at negative $V_\phi$ while the eccentric stars lie on the $V_{\rm{\phi}}\sim0$ locus coincident with GSE. The in-situ halo extends all the way out to a remarkable $r_{\rm{gal}}\approx 25$ kpc and $|Z_{\rm{gal}}|\approx20$ kpc.}
\label{fig:insitu}
\end{figure*}

\begin{figure*}
\centering
\includegraphics[width=\linewidth]{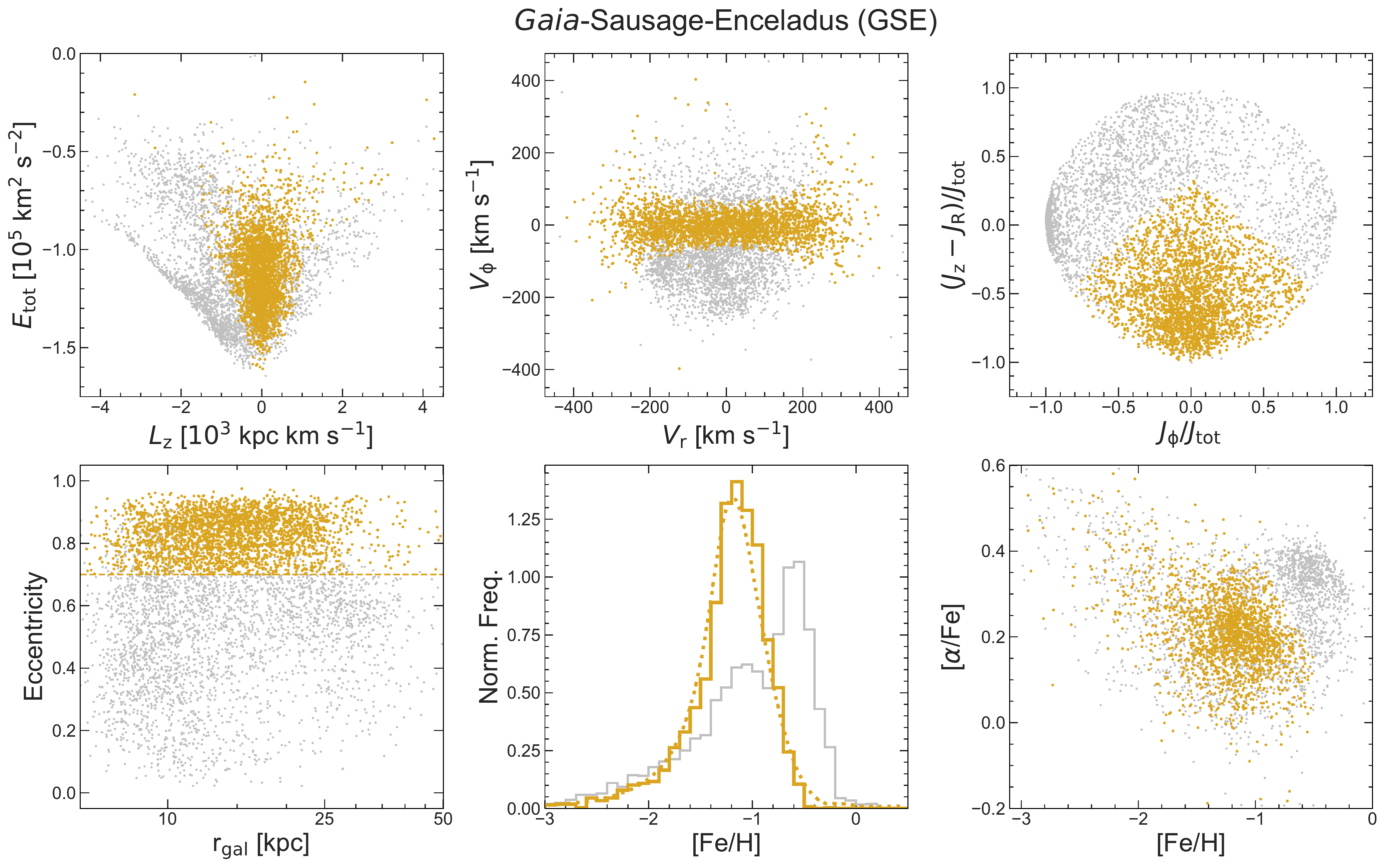}
\caption{\textit{Gaia}-Sausage-Enceladus (GSE, gold) in chemodynamical space. Panels are as in Figure \ref{fig:data}, except for the top-right, which shows the distribution of stars in action space.  GSE is selected on eccentricity ($e>0.7$) motivated by the dense population of stars in the bottom-left panel. The smooth, unimodal MDF is well-fit by a simple chemical evolution model (dotted line in MDF panel) that also reproduces the tail to low [Fe/H]. The highly eccentric GSE stars map to various projections of phase-space as overdensities at $L_{\rm{z}}\sim0$, $V_{\rm{\phi}}\sim0$, and $J_{\rm{z}}-J_{\rm{R}}<0$.}
\label{fig:ge}
\end{figure*}

Aleph\footnote{Named for its prominence in \afe$-$\feh; see Figure \ref{fig:confusogram_feh}.} is a hitherto unknown prograde substructure. We discovered Aleph in \afe$-$[Fe/H] as a sequence below the high-$\alpha$ disk at similar [Fe/H]. Examining the dynamics of these stars, we found them to be highly coherent and on circular orbits (Figure \ref{fig:confusogram_feh}), comprising the most prograde stars of our sample at higher-energy than the high-$\alpha$ disk in \elzs (Fig. \ref{fig:feh_orbit_slices}). Another characteristic feature of Aleph that clearly differentiates it from the canonical disk populations is its significant vertical action, which is seen prominently in Fig. \ref{fig:confusogram_feh}, where Aleph is completely confined to the top-left panel depicting prograde, $J_{\rm{z}}>J_{\rm{R}}$ stars. The classic $\alpha$-rich and $\alpha$-poor disk sequences typically have $J_{\rm{z}}<J_{\rm{R}}$ \citep[e.g.,][]{Sanders16,Beane19,Ting19}.

We define Aleph stars as follows:
\begin{equation}
\begin{aligned}
(V_{\rm{\phi}} < -175\ \rm{km\ s^{-1}})\ \land\ (V_{\rm{\phi}} > -300\ \rm{km\ s^{-1}})\\
\land\ (|V_{\rm{r}}|<75\ \rm{km\ s^{-1}})\\
\land\ ([\rm{Fe/H}]>-0.8) \land ([\alpha/\rm{Fe}]<0.27)\\
\land\ (\rm{excluding\ all\ previously\ defined\ structures}).
\end{aligned}
\end{equation}

\noindent
The resulting population is shown in Figure \ref{fig:aleph}. In our sample Aleph is localized spatially ($r_{\rm{gal}}=11.1^{+5.7}_{-1.6}$ kpc), with stars extending to 25 kpc. It is a metal-rich ([Fe/H]$=-0.51$), relatively alpha-poor ([$\alpha$/Fe]$=0.19$), rapidly rotating ($V_{\rm{\phi}}\approx-210$ km s$^{-1}$) structure on a highly circular orbit ($e=0.13\pm0.06$) with a strong vertical action ($\langle J_{z}\rangle\approx190$ kpc km s$^{-1}$) and orbits that rise to $|Z_{\rm{gal}}|\approx10$ kpc. All quoted values have been weighted by the selection function.

The low eccentricity and chemistry of Aleph suggest an origin within the Galactic disk. Interestingly, in our sample Aleph is mostly confined to the Galactic anti-center, where several overdensities linked to the excitation of the outer disk (e.g., Monoceros, A13, TriAnd1, TriAnd2) have been observed \citep[e.g.,][]{Newberg02,Ivezic08,Price-Whelan15,Li17,Bergemann18}, though our sample is at $|b|>40^{\circ}$, at slightly higher latitudes than these features. Several of Aleph's properties -- the radial extent, chemical nature, rotational velocity -- are also similar to recently reported features of outer disk stars in \citet[][]{Lian20}. It is possible that the \citet[][]{Lian20} APOGEE sample is the in-plane view of Aleph, while we are sampling it at higher latitudes. A detailed exploration of Aleph's nature is the subject of ongoing work. 

Aleph is coincident with the enigmatic GC Palomar 1 (Pal 1, red star in Fig \ref{fig:aleph}) in integrals of motion, is at very similar elevation ($Z=3.6$ kpc) and similar metallicity ([Fe/H]$\approx-0.5$), but is less $\alpha$-enhanced (Pal 1: \afe$\approx0$, Aleph: \afe$\approx0.2$). We adopt Pal 1 phase-space coordinates from \citet[][]{Baumgardt19} and abundances from \citet{Sakari11}. Since its discovery Pal 1 has been recognized as a curiosity -- its high elevation resembles halo GCs but its young age and high metallicity have proven puzzling ($4-7$ Gyrs, and among the youngest, most metal rich, and faintest of MW GCs, e.g., \citealt{vanderbergh04,Sakari11,Sarajedini11}). 

For decades authors have speculated about its origin, wondering whether it may be an unusually old open cluster, may have a peculiar IMF, or may have been accreted with a dwarf galaxy \citep[e.g.,][]{Rosenberg98, Rosenberg98b, Niederste-Ostholt10}. In recent years the accretion origin has gained currency. Other young ($5-8$ Gyrs), low surface brightness GCs  (e.g,. Terzan 7, Pal 12, and Whiting 1) have been associated to Sgr, i.e, they are of extragalactic origin \citep[e.g.,][]{Carraro07,Koposov07, LM10, Johnson20}. \citet[][]{Sakari11} analyzed neutron capture elements in four stars in Pal 1, and found them to be distinct from MW field stars, which led them to argue Pal 1 was accreted along with a dwarf galaxy. Pal 1 also does not lie on the in-situ branch of MW GCs in the age-metallicity relation (\citealt{Forbes10, Forbes20}; \citealt{Kruijssen20}; but see \citealt{Massari19}).  Whether or not Pal 1 was accreted or born in-situ, the similarity between Pal 1 and Aleph in chemodynamical space suggests a common origin.

\subsubsection{High-$\alpha$ Disk and In-situ Halo}
\label{subsec:insitu}

It has long been known that stars on disk-like orbits lie on one of two chemical sequences characterized by low or high values of \afe\ \citep[e.g.,][]{Edvardsson93,Fuhrmann98,Chen00,Bensby03,Adibekyan12}. With \textit{Gaia} data it was realized that the high-$\alpha$ population extends to higher eccentricities than a conventional disk-like population.  This high-$\alpha$, high eccentricity population has been dubbed the ``in-situ halo" and later as the ``Splash" \citep[e.g.,][]{Bonaca17, Bonaca20, Haywood18, DiMatteo19,Amarante20,Belokurov20}. Simultaneously, a link between the accretion of GSE, the formation of the high-$\alpha$ disk, and the creation of the in-situ halo has been proposed \citep[e.g.,][]{Helmi18, Gallart19, Belokurov20, Bonaca20}. Characterizing this component of the halo is thus critical to understanding the Galaxy's earliest epoch.

We define the high-$\alpha$ disk and in-situ halo stars relying purely on chemistry:
\begin{equation}
\label{eq:td}
\begin{aligned}
\mathrm{[\alpha/Fe]} > 0.25-0.5\ (\mathrm{[Fe/H]}+0.7)\\
\land\ (\rm{excluding\ all\ previously\ defined\ structures}).
\end{aligned}
\end{equation}

In Figure \ref{fig:insitu} we see these stars form a kinematic population that extends continuously from rotationally supported orbits (forming a locus at lower $|V_{\rm{\phi}}|$ than Aleph) to highly eccentric ones. In \elzs the high-$\alpha$ disk forms a more diffuse track slightly steeper than Aleph, which extends into high-eccentricity orbits with $L_{\rm{z}}\sim0$. The continuity of the distribution in phase-space supports scenarios in which the ancient, rotationally supported high-$\alpha$ disk was dynamically heated, perhaps by a merger. The in-situ halo extends to $r_{\rm{gal}}\approx25$ kpc, but we caution that the stars at $|Z|>15$ kpc lie very close to the selection boundary in chemistry, and may belong to other structures. We discuss the physical origin of the in-situ halo further in \S\ref{subsec:origin}.

This selection excludes the metal-poor tail of the high-$\alpha$ disk and the in-situ halo \citep[e.g.,][]{Carollo19} which lies in a region of [Fe/H] vs [$\alpha$/Fe] coincident with GSE and other accreted structures. We will return to these stars in the sections dealing with the metal-weak thick disk (\S\ref{subsec:mwtd}) and the unclassified debris (\S\ref{subsec:leftovers}).

\subsubsection{Gaia-Sausage-Enceladus (GSE)}
\label{subsec:ge}
We define “GSE” as the highly radial population that comprises the bulk of the accreted local halo. This population was identified in various ways by \citet[][]{Belokurov18, Koppelman18,Myeong18,Haywood18, Helmi18,Mackereth19,Koppelman19, Helmi20}. Different selections result in differing degrees of contamination with overlapping structures \citep[see discussion in][]{Evans20}. 

We select GSE stars by excluding the previously defined structures and requiring $e>0.7$. The eccentricity selection is motivated by the dense cloud of stars at $e>0.7$ seen in eccentricity vs $r_{\rm{gal}}$, whose density sharply drops off at $\approx30$ kpc (corresponding to the proposed apocenter of GSE; \citealt{Deason18,Lancaster19}).  Our GSE selection is therefore simply:
\begin{equation}
\begin{aligned}
(e>0.7)\\ \land\ (\rm{excluding\ all\ previously\ defined\ structures}).
\end{aligned}
\end{equation}

This selection is very similar in spirit to the $V_{\rm{r}}-V_{\rm{\phi}}$ selection in \citet{Belokurov18}, where this structure was discovered, as borne out by the second panel of Figure \ref{fig:ge}. This selection is by no means perfect -- it is incomplete in that it misses the low-eccentricity tail of GSE at $e<0.7$ that manifests as a strong peak at [Fe/H]$\approx-1.2$ in subsequent plots. And it is impure, as suggested by the structure along the margins of GSE in \elzs-- for instance, $e>0.7$ stars from Wukong (discussed in \S\ref{subsec:wukong}) are apparent at $L_{\rm{z}}/[\rm{10^{3} kpc\ km\ s^{-1}}]\sim-0.5$. A subtle sequence corresponding to residue from Wukong also appears under the GSE sequence in [Fe/H] vs [$\alpha/\rm{Fe}$]. However, the very well-behaved, unimodal MDF inspires confidence that this selection is overwhelmingly comprised of GSE stars.

The MDF is narrow -- corrected for the selection function, $85\%$ of stars are contained within 0.9 dex in [Fe/H] -- and reminiscent of some local dwarfs (e.g., Leo I and Fornax, \citealt{Kirby13}). Like Leo I and Fornax, the GSE MDF is well-fit by a simple, analytical, chemical evolution model, namely the ``Best Accretion Model" \citep{Lynden-Bell75} used in \citet{Kirby11,Kirby13}, that explains all features, including the extended metal-poor tail (dotted line in MDF panel of Figure \ref{fig:ge}). This model is a generalization of traditional leaky box models, allowing for the accretion of fresh gas, and has two parameters -- $M$, the ratio between the final mass and initial gas mass of the system, and $p$, the effective yield (i.e., a measure of the fraction of metals produced by stars the system retains) -- for which we find best-fit values of $p=0.085, M=3.28$ (slightly different from \citet[][]{Conroy19b} who found $p=0.08, M=2.1$ for a differently selected, SNR$>5$ kinematic halo sample at $-0.5<L_{\rm{z}}/[10^{3}\ \rm{kpc\ km\ s^{-1}}]<1$). As in the case of both Fornax and Leo I in \citet[][]{Kirby13}, the data falls off more steeply than the model on the metal-rich side of the MDF.

Our estimate of GSE's metallicity ([Fe/H]$=-1.15^{+0.24}_{-0.33}$, weighted) is $\approx0.1-0.2$ dex higher than most of the literature \citep[e.g.,][]{Helmi18,Matsuno19,Sahlholdt19,Vincenzo19, Mackereth19} and more in line with the recent [Fe/H]$=-1.17\pm0.34$ estimate of \citet[][]{Feuillet20}. To convert [Fe/H]$=-1.15$ to a mass estimate, we use the mass-metallicity relation from local dwarfs \citep{Kirby13} and account for the redshift evolution of the relation -- i.e., higher masses at higher redshift at fixed [Fe/H] \citep[e.g.,][]{Zahid14,Steidel14,Sanders15mosdef,Ma16MZR,Torrey19}. Assuming the trend from the FIRE simulations \citet[][]{Ma16}, which agrees well with observations out to $z\sim3$, produces $M_{\star}=4-7\times10^{8}\,M_{\odot}$ for accretion redshifts between $z=1.3$ \citep{Kruijssen20} and $z=2$ \citep{Bonaca20}. This is in excellent agreement with recent estimates from GSE's GC age-metallicity relation ($\approx2-4\times10^{8}\,M_{\odot}$, \citealt[][]{Kruijssen20}), star counts of [Fe/H]$<-1$, $e>0.7$ APOGEE red giants ($\approx2-5\times10^{8}\,M_{\odot}$, \citealt[][]{Mackereth20}), and the integrated SFR of a chemical evolution model ($\approx6\times10^{8}\,M_{\odot}$, \citealt[][]{Helmi18, Fernandez-Alvar18}).

We are also in a position to address the mean rotational velocity of GSE, which is of interest because it informs the initial configuration of the merger. We find $\langle V_{\rm{\phi}}\rangle = 1.04^{+1.26}_{-1.25}\ \rm{km\ s^{-1}}$, $\langle L_{\rm{z}}\rangle = 4.7^{+20.1}_{-10.5}\ \rm{kpc\ km\ s^{-1}}$ with errors estimated via bootstrap resampling including fully propagated errors from distance and PM samples, and weighting for the selection function.  This measurement places a very strong constraint on the lack of net rotation of GSE. This conclusion is in excellent agreement with \citet[][]{Belokurov20} who report $\langle V_{\rm{\phi}}\rangle\sim0$ for the ``Sausage" component in their velocity ellipsoid fits for a local sample drawn from \citet[][]{Sanders18} with $6.5<R_{\rm{gal}}<10$ kpc. The magnitude of rotation we measure is much lower than \citet[][]{Mackereth19} who find $L_{\rm{z}}= 176\ \rm{kpc\ km\ s^{-1}}$ using 673 APOGEE stars at $|Z|<10$ kpc with spectrophotometric distances uncertain on the $\sim15\%$ level, and \citet[][]{Helmi20} who report $\langle V_{\rm{\phi}}\rangle (d_{\rm{helio}}<1\ \rm{kpc}) = 21.1\pm1.8\ \rm{km\ s^{-1}}$ using 6 stars and $\langle V_{\rm{\phi}}\rangle (d_{\rm{helio}}<2\ \rm{kpc}) = 16.1\pm2.8\ \rm{km\ s^{-1}}$ using 23 stars from the \textit{Gaia} RVS sample cross-matched with APOGEE [Fe/H]$\geq-1.3$ stars. We caution that this measurement is sensitive to the assumed solar motion, and that the mentioned GSE samples have all been selected differently. We also observe that including Sequoia, Arjuna, or Arjuna and Sequoia (retrograde structures discussed in \S\ref{subsec:retrograde}) in GSE results in $\langle V_{\rm{\phi}}\rangle = [3.7^{+1.3}_{-1.3}, 7.0^{+1.6}_{-1.5}, 9.4^{+1.7}_{-1.6}]\ \rm{km\ s^{-1}}$ respectively. Despite these caveats, it is clear that GSE is far from highly retrograde. We further discuss the rotation of the halo in \S\ref{subsec:protretro}.

\subsubsection{Helmi Streams}
\label{subsec:hs}

\begin{figure*}
\centering
\includegraphics[width=\linewidth]{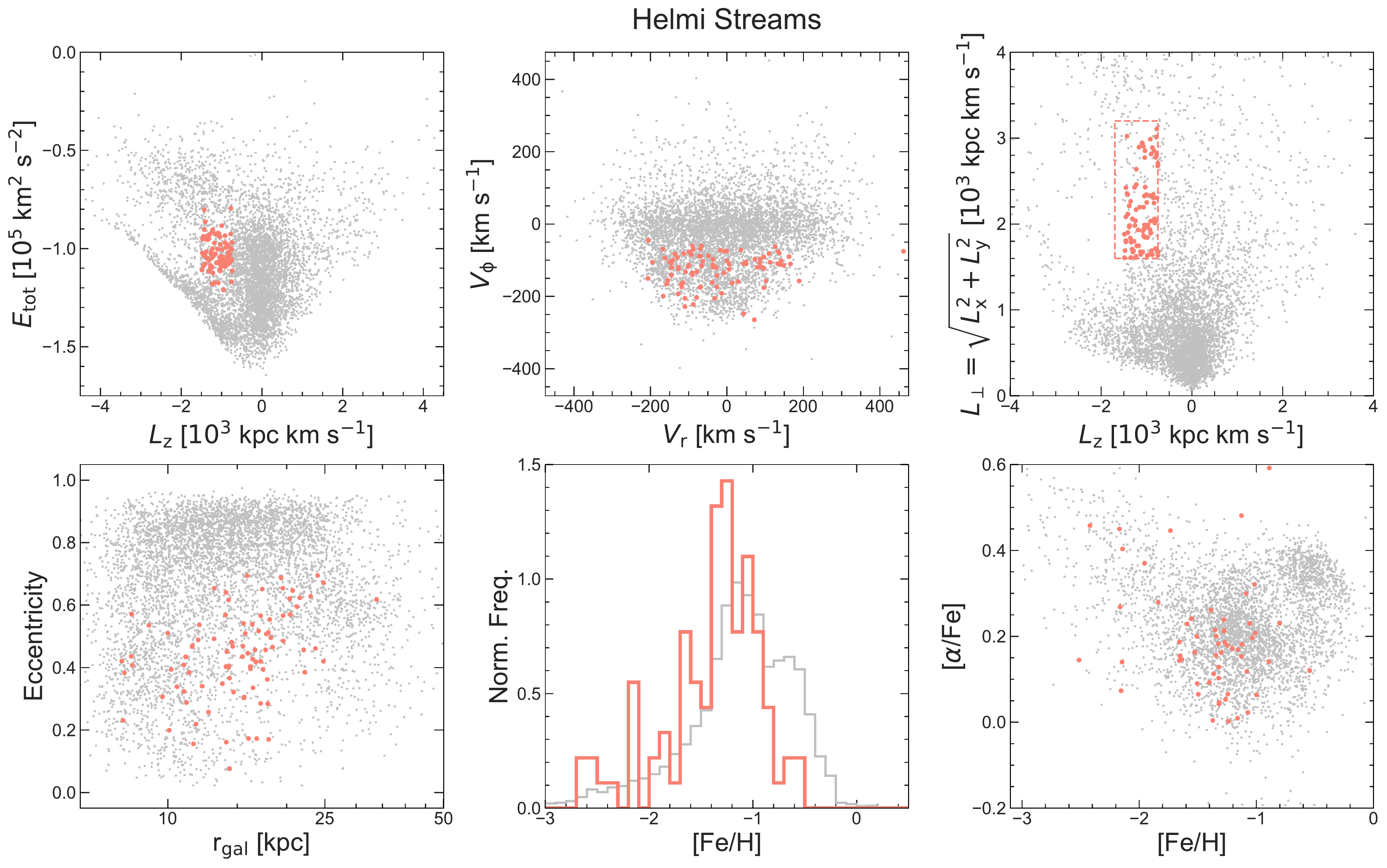}
\caption{The Helmi Streams (salmon pink) in chemodynamical space. Panels are as in Figure \ref{fig:data}, except for the top-right, which shows the HS selection plane $L_{\rm{z}}-L_{\rm{\perp}}$.  We follow \citet[][]{Koppelman19HS} to define our selection. The Helmi Streams extend to $\approx25$ kpc, in line with expectations from orbit integration of high-energy stars in local studies \citep{Helmi20}. The multi-modal MDF and complex [Fe/H] vs [$\alpha$/Fe] morphology suggest a complex star-formation history and/or more than one sub-population.}
\label{fig:helmi}
\end{figure*}

\begin{figure*}
\centering
\includegraphics[width=\linewidth]{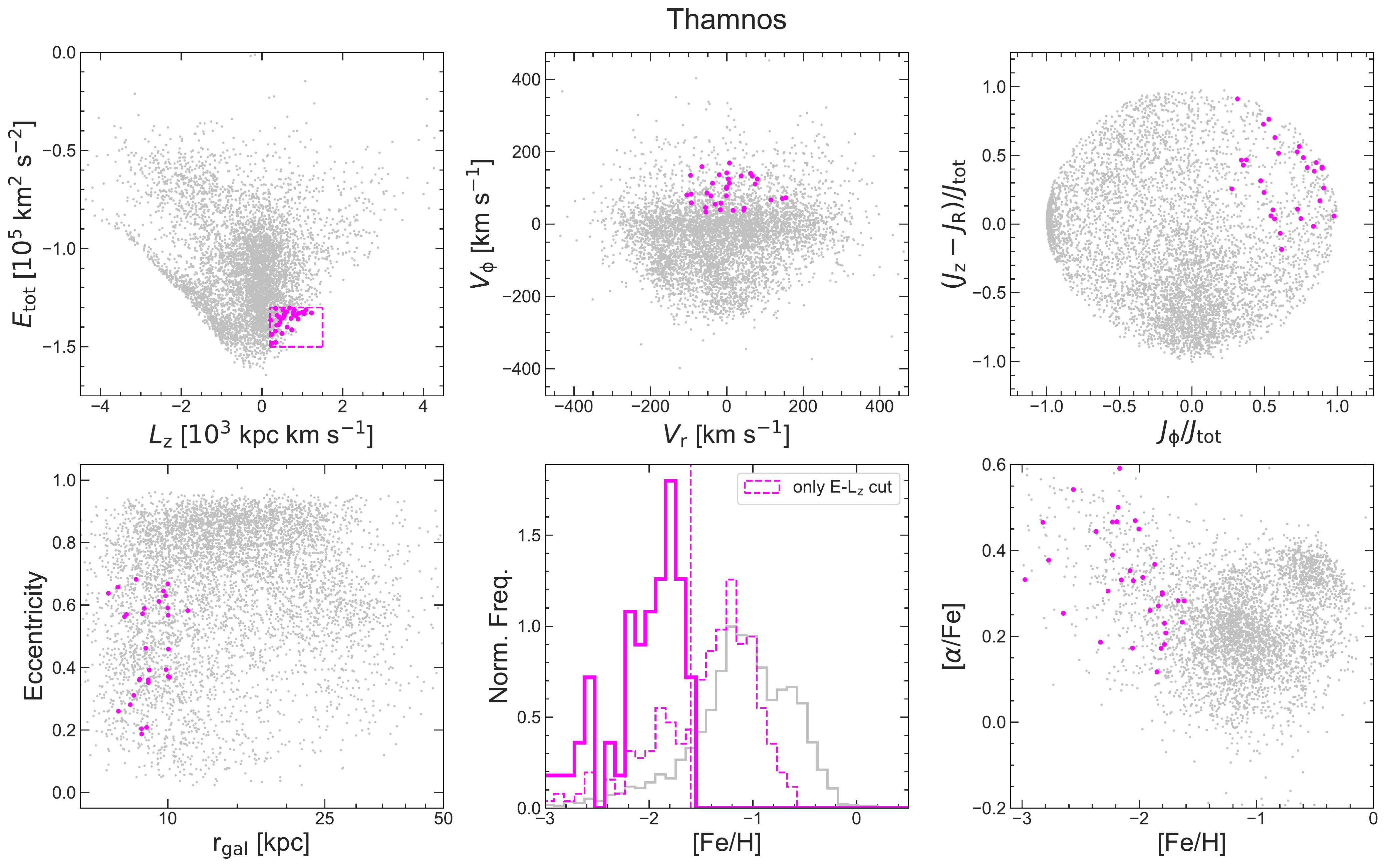}
\caption{Thamnos (fuchsia) in chemodynamical space. Panels are as in Figure \ref{fig:ge}.  The initial selection is in \elzs with a further cut in [Fe/H] to refine the sample. The MDF for only the \elzs cut is shown in dashed lines -- two populations corresponding to GSE ([Fe/H]$=-1.2$) and Thamnos ([Fe/H]$=-1.9$) are visible. In the final panel we see a metal-poor, $\alpha$-rich sequence as expected for a low-mass dwarf galaxy accreted at high-redshift.}
\label{fig:thamnos}
\end{figure*}

The Helmi Streams were among the first bona fide accreted substructures discovered in the halo via integrals of motion as opposed to on-sky streams \citep[][]{Helmi99}. \citet{Koppelman19, Koppelman19HS, Koppelman20} provide an updated view of these streams using \textit{Gaia} DR2 data.

While a prominent \elzs overdensity corresponding to the Helmi Streams appears among the H3 dwarfs, it is not as readily apparent in the giants, though there is a hint of a vertical spur at $L_{\rm{z}}/[\rm{10^{3} kpc\ km\ s^{-1}}]\sim-1.5$ in e.g., Figure \ref{fig:elz_opener}. To select the Helmi Streams we rely on the $L_{\rm{z}}-L_{\rm{\perp}}$ selection in \citet[][``Box B", their Fig. 2]{Koppelman19}:
\begin{equation}
\label{eq:HS}
\begin{aligned}
(-1.7<L_{\rm{z}}/[\rm{10^{3} kpc\ km\ s^{-1}}]<-0.75)\\ 
\land\ (1.6<L_{\rm{\perp}}/[\rm{10^{3} kpc\ km\ s^{-1}}]<3.2)\\\land\ (\rm{excluding\ all\ previously\ defined\ structures}).
\end{aligned}
\end{equation}

The selected stars are shown in Figure \ref{fig:helmi}. Orbits of the Helmi Streams in local samples \citep[][their Figure 12]{Helmi20} rise to high latitudes and extend out to $R_{\rm{gal}}\approx25$ kpc -- this is borne out in Figure \ref{fig:helmi}. The large spread in eccentricity mirrors the large spread in eccentricties of GCs attributed to HS from considerations of the GC age-metallicity relation \citep{Massari19,Kruijssen20,Forbes20}. In the MDF we see a broad distribution, consistent with the complex population with an extended star-formation history modeled in \citet{Koppelman19HS}. The distribution in [Fe/H] vs [$\alpha$/Fe] traces the typical trend of decreasing [$\alpha$/Fe] with increasing [Fe/H] expected in halo populations. Assuming accretion redshifts of $z\sim0.5-1.1$, i.e., 5-8 Gyrs ago \citep{Koppelman19HS}, and a (weighted) mean metallicity of [Fe/H]$\approx-1.3$, we estimate the Helmi Streams to have a stellar mass of $M_{\star}\approx0.5-1\times10^{8}M_{\rm{\odot}}$ via the \citet[][]{Kirby13} MZR and its expected evolution to higher redshifts \citep{Ma16}, in excellent agreement with \citet{Koppelman19HS}.

\subsubsection{Thamnos}
\label{subsec:thamnos}

The Thamnos structure was recently discovered in \citet{Koppelman19}. These authors found two overdensities in chemodynamical space (``Thamnos 1" and ``Thamnos 2") that they attribute to the same progenitor. An overdensity corresponding to their proposed structure appears in our \elzs diagrams as a jagged ridge along the retrograde edge of GSE (at $E_{\rm{tot}}/[10^{5}\ \rm{km}^{2} \ \rm{s}^{-2}]\approx-1.4$). This ridge resembles the corrugations of a single massive satellite producing multiple over-densities in \elzs and other phase-space diagrams \citep[][]{Jean-Baptiste17}. To distinguish whether Thamnos is a remnant of a distinct satellite, or a part of GSE, it is important to verify that the chemistry of Thamnos is distinct from GSE, particularly because of the small sample ($\sim$20) of Thamnos stars with abundances in \citealt{Koppelman19}, the majority of which overlap with GSE within error-bars (their Figure 4).

To define our Thamnos selection we begin by selecting all stars at $(L_{\rm{z}}>0.2) \land (-1.5<E_{\rm{tot}}<-1.3)$. The \elzs selection is motivated by the contours for Thamnos provided in \citet{Koppelman19} as well as the overdensity we see in that region. In this energy range we expect high contamination primarily from GSE. The resulting MDF depicted with a dashed line in Figure \ref{fig:thamnos} shows a strong peak at [Fe/H]$=-1.9$ that we attribute to Thamnos as well as a second peak corresponding to GSE's metallicity of [Fe/H]$\approx-1.2$. We further refine the Thamnos selection informed by this MDF by restricting the selection to [Fe/H]$<-1.6$, where we see a break. This results in a final selection of 32 stars that produces a clean [Fe/H]-[$\alpha$/Fe] sequence. To summarize, the Thamnos selection is:

\begin{equation}
\begin{aligned}
(0.2<L_{\rm{z}}/[\rm{10^{3}\ kpc\ km\ s^{-1}}]<1.5) \\
\land\ (-1.5<E_{\rm{tot}}/[10^{5}\  \rm{km}^{2}\ \rm{s}^{-2}]<-1.3)\\
\land\ (\rm{[Fe/H]}<-1.6)\\\land\ (\rm{excluding\ all\ previously\ defined\ structures}).
\end{aligned}
\end{equation}

\citet{Koppelman19} estimate Thamnos' stellar mass to be $<5\times10^{6}M_{\odot}$ by comparing its extent in \elzs against a suite of simulations. Thamnos lies at lower energy than GSE, which is unexpected for a system of such low stellar mass if it were accreted at $z\sim0$, as its would be shredded in the outer reaches of the halo \citep[e.g.,][]{Amorisco17,Pfeffer20}. This suggests that it was accreted very early when the Galaxy was not very massive, or perhaps simultaneously with GSE at $z\sim1.3-2$ \citep{Kruijssen20,Bonaca20}. Now that we have a good handle on the metallicity ([Fe/H]=$-1.9$) we can provide a complementary mass-metallicity relation (MZR) constraint on the mass. Using the $z=0$ \citet{Kirby13} relation for [Fe/H] produces a mass $2\times10^{5}M_{\odot}$, but this is a strict lower limit. This is because we must account for the redshift evolution of the MZR. Assuming the trend from the FIRE simulations \citet[][]{Ma16}, which agree well with observations out to $z\sim3$, and predict a $\sim1$ dex increase in mass for an [Fe/H]$=-1.9$ galaxy between $z=0$ and $z=1.5$, we find $M_{\star}\approx2\times10^{6}M_{\odot}$ for Thamnos, in good agreement with \citet{Koppelman19}.

Thamnos is potentially a very exciting object because of its very low stellar mass that we estimate here, and because it lies so deep in the potential. Put another way, the debris from Thamnos lies only at $d_{\rm{helio}}\approx6$ kpc (weighted), can be easily targeted using our clean sample, and thus offers a unique view of galaxy evolution (e.g., stellar abundances) in a mass regime that will be out of reach even for the \textit{James Webb Space Telescope} at high redshift \citep[e.g.,][]{Boylan-Kolchin15,Boylan-Kolchin16,Weisz14}.

\subsubsection{The High-Energy Retrograde Halo: Arjuna, Sequoia, and I'itoi}
\label{subsec:retrograde}

\begin{figure*}
\centering
\includegraphics[width=\linewidth]{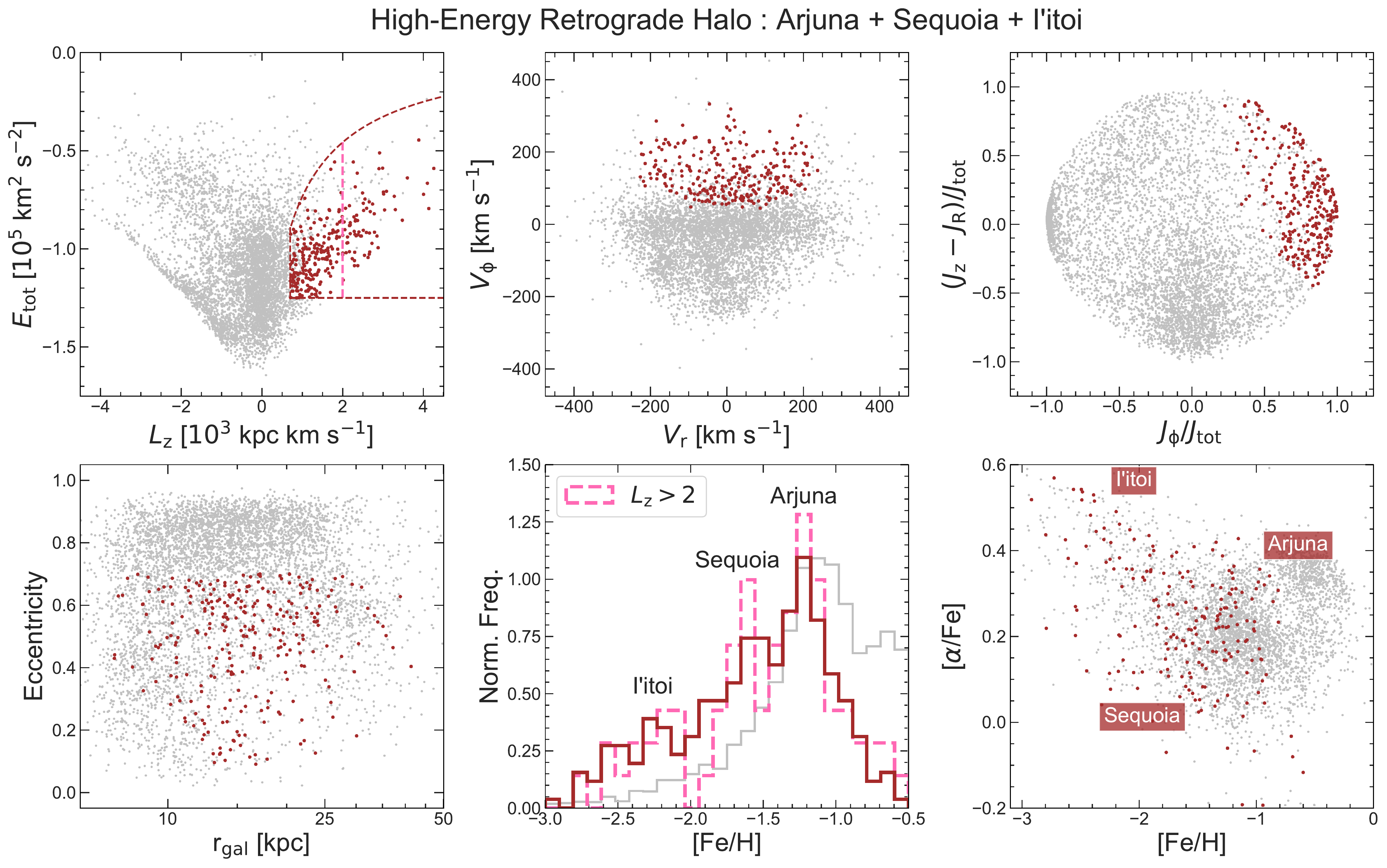}
\caption{The high-energy retrograde halo (brown) in chemodynamical space.  Panels are as in Figure \ref{fig:ge}.  Brown dashed lines in the \elzs panel depict the selection (see Eq. \ref{eq:retro}). The resulting MDF shows three peaks at [Fe/H] of $-1.2,-1.6,<-2$ that we identify as Arjuna, Sequoia, and I'itoi respectively. Imposing a conservative $L_{\rm{z}}/[10^{3}\rm{\ kpc\ km\ s^{-1}}]>2$ selection also produces a similarly shaped MDF with a more pronounced I'itoi peak (dashed pink). We display a summary of actions (top right) similar to the diagram in which Sequoia was discovered \citep{Myeong19}. Those authors identified Sequoia as belonging to the bottom-right quadrant of this action diagram. In the bottom left panel we see Arjuna, Sequoia, and I'itoi are quite eccentric ($\langle e\rangle\approx0.5-0.6$) and extend to $\approx 40$ kpc.}.
\label{fig:retrograde}
\end{figure*}

\begin{figure*}
\centering
\includegraphics[width=\linewidth]{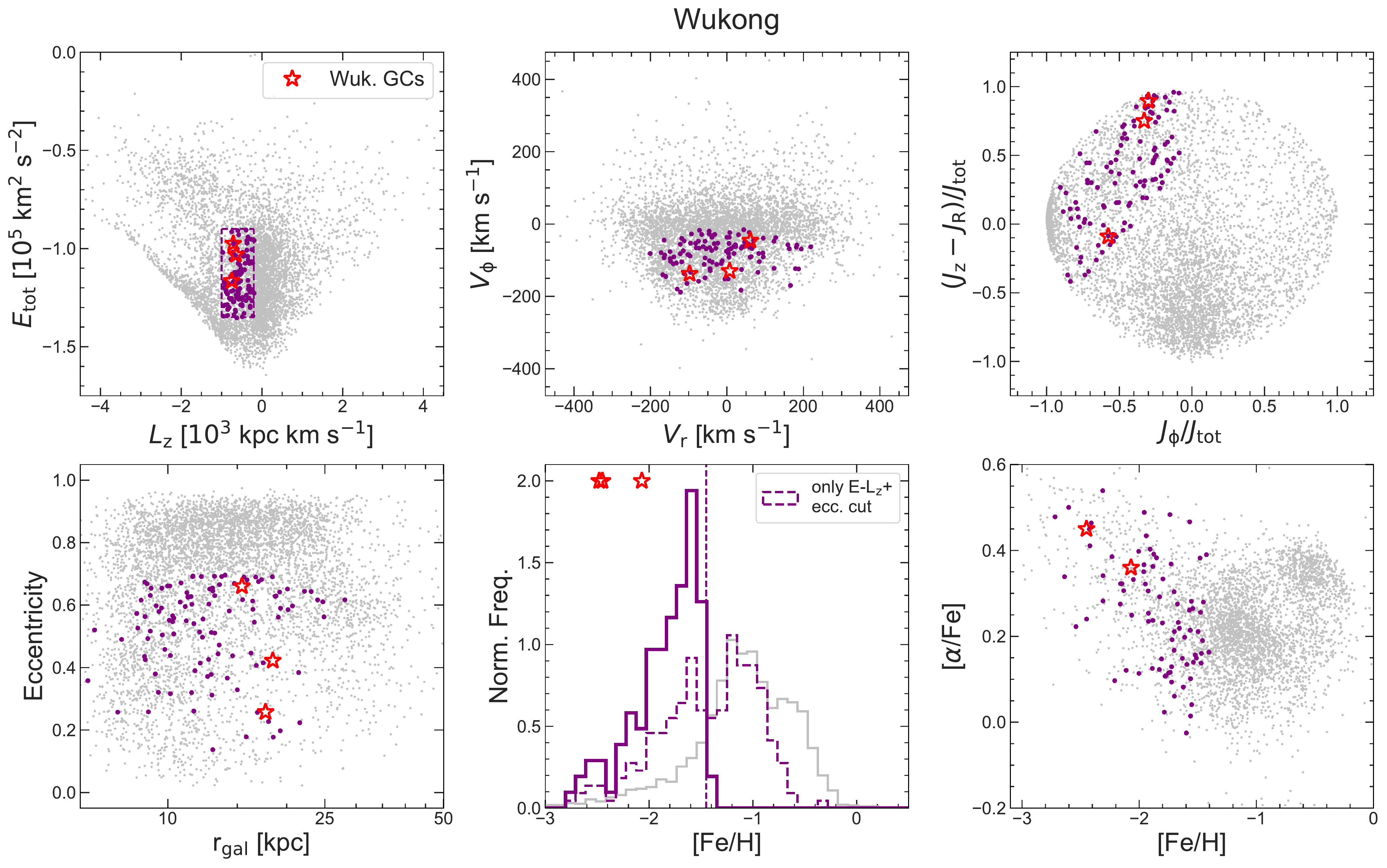}
\caption{Wukong (purple) in chemodynamical space.  Panels are as in Figure \ref{fig:ge}. Wukong is initially selected in \elz (box in top-left panel). A further cut is made in [Fe/H] based on the break in the resulting MDF (dashed histogram) at [Fe/H]$\approx-1.45$. Wukong stars have high $J_{\rm{z}}$, show a broad spread in eccentricity characteristic of more massive accreted structures like Sequoia and the Helmi Streams, and are contained within $\approx25$ kpc. Three metal-poor GCs -- ESO 280-SC06 ([Fe/H]$=-2.5$), NGC 5024 ([Fe/H]$=-2.1$), NGC 5053 ([Fe/H]$=-2.5$) -- satisfy our Wukong selection and are shown as red stars in all panels ([$\alpha$/Fe] is not available for ESO 280-SC06).}
\label{fig:wukong}
\end{figure*}

Studies of the local stellar halo have found a wealth of retrograde substructure \citep[e.g.,][]{Helmi17,Myeong18d,Myeong18b,Myeong18c,Myeong19,Matsuno19, Koppelman19, Yuan20}. Debris of the Sequoia dwarf galaxy \citep[][]{ Myeong18b,Myeong19,Matsuno19} dominates the local retrograde halo at higher energies while Thamnos \citep[][]{Koppelman19} resides at lower energy. In this section we turn our attention to the high-energy retrograde halo (i.e., at higher energies than Thamnos) that we select as follows:

\begin{equation}
\begin{aligned}
\label{eq:retro}
(\eta>0.15) \land (L_{\rm{z}}/[\rm{10^{3} kpc\ km\ s^{-1}}]>0.7)\\ 
\land\ (E_{\rm{tot}}/[10^{5} \rm{km}^{2}\ \rm{s}^{-2}]>-1.25)\\\land\ (\rm{excluding\ all\ previously\ defined\ structures}).
\end{aligned}
\end{equation}

The circularity, $\eta=L_{\rm{z}}/|L_{\rm{z,max}}(E_{\rm{tot}})|$, where $L_{\rm{z,max}}(E_{\rm{tot}})$ is the maximum $L_{\rm{z}}$ achievable for an orbit of energy $E_{\rm{tot}}$. We compute $L_{\rm{z,max}}(E_{\rm{tot}})$ by assuming a perfectly circular orbit with the star's $r_{\rm{gal}}$ and total 3D velocity. The circularity condition, $\eta>0.15$, ensures a generous selection of retrograde orbits, $L_{\rm{z}}/[\rm{10^{3}\ kpc\ km\ s^{-1}}]>0.7$ reduces contamination from GSE, and the energy limit avoids Thamnos. The stars satisfying this selection are shown in Figure \ref{fig:retrograde}.

A prominent peak in the high-energy retrograde MDF appears exactly where expected for Sequoia at [Fe/H]$\approx-1.6$ \citep[][]{Matsuno19, Myeong19, Monty19}. More surprisingly, two other distributions are apparent in the MDF -- one centerd at [Fe/H]$\approx-1.2$, and another spanning very low metallicity at [Fe/H]$<-2$. Furthermore, these stars occupy a complex distribution in [Fe/H]-[$\alpha$/Fe] that is suggestive of multiple populations. We name the metal-rich population ``Arjuna"\footnote{Arjuna is named for the legendary archer from the Indian epic, the Mahabharata. Arjuna extends to high-energy, mirroring Sagittarius (Latin for archer) on the prograde side.} and the metal-poor sequence ``I'itoi"\footnote{I'itoi (pronounced ``ee ee thoy") is named for the ``man in the maze" who features in creation legends of the Tohono O'odham people. Our survey telescope, the MMT Observatory, stands on the ancestral lands of the Tohono O'odham. Further, I'itoi is said to reside in a cave adjacent to a mountain, paralleling the location of I'itoi in \elzs with respect to GSE (Enceladus is entombed within Mt. Etna in Sicily).}. Based on the peaks and breaks in the MDF we define Arjuna stars as those with [Fe/H]$>-1.5$, Sequoia stars as those with $-2<$[Fe/H]$<-1.5$, and the remaining stars at [Fe/H]$<-2$ as belonging to I'itoi.

Contrary to expectations from local studies \citep[e.g.,][]{Myeong19,Koppelman19}, Sequoia is not the dominant component of the high-energy retrograde halo -- it has fewer than half as many stars as Arjuna. This raises the question as to why Arjuna was missed in the local studies that found Sequoia. The answer may lie in its spatial extent -- Arjuna lies at larger distances compared to Sequoia (median $r_{\rm{gal}}\sim25$ kpc vs. $\sim15$ kpc, weighted), and stars with apocenters of 25 kpc are relatively rare in the solar neighborhood compared to those with apocenters of 15 kpc. 

Given that Arjuna may prove to be a massive component of the halo, and because of its similarity in [Fe/H] to GSE, it is important to discuss possible connections to GSE. Recent work has also cast doubt on the status of Sequoia as a dwarf galaxy and argued that it may be debris from the outer reaches of GSE \citep{Koppelman19,Helmi20}. We explore this issue in Figure \ref{fig:retrograde}. We show that even when restricting the retrograde selection to very high $L_{\rm{z}}/\rm{[10^{3}\ kpc\ km\ s^{-1}]}>2$, far from the $L_{\rm{z}}\approx0$ overdensity defined by GSE, the peaks in the MDF associated with Arjuna and Sequoia remain. None of the existing studies of GSE show a significant [Fe/H]$\approx-1.2$ population at such high $L_{\rm{z}}$, with stars outnumbering those from Sequoia \citep[e.g.,][]{Belokurov18,Mackereth19,Myeong19,Koppelman19,Helmi20}. So Arjuna may either be a hitherto unknown extension of GSE to highly retrograde orbits (such high $L_{\rm{z}}$ extensions exist for the largely eccentric debris of massive, $M_{\rm{\star}}>10^{8}M_{\rm{\odot}}$, accreted galaxies in the \citealt[][]{bj05_1} halos), or it may be the debris of a distinct dwarf galaxy. As for Sequoia, any attempt to tie both Arjuna and Sequoia to GSE must account for $L_{\rm{z}}/\rm{[10^{3} kpc\ km\ s^{-1}]}>2$ components of GSE as well as the spread in abundances (e.g., by appealing to a steep metallicity gradient or multiple populations). More detailed modeling of a GSE-like merger \citep[in the vein of][]{Bignone19,Vincenzo19,Elias20} is required to understand if it is possible for a single progenitor to simultaneously populate such disparate regions of phase-space as well as chemistry.

The proximity of I'itoi and Thamnos in [Fe/H] vs [$\alpha$/Fe] may indicate these structures are related -- however, I'itoi is prominent at $L_{\rm{z}}/\rm{[10^{3} kpc\ km\ s^{-1}]}>2$ and Thamnos' mass argues against such a wide extent in \elz \citep[][their Fig. 5]{Koppelman19}. This is because low-mass structures are typically compact, and experience similar dynamical friction across all their stars, compared to larger structures like GES. I'itoi may also be the metal-poor tail of Arjuna and/or Sequoia -- more work needs to be done to differentiate these three structures in phase space as well. This will be particularly challenging due to the error-vector in \elzs and similar spaces that dramatically scatters structures with high angular momentum (see Appendix \ref{appendix:errorvec}), highlighting the importance of leveraging chemistry in this region of phase-space. We defer detailed characterization of these structures to forthcoming work.

\subsubsection{Wukong}
\label{subsec:wukong}

\begin{figure*}
\centering
\includegraphics[width=\linewidth]{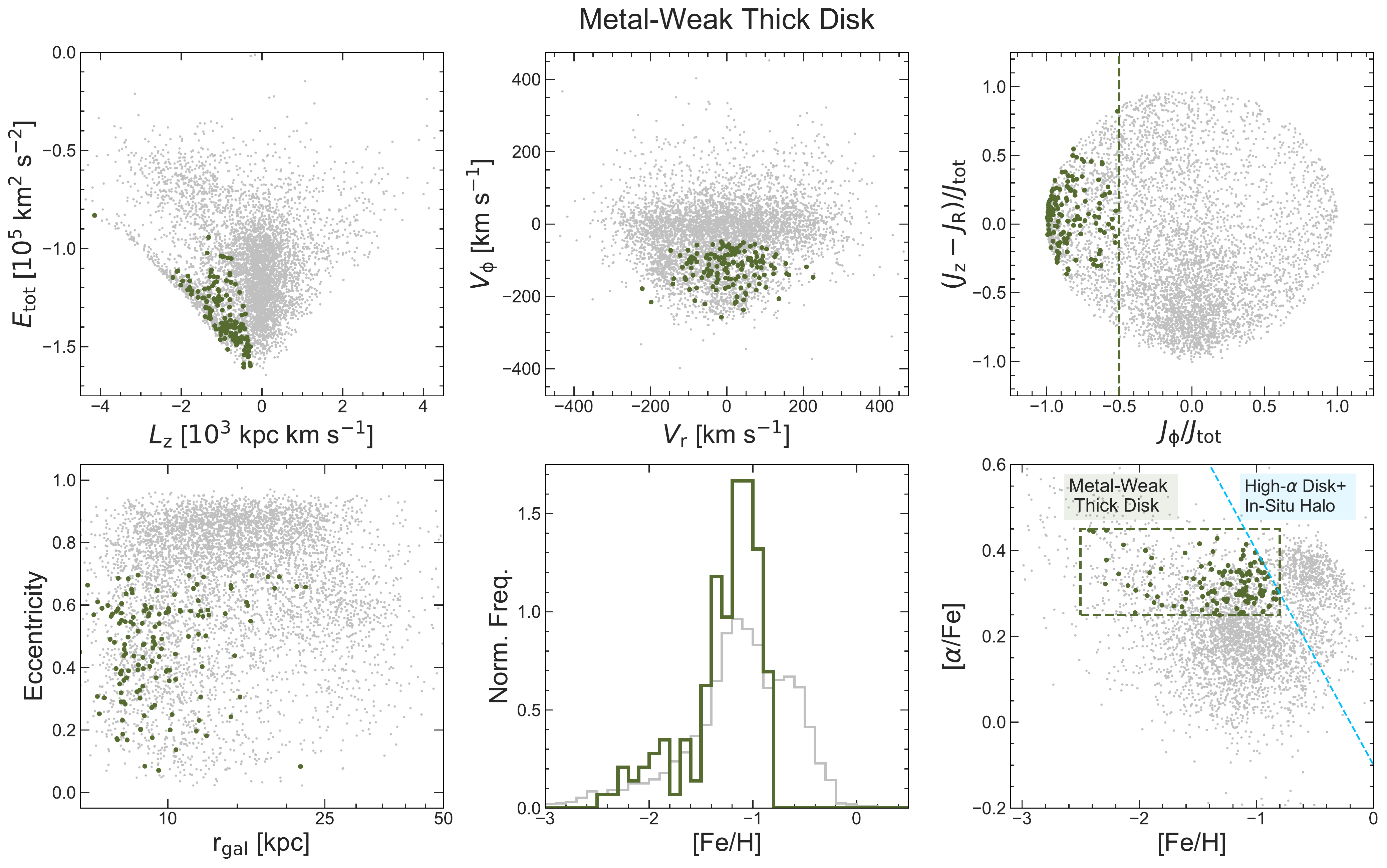}
\caption{The metal-weak thick disk (MWTD, olive green) in chemodynamical space.  Panels are as in Figure \ref{fig:ge}. MWTD stars are selected to be rotationally supported (top-right) and to lie at similar [$\alpha$/Fe] but lower [Fe/H] than the high-$\alpha$ disk (bottom-right) following \citet[][]{Carollo19}. The resulting stars show a broad range of eccentricities and follow the locus of the high-$\alpha$ disk and in-situ halo in \elzs and $V_{\rm{r}}-V_{\rm{\phi}}$.}
\label{fig:MWTD}
\end{figure*}

\begin{figure*}
\centering
\includegraphics[width=\linewidth]{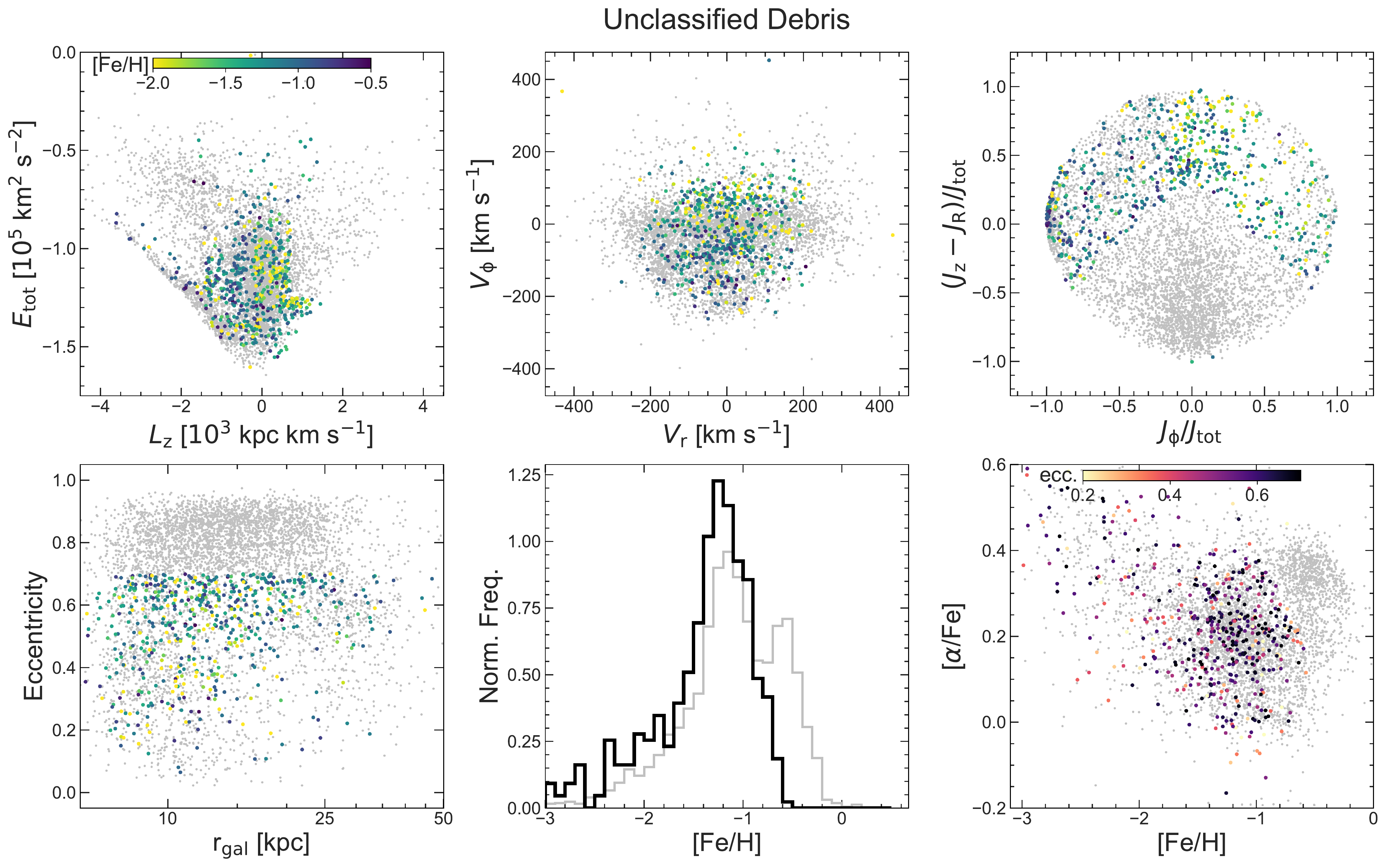}
\caption{Unclassified debris in chemodynamical space.  Panels are as in Figure \ref{fig:ge}.  The projections of phase-space are color-coded by [Fe/H] except for the final panel, which is color-coded by eccentricity. $\approx30\%$ of these stars have very negative $J_{\rm{\phi}}/J_{\rm{tot}}$ indicating some degree of rotational support.  These stars are also clustered at $E<-1.3$ in \elzs, following the contours of the high-$\alpha$ disk and in-situ halo (see Figure \ref{fig:insitu}). These disk-like stars are likely metal-poor members of the high-$\alpha$ disk and in-situ halo that evaded our earlier selection of those populations. $70\%$ of the remaining unclassified stars have eccentricities between $0.5-0.7$. These high eccentricity stars are similar to GSE (which was defined to have $e>0.7$), Arjuna ($e\approx0.6$), and Sequoia ($e\approx0.6$) in [Fe/H] vs [$\alpha$/Fe], and have an MDF similar to GSE (bottom center) which strongly suggests these are stars from these structures that were missed in our earlier selections. There are overdensities in \elzs at the locations of Thamnos and Wukong -- these stars satisfy the phase-space selections for these structures but not the chemistry cuts, and have MDFs resembling GSE, i.e., they may represent the lower eccentricity tail of GSE (see dashed MDFs in Figures \ref{fig:thamnos}, \ref{fig:wukong}). There is also a set of stars at $L_{\rm{z}}\approx-1.5$ in \elzs that closely follows the contours of the Helmi Streams (Figure \ref{fig:helmi}) but did not satisfy the $L_{\perp}$ selection.}
\label{fig:leftovers}
\end{figure*}

Here we present Wukong\footnote{Named for Sun Wukong, the celestial Monkey King from the \textit{Journey to the West}. Sun Wukong is imprisoned under a mountain by the Buddha for his uprising against Heaven and is later set free by the scholar Tripitaka. We play the role of the scholar here, setting Wukong free from underneath \textit{Gaia}-Sausage-Enceladus (Enceladus is entombed within Mt. Etna in Sicily).}, a hitherto unknown prograde structure, that appears as a pair of overdensities in \elzs ($E_{\rm{tot}}/[10^{5} \rm{km}^{2}/\rm{s}^{2}]$ = -1.1, -1.3), lining the prograde margin of GSE. In Appendix \ref{appendix:pot} we show these clumps to be even more pronounced in the \citet[][]{McMillan17} potential. We select Wukong as follows:
\begin{equation}
\label{eq:wukong}
\begin{aligned}
(-1<L_{\rm{z}}/[\rm{10^{3} kpc\ km\ s^{-1}}]<-0.2) \\
\land\ (-1.35<E_{\rm{tot}}/[10^{5} \rm{km}^{2}/\rm{s}^{2}]<-0.9)\\
\land\ (\rm{[Fe/H]}<-1.45)\\
\land\ (\rm{excluding\ all\ previously\ defined\ structures}).
\end{aligned}
\end{equation}

\noindent
The conditions in \elzs draw a box around the overdensities and extend it inwards towards GSE. The [Fe/H] selection is motivated by the MDF resulting from the \elzs cut (dashed histogram in MDF panel of Figure \ref{fig:wukong}) that shows multiple peaks at [Fe/H]$=-1.2$ (GSE),$-1.6$,$-1.9$, and a break at $\approx-1.45$. This leaves us with a sample of 111 Wukong stars that form a sequence in [Fe/H] vs [$\alpha$/Fe].

Three metal-poor GCs -- NGC 5024 ([Fe/H]$=-2.1$), NGC 5053 ([Fe/H]$=-2.5$), ESO 280-SC06  ([Fe/H]$=-2.5$) -- satisfy all the selection criteria in Eq. \ref{eq:wukong}, and may have been accreted along with Wukong (phase-space parameters from \citealt[][]{Baumgardt19} and abundances from \citealt[][]{Boberg15,Boberg16,Simpson19b}). \citet[][]{Massari19} attribute NGC 5024, NGC 5053 to the Helmi Streams and ESO 280-SC06 to GSE. We note however, that none of these GCs satisfy the Helmi Streams selection from \citet{Koppelman19HS} that we also use in this work (Eq. \ref{eq:HS}), and that ESO 280-SC06 ([Fe/H]$=-2.5$, $e=0.66$) has properties only marginally consistent with the GSE stars in the sample. That multiple GCs with metallicities consistent with Wukong are aligned with it in phase-space is a promising sign that it is a genuine structure.

\subsubsection{Metal-Weak Thick Disk}
\label{subsec:mwtd}

The metal-poor tail of the high-$\alpha$ disk was not included in our earlier selection of the high-$\alpha$ disk and in-situ halo in chemical space (Figure \ref{fig:insitu}). Metal-weak thick disk (MWTD)\footnote{Following previous work, we refer to this population as the metal-weak thick disk.  However, given our selection it might be more appropriate to refer to this structure as the ``metal-weak high-$\alpha$ disk".} stars are expected to fall right next to the high-$\alpha$ disk in [Fe/H] vs [$\alpha$/Fe]. They are more metal poor than the high-$\alpha$ disk but are at similar $\alpha$ and are rotationally supported -- i.e., prograde and with strong $J_{\rm{\phi}}$ (top-left panel of Figure \ref{fig:confusogram_feh}). This motivates the MWTD selection:
\begin{equation}
\begin{aligned}
(-2.5<\rm{[Fe/H]}<-0.8) \land (0.25<\rm{[\alpha/Fe]}<0.45)\\
\land\ (J_{\phi}/J_{\rm{tot}}<-0.5)\\
\land\ (\rm{excluding\ all\ previously\ defined\ structures}).
\end{aligned}
\end{equation}

The stars that satisfy these cuts are shown in Figure \ref{fig:MWTD}. They are mostly clustered very close to the high-$\alpha$ disk in [Fe/H] vs [$\alpha$/Fe] at [Fe/H]$\approx-0.8$. These stars support the finding of \citet[][]{Carollo19} that while the MWTD may be a prominent component of the $|Z_{\rm{gal}}|<3$ kpc Galaxy, it is only a minor component at larger distances (weighted fraction of $<5\%$ at $|Z_{\rm{gal}}|>3$ kpc).

\subsubsection{Unclassified Debris}
\label{subsec:leftovers}

We have assigned $92\%$ (weighted) of our sample to the aforementioned structures. This leaves us with $8\%$ (weighted) that we designate as ``unclassified debris". The unclassified debris is depicted in Figure \ref{fig:leftovers}. 

A prograde population that closely follows the contours of the high-$\alpha$ disk/in-situ halo in \elzs is evident at  $E_{\rm{tot}}<-1.3, |L_{\rm{z}}|<1$ (compare with Figure \ref{fig:insitu}). There is also an overlapping population, extending to higher energies ($E_{\rm{tot}}\approx-0.8$) with a high degree of rotational support ($J_{\rm{}\phi}/J_{\rm{tot}}<-0.75$). These disk-like stars did not meet the high-$\alpha$ disk/in-situ halo and MWTD cuts. These stars are either (i) the eccentric, metal-poor tail of the high-$\alpha$ disk (or the low eccentricity tail of GSE) that did not meet the rotational support criteria of the MWTD, or ii) rotationally supported stars that fall outside our high-$\alpha$ disk and MWTD chemistry selection boxes. We designate these stars as ``disk-like" unclassified debris and they constitute $2\%$ (weighted) of the total sample.

Then there are ``halo-like" stars at higher-energy clustered around various selection boxes. Most of these stars have eccentricities between 0.6 and 0.7 ($5\%$ of the total sample, weighted). This concentration in eccentricity is noteworthy since we select GSE stars with a hard cut at $e>0.7$. This, and the prominent peak in the MDF at GSE's metallicity strongly suggests these stars are $e<0.7$ members of GSE. The clumps of unclassified debris stars that appear at the locations of Thamnos and Wukong in \elzs bear this out: these stars satisfied the phase-space selections for these structures, but had GSE-like metallicity and were excluded via cuts on the MDF (see dashed MDFs in Figures \ref{fig:thamnos}, \ref{fig:wukong}). This aspect of our work may be improved with more probabilistic methods of assigning membership that do not impose discontinuous selection boxes as we have done here \citep[e.g.,][]{Yuan20}.

Some of the prograde, high-energy ``halo-like" stars ($<1\%$ of the entire sample, weighted) are also clustered in a selection box corresponding to the Helmi Streams. These stars have similar $L_{\rm{z}}$ and energies but do not satisfy the $L_{\rm{\perp}}$ condition we imposed. These stars are plausible members of the Helmi Streams.

Taking into account these likely associations, we are left with $\approx1\%$ (weighted) of the total sample as being either unclassified or unassociated. These stars may belong to low-mass structures that we sample too few stars from to detect coherent features. Or these stars may simply have bad stellar or orbital parameters.  Either way, it is clear that we have identified the vast majority of structure in the halo as viewed by H3.

\begin{figure*}
\centering
\includegraphics[width=0.85\linewidth]{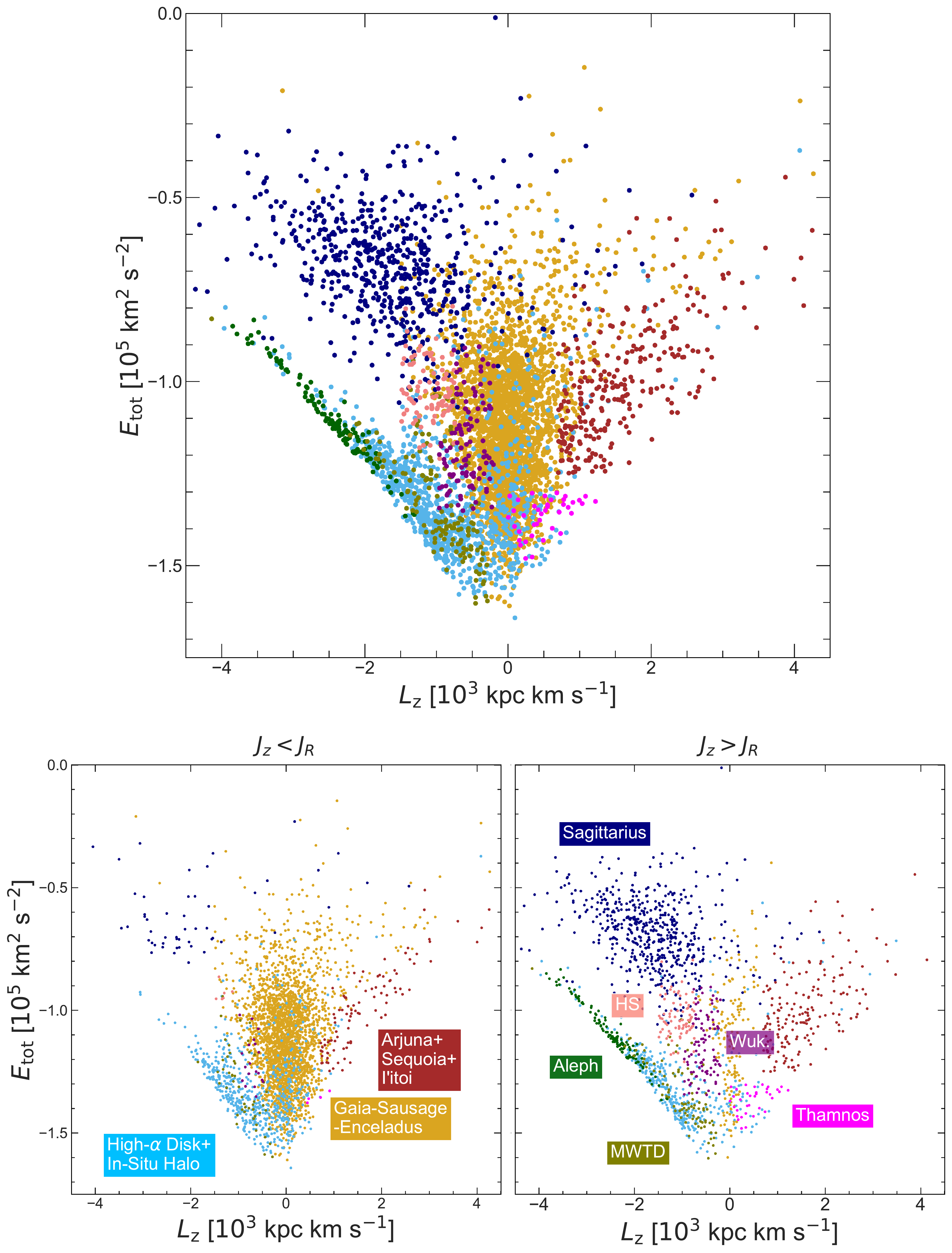}
\caption{\textbf{Top:} \elzs diagram depicting all the structures identified in this work. \textbf{Bottom:} \elzs split by actions with stars on radial ($J_{\rm{z}}<J_{\rm{R}}$) orbits on the left and polar, circular ($J_{\rm{z}}>J_{\rm{R}}$) orbits on the right. We abbreviate the Helmi Streams as ``HS" and Wukong as ``Wuk." to avoid crowding.}
\label{fig:summary2}
\end{figure*}

\begin{deluxetable*}{lrrrrrrrrrr}
\label{table:summary}
\tabletypesize{\footnotesize}
\tablecaption{Summary of Substructure in the $|b|>40^{\circ}$, $|Z_{\rm{gal}}|>2$ kpc Milky Way}
\tablehead{
\colhead{Substructure} & \colhead{$N_{\rm raw}$} & \colhead{frac.} & \colhead{[Fe/H]} & \colhead{[$\alpha$/Fe]} & \colhead{ecc.} & \colhead{$r_{\rm{gal}}$} & \colhead{$|Z_{\rm{gal}}|$} &
\colhead{$(J_{\rm{z}}-J_{\rm{R}})/J_{\rm{tot}}$} &
\colhead{$E_{\rm{tot}}$} & 
\colhead{$L_{\rm{z}}$} \vspace{-0.2cm} \\ \colhead{} & \colhead{} & \colhead{} & \colhead{} & \colhead{} & \colhead{} & \colhead{[kpc]} & \colhead{[kpc]}  &
\colhead{} &
\colhead{[$10^{5}\ \rm{km}^{2}\ \rm{s}^{-2}$]} & 
\colhead{[$10^{3}\ \rm{kpc}\ \rm{km\ s^{-1}}$}]}
\startdata
\vspace{-0.2cm}  \\
\textit{Gaia}-Sausage-Enceladus & 2684 & 0.42 & -1.15 & 0.21 & 0.84 & 17.72 & 12.88 & -0.50 & -1.04 & -0.01 \\
Sagittarius & 675 & 0.24 & -0.96 & 0.12 & 0.54 & 32.31 & 24.13 & 0.51 & -0.67 & -1.51 \\
High-$\alpha$ Disk + In-Situ Halo & 950 & 0.15 & -0.54 & 0.34 & 0.48 & 9.00 & 3.53 & -0.03 & -1.34 & -0.94 \\
Arjuna & 139 & 0.02 & -1.20 & 0.24 & 0.55 & 22.91 & 16.66 & 0.16 & -0.91 & 1.73 \\
Metal-Weak Thick Disk & 144 & 0.02 & -1.12 & 0.32 & 0.47 & 8.60 & 4.25 & 0.04 & -1.38 & -0.9 \\
Aleph & 122 & 0.02 & -0.51 & 0.19 & 0.13 & 11.06 & 3.51 & 0.07 & -1.13 & -2.36 \\
Wukong & 111 & 0.01 & -1.58 & 0.24 & 0.56 & 12.75 & 9.55 & 0.42 & -1.18 & -0.59 \\
Helmi Streams & 91 & 0.01 & -1.28 & 0.15 & 0.46 & 17.17 & 13.55 & 0.47 & -1.03 & -1.14 \\
Sequoia & 72 & 0.01 & -1.59 & 0.14 & 0.56 & 15.55 & 11.02 & 0.16 & -1.02 & 1.31 \\
I'itoi & 65 & 0.01 & -2.39 & 0.38 & 0.47 & 12.37 & 7.46 & 0.09 & -1.04 & 1.35 \\
Thamnos & 32 & 0.01 & -1.90 & 0.29 & 0.46 & 8.68 & 5.11 & 0.41 & -1.35 & 0.46 \\
Unclassified Debris (disk-like) & 208 & 0.02 & -1.16 & 0.22 & 0.53 & 8.63 & 4.72 & 0.13 & -1.36 & -0.52 \\
Unclassified Debris (halo-like) & 463 & 0.06 & -1.20 & 0.19 & 0.60 & 18.25 & 14.19 & 0.40 & -1.04 & -0.03 \\
\enddata
\tablecomments{All quantities -- with the exception of $N_{\rm raw}$ -- are corrected for the selection function (see \S\ref{subsec:magcorr}). The reported values are medians of their respective distributions. The listed substructures were defined via selections in phase-space and chemistry, and so the reported values for those quantities used in the selection must be interpreted with care.  For example, [Fe/H] and [$\alpha$/Fe] reported for Arjuna, Sequoia, Wukong, and Thamnos are the peaks of their distributions, and not the medians, since these structures are selected using their MDFs in a way that biases the median. }
\end{deluxetable*}

\begin{figure*}
\centering
\includegraphics[width=\linewidth]{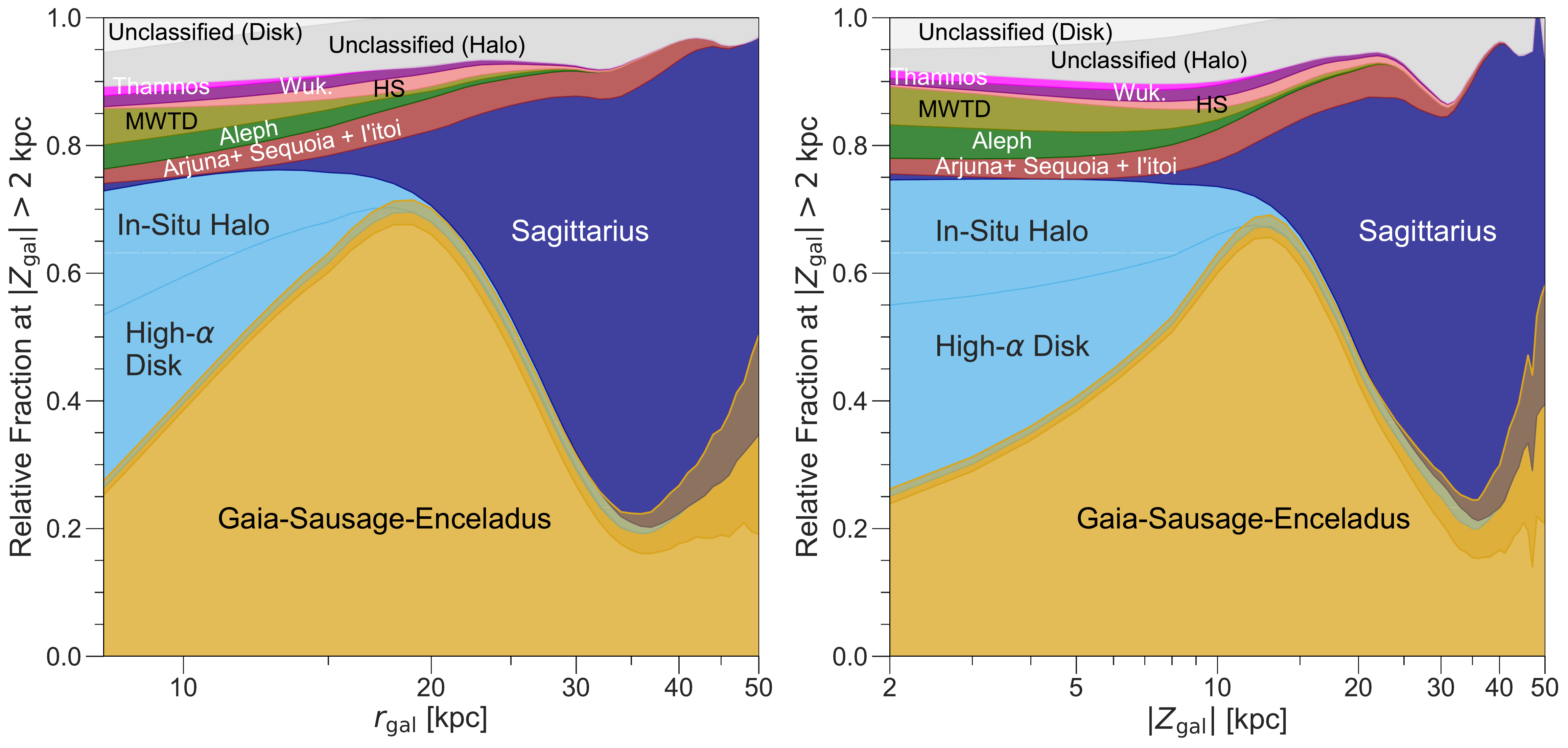}
\caption{Relative fractions of structures as a function of total Galactocentric distance, $r_{\rm{gal}}$ (\textbf{left}) and distance from the plane, $|Z_{\rm{gal}}|$ (\textbf{right}). Fractions have been corrected for the H3 Survey selection function.  The high-$\alpha$ disk is defined to lie at $e<0.5$ and the in-situ halo at $e>0.5$, though note these stars define a continuous distribution in eccentricity. Poisson error intervals are shown only for GSE to minimize crowding. GSE, the high-$\alpha$ disk \& in-situ halo, and Sgr together account for $\gtrsim75\%$ of all stars at all distances. The high-$\alpha$ disk and in-situ halo are prominent in the solar neighborhood and close to the plane, but their fraction rapidly declines to $<5\%$ by $|Z_{\rm{gal}}|\approx$10 kpc ($r_{\rm{gal}}\approx$15 kpc). GSE is the dominant component within $|Z_{\rm{gal}}|\approx10-20$ kpc ($r_{\rm{gal}}\approx15-25$ kpc) while the majority of stars at larger distances belong to Sgr. None of the other components contribute more than $\approx5\%$ at any distance. Note that the unclassified halo debris is mostly comprised of the low-eccentricity tail ($e<0.7$) of GSE (\S\ref{subsec:leftovers}).}
\label{fig:summary1}
\end{figure*}

\begin{figure*}
\centering
\includegraphics[width=0.95\linewidth]{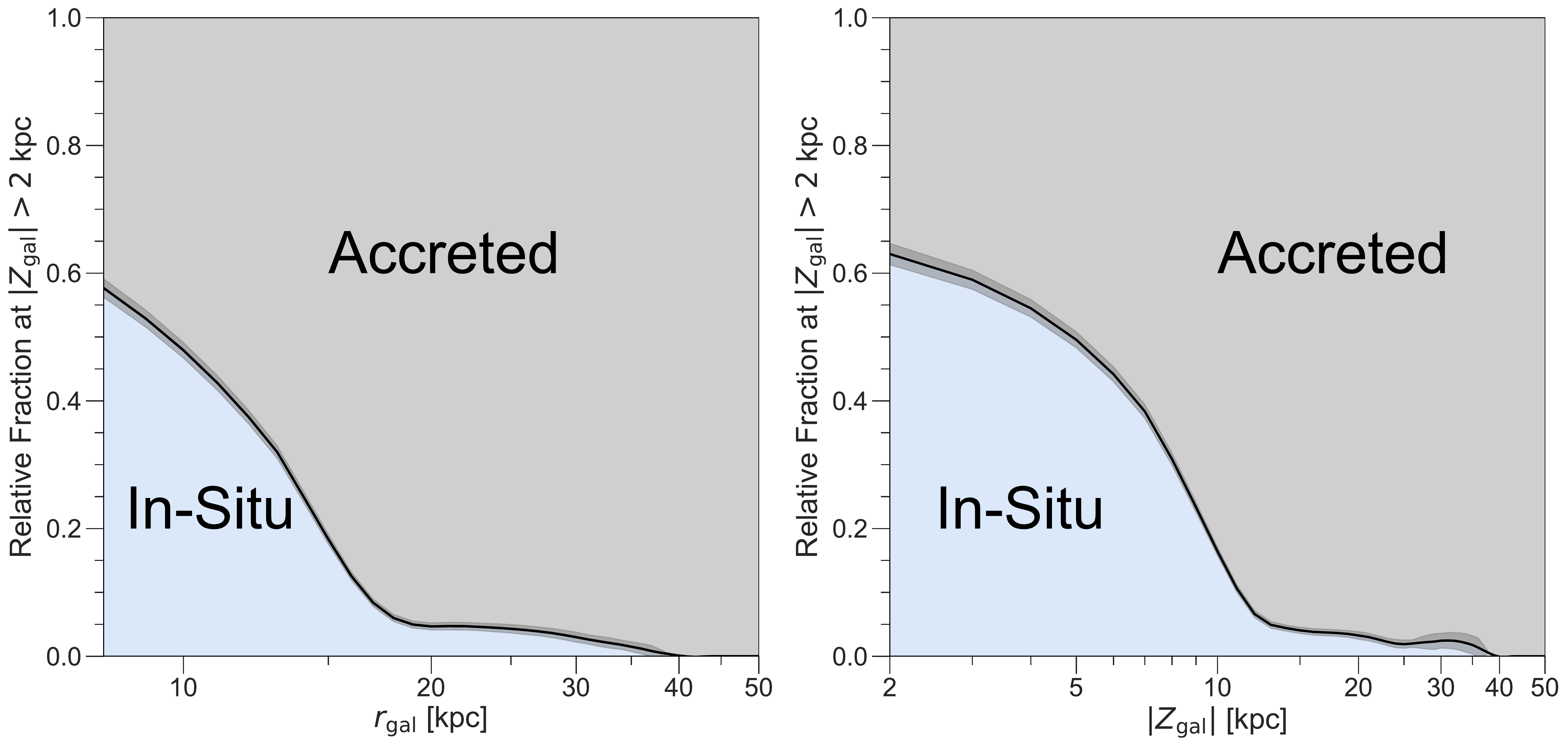}
\caption{Relative fractions of accreted and in-situ components as a function of total Galactocentric distance, $r_{\rm{gal}}$ (\textbf{left}) and distance from the plane, $|Z_{\rm{gal}}|$ (\textbf{right}). The high-$\alpha$ disk \& in-situ halo, the metal-weak thick disk, Aleph, and unclassified disk debris are classed as ``in-situ" while the other components (including the unclassified halo debris) are classed as ``accreted". Poisson errors are shown as a gray envelope. The in-situ components are largely confined within 20 kpc of the Galactic center and the Galactic plane. We caution that the in-situ halo stars at $|Z|>15$ kpc lie close to the selection boundary in Figure \ref{fig:insitu}, and may belong to other structures, so the in-situ relative fraction at these distances should be considered an upper limit. Simulations suggest that a high in-situ fraction around the solar circle that rapidly tapers off within $\sim30$ kpc suggests a quiescent recent accretion history, and that the halo was largely built at early times (see \S\ref{subsec:insitu}).}
\label{fig:accretedinsitu}
\end{figure*}

\begin{figure*}
\centering
\includegraphics[width=0.95\linewidth]{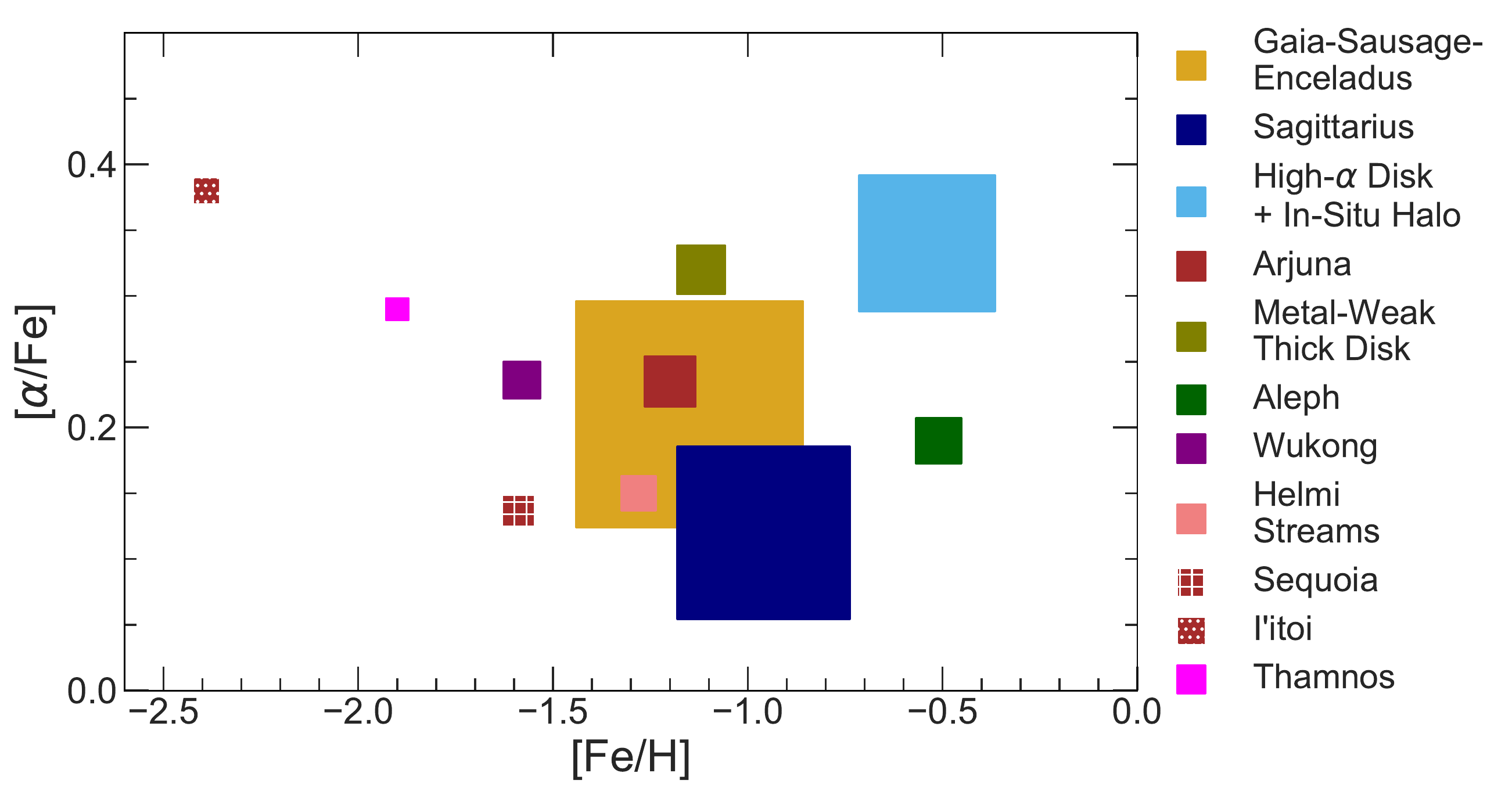}
\caption{Median [$\alpha$/Fe] vs median [Fe/H] for all structures identified in this work. The symbols are scaled linearly by the fractions in Table \ref{table:summary}, and the legend is sorted by decreasing fraction. The three main components of the $|Z_{\rm{gal}}|>2$ Galaxy -- GSE, Sgr, the high-$\alpha$ disk and in-situ halo -- all lie at [Fe/H]$>-1.5$, producing a halo that is metal-rich to at least 50 kpc. Sgr has the lowest [$\alpha$/Fe] consistent with it being accreted relatively recently. On the other hand, I'itoi and Thamnos are highly $\alpha$-enhanced and were perhaps accreted quite early in the history of the Galaxy.}
\label{fig:FeHaFe}
\end{figure*}

\subsection{Summary of Structure}
\label{subsec:structuresummary}
In this section we present a synopsis of the structures identified in this work.
\begin{enumerate}
    \item \textit{Gaia-Sausage-Enceladus} ($f=0.42$\footnote{Fraction of stars assigned to the structure, weighted to account for the selection function.}): The radial, head-on merger that dominates the metal-poor local halo, GSE is largely contained within $\approx30$ kpc, and displays a narrow MDF ([Fe/H]$=-1.15^{+0.24}_{-0.33}$, weighted) reminiscent of the local dwarfs Leo I and Fornax. Its metallicity and proposed accretion redshifts imply a stellar mass of $4-7\times10^{8}M_{\rm{\odot}}$. Its mean rotation is consistent with zero: $\langle V_{\rm{\phi}}\rangle = 1.04^{+1.26}_{-1.25}\ \rm{km\ s^{-1}}$, $\langle L_{\rm{z}}\rangle = 4.7^{+20.1}_{-10.5}\ \rm{kpc\ km\ s^{-1}}$. [\S\ref{subsec:ge}, Figure \ref{fig:ge}]
    \item \textit{Sagittarius} ($f=0.24$): One of the first known streams, Sgr displays a uniquely high $|L_{\rm{y}}|$ that allows for a clean selection leveraging full 6D phase-space. Its MDF is multi-peaked with a pronounced metal-poor tail. [\S\ref{subsec:sgr}, Figure \ref{fig:sgr}]
    \item \textit{High-$\alpha$ Disk and In-situ Halo} ($f=0.15$): A major component of the local Galaxy, the high-$\alpha$ disk, and its high-eccentricity  tail (the ``in-situ" halo) extend out to $|Z_{\rm{gal}}|\approx15$ kpc. Their eccentricity distribution is \textit{continuous}, ranging from very circular to highly eccentric, supporting scenarios in which the primordial high-$\alpha$ disk was disturbed by a merger event (likely GSE). [\S\ref{subsec:insitu}, Figure \ref{fig:insitu}]
    \item \textit{Arjuna, Sequoia, I'itoi} ($f=0.02,\ 0.01,\ 0.01$): The constituents of the high-energy retrograde halo -- Arjuna ([Fe/H] $= -1.2$), Sequoia ([Fe/H] $= -1.6$), and I'itoi ([Fe/H] $<-2$) -- are eccentric ($e\approx0.5-0.6$) and extend to highly retrograde orbits ($L_{\rm{z}}/[10^{3}\rm{\ kpc\ km\ s^{-1}}]>2$). Arjuna is the dominant component with $\gtrsim$2$\times$ the stars as Sequoia, and may be a distinct accreted structure or a hitherto unknown, highly retrograde extension of GSE. [\S\ref{subsec:retrograde}, Figure \ref{fig:retrograde}]
    \item \textit{Metal-weak thick disk} ($f=0.02$): The metal-poor extension of the high-$\alpha$ disk is only a minor component of the $|Z|>3$ kpc halo ($\lesssim5\%$) as suggested by local studies. [\S\ref{subsec:mwtd}, Figure \ref{fig:MWTD}]
    \item \textit{Aleph} ($f=0.02$): A highly circular structure ($e=0.13\pm0.06$) that rises $\approx$10 kpc off the plane. It is significantly enriched compared to typical halo structures ([Fe/H]$=-0.5$, [$\alpha$/Fe]$=0.2$), and may be associated with the enigmatic globular cluster Palomar 1. Whether it is an in-situ or ex-situ structure is under investigation. [\S\ref{subsec:aleph}, Figure \ref{fig:aleph}]
    \item \textit{Wukong} ($f=0.01$):  Comprising the ``prograde shards" of the halo, Wukong ([Fe/H$=-1.7$]) spans a wide range in energy and eccentricity reminiscent of massive structures like Sequoia and the Helmi Streams and is likely associated with the metal-poor GCs NGC 5024, NGC 5053, and ESO 280-SC06. [\S\ref{subsec:wukong}, Figure \ref{fig:wukong}]
    \item \textit{Helmi Streams} ($f=0.01$): Among the first halo structures discovered in integrals of motion, the Helmi Streams show a complex chemical population consistent with an extended star-formation history, have a stellar mass of $\approx0.5-1\times10^{8}M_{\rm{\odot}}$, and rise $\approx25$ kpc off the plane, as expected from local samples. [\S\ref{subsec:hs}, Figure \ref{fig:helmi}]
    \item \textit{Thamnos} ($f=0.01$): A recently discovered structure whose existence we confirm, Thamnos is among the most metal-poor structures in the halo ([Fe/H]=$-1.9$, [$\alpha$/Fe]=$0.3$). With a stellar mass that we estimate to be $\approx2\times10^{6}M_{\rm{\odot}}$, it is remarkable that Thamnos lies so deep in the potential ($r_{\rm{gal}}=9$ kpc), which makes it an exciting and accessible ($d_{\rm{helio}}=6$ kpc) target for near-field cosmology.
    [\S\ref{subsec:thamnos}, Figure \ref{fig:thamnos}]
\end{enumerate}

We have assigned $92\%$ (weighted) of our sample to these structures. Their properties are summarized in Table \ref{table:summary}. In Figure \ref{fig:summary2} we depict the various structures we have identified in \elzs, and in \elzs split by actions. As we have shown in this section, these populations that are clumped in \elzs not only occupy similar regions of phase-space, but also often define distinct chemical populations. Almost the entire halo can be accounted for as the superposition of these populations. 

Examining the remaining $8\%$, the ``unclassified debris" (\S\ref{subsec:leftovers}, Figure \ref{fig:leftovers}), $2\%$ are ``disk-like" and likely eccentric, metal-poor members of the high-$\alpha$ disk and in-situ halo. The remaining $5\%$ have higher-energy ``halo-like" orbits. A large fraction of these ($\approx4\%$ of the sample) are plausible members of GSE and the Helmi Streams. Within the survey footprint, any remaining unidentified systems must comprise, in aggregate, no more than $\approx1\%$ of the high-latitude Galaxy within 50 kpc.

\section{Discussion}
\label{sec:discussion}

\subsection{Relative Fractions of Substructure and the Mass Function of Accreted Structure}
\label{subsec:relfrac}

We are now in a position to examine which components of the halo are dominant at different distances. We depict the relative fractions of substructure, corrected for the selection function, as a function of Galactocentric distance and distance from the plane in Figure \ref{fig:summary1}. Fractions are computed in running 5 kpc bins and smoothed with a Savitzky-Golay filter (10 kpc window, second-order polynomial) for clarity. The unclassified debris are included as grey bands on the top -- note that we argued in \S\ref{subsec:leftovers} that a majority of these stars can be reasonably attributed to GSE. 

In agreement with local studies \citep[e.g.,][]{Bonaca17,Haywood18,DiMatteo19,Amarante20,Belokurov20} we find the high-$\alpha$ disk and its heated high-eccentricity tail (referred to as the ``in-situ halo" in this work, and ``Splash" in \citealt[][]{Belokurov20}) contribute the majority of stars at $|Z_{\rm{gal}}|\approx2$ kpc. We separate the high-$\alpha$ disk from the in-situ halo based on eccentricity ($e>0.5$) -- this is an arbitrary cut, since these populations define a continuous distribution in eccentricity (Figure \ref{fig:insitu}). The eccentric in-situ halo (blue hatched region) extends farther and rises to larger elevation than high-$\alpha$ disk. The high-$\alpha$ disk and in-situ halo fraction falls rapidly from $\sim50\%$ at $|Z_{\rm{gal}}|=2$ kpc to $<5\%$ beyond $|Z_{\rm{gal}}|=15$ kpc. At $|Z_{\rm{gal}}|\approx10$ kpc, GSE takes over and comprises $>50\%$ of the stars, and at farther distances, between $|Z_{\rm{gal}}|\approx25-50$ kpc, the majority of stars belong to Sagittarius. Similar trends are observed in the relative fractions as a function of $r_{\rm{gal}}$ as well. The span of GSE, largely contained within $35$ kpc, is in excellent agreement with observational estimates of its spatial extent \citep[e.g.,][]{Deason18,Lancaster19}. Figure \ref{fig:summary1} shows an uptick in the GSE relative fraction past 35 kpc -- this is mostly due to the fractions of all non-Sgr structures falling off, and the large Poisson noise at these distances (depicted in Figure \ref{fig:summary1} as a golden envelope).

At all distances the other structures comprise $<25\%$ of the sample. Our fractional budget clearly confirms the prediction of various simulations that at $r_{\rm{gal}}<50$ kpc the accreted halo is built by a handful of massive ($M_{\rm{star}}=10^{8}-10^{9} M_{\rm{\odot}}$) progenitors (GSE and Sgr in the MW's case) with a subdominant contribution from lower-mass galaxies and ultra-faints ($M_{\rm{star}}<10^{5} M_{\rm{\odot}}$) \citep[e.g.,][]{Deason16, Santistevan20, Fattahi20}.

The three structures that together comprise $\gtrsim75\%$ of the sample at all distances (the high-$\alpha$ disk + in-situ halo, GSE, Sgr) have particularly secure selections. Sgr due to its recent accretion is highly coherent in phase-space, the high-$\alpha$ disk and in-situ halo occupy a unique location in [Fe/H] vs [$\alpha$/Fe], and the GSE MDF shows one clear component that is well-fit by a simple analytical model and is reminiscent of the narrow MDFs of local dwarfs like Fornax and Leo I \citep{Kirby13}. The robustness of these selections inspire confidence in our conclusion that the mass function of accreted material is indeed ``top-heavy". In fact, the dominance of these three components is even more pronounced ($>80\%$) if a large fraction of the ``unclassified debris", as we have argued in \S\ref{subsec:leftovers}, is allocated to GSE.

\subsection{The Origin of the Stellar Halo}
\label{subsec:origin}
As per our accounting of structure, the halo is built almost entirely by the accretion of dwarf galaxies (e.g., GSE, Sgr, Arjuna), and the response of the Galaxy to their accretion (e.g., the heating of the high-$\alpha$ disk). The in-situ halo is an important component at $r_{\rm{gal}}<10$ kpc, but its relative fraction rapidly falls off at larger radii (Figure \ref{fig:summary1}). This can also be seen in Figure \ref{fig:accretedinsitu} where we combine all the components that likely originated in the Galaxy (high-$\alpha$ disk and in-situ halo, the metal-weak thick disk, Aleph, the unclassified disk debris), and compare their extent to the accreted components. As we detail in the remainder of this section, our inventory of structure leaves little room for other proposed in-situ (e.g., a smooth ``collapsed halo", outflows that deposit stars in the halo) or ex-situ components (e.g., dissolved globular clusters, an [Fe/H]$\sim-2.2$ spherical ``outer halo").

\subsubsection{The In-Situ Halo}
\label{subsubsec:insitu}

The mass budget and origin of the in-situ halo -- not just the heated disk, but also stars forming from gas that is smoothly accreted, stripped from infalling galaxies, or ejected in outflows -- is debated across simulations and may help constrain sub-grid physics such as star-formation and feedback prescriptions \citep[e.g.,][]{Font11, Cooper15, Pillepich18, Fattahi20,Yu20, Font20}. The extent and relative fraction of the in-situ halo (in particular, the heated disk) are also sensitive probes of the accretion history of the MW \citep[e.g.,][]{Zolotov09,Monachesi19}.

We find that other than the heated high-$\alpha$ disk, the $r_{\rm{gal}}<50$ kpc halo does not contain any other in-situ populations (Figures \ref{fig:summary1}, \ref{fig:accretedinsitu}). Aleph, which may have been heated or kicked from the disk may be an exception, but more work needs to be done to ascertain its nature. In any case, the high-$\alpha$ disk + in-situ halo and Aleph combined form a significant fraction of the halo only at $r_{\rm{gal}}\lesssim15$ kpc. We do not see a substantial fraction of eccentric stars with low-$\alpha$ or high-$\alpha$ disk-like chemistry in the distant halo (see hatched blue band in Figure \ref{fig:summary2}) as might be expected if stars in outflows formed a significant component of the halo \citep[e.g.,][]{Yu20}. Nor do we see a significant smoothly accreted component built out of cooling gas from the circumgalactic medium and cosmological inflows whose relative fraction is comparable to accreted material \citep[e.g.,][]{Cooper15}. These stars would resemble a smooth, isotropic, relatively metal-poor halo from monolithic collapse and not show the cogent structure associated with chemical evolution in dwarf galaxies in [Fe/H] vs [$\alpha$/Fe]. The final category of in-situ halo stars, stars formed from stripped gas from a dwarf galaxy, would be very difficult to tell apart from the debris of the dwarf, since they are likely to be aligned in phase-space as well as chemically -- progress could be made with precise ages to isolate stars that formed after the satellite's infall, as has been done for Sgr \citep[e.g.,][]{Siegel07,deboer15,Alfaro-Cuello19}.

The in-situ halo can also be used to constrain the MW's merger history. \citet[][]{Monachesi19} found that galaxies in the Auriga simulation suite with high in-situ fractions ($>50\%$) beyond their optical radii ($\gtrsim30$ kpc) either underwent a recent violent merger, or a very early ($>8$ Gyr ago) merger that ejected disk stars to large radii. Similarly \citet{Zolotov09} noted that simulated galaxies with quiescent merger histories (most of the mass in their halos was in place at $\sim9$ Gyrs) have a higher fraction of in-situ stars in their inner halo ($\sim20-50\%$) that rapidly tapers off by $\sim30$ kpc. This is very similar to what we find (Figure \ref{fig:accretedinsitu}), suggesting the bulk of the halo was already in place at early times and that the MW's recent growth has largely been quiescent (modulo Sgr, which is an important perturber of the disk, but due to its polar orbit, not a major contributor to the in-situ halo). This is a completely complementary way of accessing the MW's merger history and is in excellent agreement with the picture of a quiet merger history at later times inferred from GCs \citep[][]{Kruijssen19, Kruijssen20}, precise ages of the MW's various components \citep[][]{Bonaca20}, and the presence of a prominent break in the density profile \citep{Deason13}. This qualitative finding can be further refined through more detailed comparisons with simulations after accounting for the simple H3 selection function.

\subsubsection{The Ex-Situ Halo}

The fact that the distant halo is clumpy and highly structured has long been interpreted as strong evidence for an accretion-origin of the halo \citep[e.g.,][]{Newberg02,Bell08,Starkenburg09,Xue11,Schlaufman12}. Here we confirm this picture, and further refine this finding by quantifying the proportions of various in-situ and ex-situ components. As described in \S\ref{subsec:relfrac} and Figure \ref{fig:summary1} we find the accreted component almost entirely arises from a handful of massive ($M_{\rm{\star}}\sim10^{8}-10^{9}M_{\rm{\odot}}$) dwarfs. In the rest of this section we examine two popular scenarios about the nature of the accreted component in the context of our findings (the ``dual halo" and disrupting GCs).

\citet{Carollo07, Carollo10} and \citet{Beers12} used a local sample ($d_{\rm{helio}}<4$ kpc) and integrated orbits to infer that the halo was best described as a ``dual halo" \citep[but see][]{Schonrich11}. In their picture the dual halo is comprised of an inner in-situ halo ($r_{\rm{gal}}\lesssim15$ kpc, [Fe/H]$=-1.6$, small net prograde motion, high eccentricity) and an outer accreted halo ($r_{\rm{gal}}\sim20-50$ kpc,[Fe/H]$=-2.2$, mean retrograde motion, wide range of eccentricities) that are two fundamentally different populations. With the benefit of a post-\textit{Gaia} perspective, \citet{Helmi20} interpret the outer retrograde halo as GSE with a steep metallicity gradient and the inner halo as the heated high-$\alpha$ disk. This is partially motivated by the \textit{Gaia} color-magnitude diagram in the local halo which shows two prominent sequences that have been attributed to GSE and the heated high-$\alpha$ disk. \citet{Belokurov20} on the other hand argue the ``inner halo" is in fact GSE. 

Figures \ref{fig:summary2} and \ref{fig:accretedinsitu} help clarify this debate. The ``inner halo" ($r_{\rm{gal}}\lesssim15$ kpc) is predominantly built by GSE and the heated high-$\alpha$ disk, with GSE contributing a larger relative fraction. This is exactly as expected from simulations that find the inner halo to be a mixture of heated disk stars and accreted material, with the proportion varying with details of the accretion history of the galaxy \citep[e.g.,][]{Zolotov09,Tissera14,Monachesi19}. As for the \citet{Carollo10} ``outer halo" ($r_{\rm{gal}}\sim20-50$ kpc), after setting Sagittarius aside since it does not pass through the local halo, GSE is still a major component, but the retrograde Arjuna, Sequoia, and I'itoi contribute a significant fraction too, and perhaps explain the finding of a net retrograde motion. Though we note that considering only these structures at $r_{\rm{gal}}\sim20-50$ kpc produces only a mildly retrograde $\langle V_{\rm{\phi}}\rangle=12.5^{+6}_{-6}\ \rm{km}\ \rm{s}^{-1}$ (weighted). Also, at no distance does a very metal-poor (e.g., [Fe/H]$<-2$) component comprise a significant fraction of the Galaxy. We conclude by noting that no single population -- neither the in-situ halo nor GSE -- neatly maps onto either component of the dual halo, and more work needs to be done to understand the effects of extrapolating the nature of the distant Galaxy from energetic local halo orbits.

Several authors have hypothesized that stars born in GCs might contribute significantly to the stellar halo mass budget -- with estimates ranging from $\lesssim10\%$ to $\lesssim50\%$ \citep[e.g.,][]{Gnedin97,Schaerer11, Martell11,Martell16,Carretta16,Schiavon17,Koch19}. GCs are attractive candidates for building up at least some fraction of the halo because several of them are in the process of being tidally disrupted \citep[e.g.,][]{Grillmair06,Myeong17, Shipp18,Malhan18}. Further, several popular scenarios for GC formation assume they underwent drastic mass-loss at some point in their early history \citep[see][for a recent review]{Bastian18}. 

Post \textit{Gaia}-DR2 almost all halo GCs (those at high-energy, that are not on disk-like orbits) have been associated with phase-space structures seen in stars, strongly suggesting they were accreted along with some dwarf Galaxy \citep{Massari19,Kruijssen19,Myeong19,Kruijssen20, Forbes20}. This complicates the evaluation of the halo fraction arising from GCs, since in their phase-space coordinates these accreted GCs resemble field stars from their parent dwarf galaxies. 

However, there may still be a contribution from GCs associated with low-mass accreted dwarfs or ancient in-situ GCs that may have dissolved in the MW halo in the distant past. We find that at least within 50 kpc such GCs play a very limited role in building the halo. Being very conservative and allowing all the unclassified ``halo-like" debris (\S\ref{subsec:leftovers}) to emanate from GCs limits their contribution to $<6\%$ (weighted) at all distances. This upper limit is in excellent agreement with high-resolution simulations \citep[][]{Reina-Campos20} that find similarly low fractions ($2-5\%$), with recent searches for second-generation GC stars in the halo \citep[][]{Koch19,Hanke20} that find a low observed fraction of $2.6\pm0.2\%$ that they adjust to $<11\pm1\%$ to account for first-generation stars, and arguments based on BHB to blue-straggler ratios that found the halo ratio resembled dwarf galaxies and not GCs \citep{Deason15}.

\subsection{Prograde vs. Retrograde, and the Net Rotation of the Halo}
\label{subsec:protretro}
In local halo studies there appears to be an asymmetry in the distribution of accreted stars, with more retrograde than prograde structure. For instance, \citet{Helmi17} find $58-73\%$ of high-energy halo stars are retrograde \citep[see also][]{Myeong18b, Myeong18c}. More recently, \citet{Yuan20} recovered six new retrograde ``dynamically tagged groups"  compared to two prograde groups in LAMOST DR3. These findings bear echoes of the \citet{Carollo07, Carollo10} ``dual halo", whose ``outer halo" is retrograde. This asymmetry may simply be a selection effect, since it is easier to avoid contamination from the disk and in-situ halo on the retrograde side, and because some structures, such as Sgr, are not represented in the local halo. It might also be physical -- some models predict that the disk is more efficient at mixing structure accreted on prograde orbits compared to retrograde orbits \citep[e.g.,][]{Quinn86,Byrd86,Norris89,donghia10}. 

We observe no significant asymmetry in the distant halo. Setting the disk populations and unclassified debris aside, we find three prograde (Sagittarius, Helmi Streams, Wukong) and four retrograde (Arjuna, Sequoia, I'itoi, Thamnos) accreted structures. GSE shows net rotation consistent with zero ( $\langle V_{\rm{\phi}}\rangle = 1.04^{+1.26}_{-1.25}\ \rm{km\ s^{-1}}$). If Sequoia and Arjuna are indeed associated with GSE, the net rotation combining these two components is weakly retrograde $\langle V_{\rm{\phi}}\rangle = 9.4^{+1.7}_{-1.6}\ \rm{km\ s^{-1}}$. In terms of relative fractions, after setting the $\langle V_{\rm{\phi}}\rangle \sim0$ GSE aside, prograde stars outnumber retrograde stars $\approx$3:2 (mostly due to Sgr, excluding it results in $\approx$1:1). The net rotation of accreted material (also counting the halo-like unclassified debris) is $\langle V_{\rm{\phi}}\rangle = -25.23^{+2.54}_{-2.71}\ \rm{km\ s^{-1}}$ with Sgr, and $\langle V_{\rm{\phi}}\rangle = 5.7^{+1.7}_{-1.6}\ \rm{km\ s^{-1}}$ when Sgr is excluded. All numbers quoted here have been weighted to correct for the selection function. Assuming an isotropic distribution of infalling satellites, the numbers in this section suggest that the prograde satellites are not more efficiently disrupted than retrograde satellites, and that there is no strong mean rotation signal. 

\subsection{Interpreting the Halo in Chemical Space}
\label{subsec:chemistry}
The locations of various structures in [Fe/H] vs [$\alpha$/Fe], as listed in Table \ref{table:summary}, are depicted in Figure \ref{fig:FeHaFe}. The markers representing the structures are sized linearly as per their weighted relative fractions. The markers are all square, and not intended to reproduce the spread in abundances -- since we make hard cuts on the MDF to select some structures we are not well-positioned to make fair estimates of the spread. Broadly, in the [Fe/H]-[$\alpha$/Fe] plane, galaxies are expected to start off in the top-left (\afe-rich and \feh-poor), and end up towards the bottom right (\afe-poor and \feh-rich) as they evolve and Type Ia supernovae take over from Type II supernovae as the chief pollutants of the ISM \citep[e.g.,][]{Tinsley80,Matteucci86,Maiolino19}. This journey is interrupted when a galaxy is accreted and shredded by the Milky Way, and to first order its abundances are frozen in place.

We highlight some features of this space, while noting that detailed modeling \citep[e.g.,][]{Fernandez-Alvar18,Vincenzo19,Lian20} is required to deduce finer details. Thamnos and I'itoi are the most $\alpha$-rich and metal-poor structures. Based on its depth in the potential, we argued Thamnos was accreted at high-redshift ($z\approx1.5$), i.e., early in its chemical evolution. This may be the case for I'itoi as well, though it occurs at higher energy, which may mean it was accreted relatively recently but is simply very low-mass and formed stars inefficiently. At the other extreme of the plot, Sagittarius is the most $\alpha$-poor and metal-rich of all the structures in the halo, likely because it was accreted very recently ($z<1$, e.g., \citealt{Laporte18, Kruijssen20, Ruiz-Lara20}), after undergoing significant enrichment by Type Ia supernovae. GSE, with mass comparable to Sagittarius, is relatively $\alpha$-enhanced and metal-poor, which fits with the recent finding that it began interacting with the Milky Way at $z\approx2$ \citep{Bonaca20}, when we expect its chemical evolution to have been interrupted. The Helmi Streams, accreted at $z\sim0.5-1$ \citep{Koppelman19HS}, i.e., at a similar epoch as Sgr, have [$\alpha$/Fe] similar to Sgr, but are about 0.3 dex more metal-poor, exactly as expected for a structure $\sim10\times$ less massive \citep[][their Figure 2]{Lee15}. Wukong, Sequoia, and Arjuna are at intermediate locations between the two extremes of I'itoi and Sagittarius, and finer estimates of their masses and accretion redshifts are required to further interpret their [Fe/H] vs [$\alpha$/Fe] locations.

This figure is another way to see that the halo, at least out to 50 kpc, as seen in our sample, is relatively metal-rich compared to several earlier studies \citep[e.g.,][]{Carollo07,Carollo10,Xue15,Das16b,Liu18}. The three main components -- GSE, Sgr, the high-$\alpha$ disk and in-situ halo (those with the largest marker sizes in Figure \ref{fig:FeHaFe}) all lie almost entirely at [Fe/H]$>-1.5$. Only a small fraction of material can be attributed to the debris of metal-poor structures like I'itoi and Thamnos ([Fe/H]$\lesssim-2$). Even setting our definitions of various structures aside, we compute $\langle\rm{[Fe/H]}\rangle=-1.18^{+0.01}_{-0.01}$ (weighted) for all the accreted material taken together -- i.e., Sgr, GSE, Arjuna, Helmi Streams, Sequoia, Wukong, I'itoi, Thamnos, the unclassified halo-like debris \citep[see also][]{Conroy19b}.

\subsection{Caveats and Limitations}
\label{subsec:caveats}

Our census of the stellar halo is incomplete owing to the H3 Survey field locations, which are currently restricted to $|b|>40^\circ$ and Dec.$>-20^\circ$ (see Figure \ref{fig:data-1}). An accreted structure completely confined to in-plane orbits or the bulge, e.g., the ``ex-situ disk" \citep{Gomez17}, ``Kraken"/``Koala" \citep{Kruijssen19, Kruijssen20, Forbes20}, would be missed.  Furthermore, we systematically under-count stars from structures on orbits that spend most of their time close to the plane or in the Southern Hemisphere. Finally, recently accreted structures that have a strong on-sky coherence may be missed or biased in our existing fields. Assessing and correcting these biases will be the subject of future work.

The number of halo stars in our current sample sets an effective limit on the lowest stellar mass system we could plausibly detect. We can estimate the mass-completeness as follows: if the halo has $\sim10^{9}$ stars \citep[e.g.,][]{Deason19, Mackereth20}, and we are tracing it with 5684 stars, for every structure with $\sim175,000$ stars we find 1 star in the survey (assuming all structures are isotropic and completely mixed -- in detail we observe more stars from nearby structures). That is, from a $10^{6} M_{\odot}$ accreted galaxy we expect $\sim10$ stars in the sample. The detectability of such a structure would depend strongly on its location in phase-space/chemistry, e.g., the $\sim10^{6}\, M_{\odot}$ Thamnos stands out due to being very metal-poor/$\alpha$-rich in a region of phase-space that is populated by metal-rich GSE stars. Even if we are currently unable to identify some low-mass structures as distinct components of the halo, it is clear that taken together they play only a subdominant role in the overall mass budget (Figures \ref{fig:summary1}, \ref{fig:FeHaFe}, \S\ref{subsec:relfrac}).

Another limitation of this work is our decision to apply hard cuts to select various structures. Due to this choice we miss the tails of various distributions. This is particularly evident for GSE, whose low-eccentricity ($e<0.7$) tail appears as a contaminant in e.g., the initial phase-space selections for Thamnos and Wukong (see dashed histogram in the MDF panels of Figure \ref{fig:thamnos}, \ref{fig:wukong}). These GSE stars end up classed as ``unclassified halo debris". On the other hand, because we attribute all $e>0.7$ stars to GSE (after excluding the in-situ halo and Sgr) we miss the high-eccentricity tails of all the structures that follow it in our inventory. These stars are likely a very insignificant fraction of the GSE sample, as our GSE MDF is smooth, well-behaved, and modeled well as a single population. However, what is a small fraction for GSE might be a significant fraction of the other low-mass structures. This aspect of our work may be improved with probabilistic methods (discussed in \S\ref{subsec:inventory}), but for now we list it as a caveat.

Finally, we caution that individual structures identified in this work may not necessarily correspond to unique, accreted dwarf galaxies. It is possible that e.g., Wukong is comprised of multiple sub-populations corresponding to the modes in its MDF, or that GSE and Arjuna are linked. Simulations show the same accreted structure can deposit stars in surprisingly disparate regions of phase-space (e.g., \citealt[][]{Jean-Baptiste17,Lilleengen20,Elias20}), though this is typically not the case \citep[e.g.,][]{bj05_1, bj05_3,Pfeffer20}. Further analysis of these individual structures is necessary in order to link them to unique accreted systems.

\section{Summary}
\label{sec:summary} 

We have used the H3 Survey in combination with \textit{Gaia} data to conduct a detailed census of substructure beyond the solar neighborhood. Our sample extends to $50$ kpc, is unbiased in metallicity, arises from a simple selection function, and has full 6D phase-space coordinates along with [Fe/H] and [$\alpha$/Fe]. We find the following:

\begin{itemize}

    \item The distant Galaxy displays a high degree of structure in integrals of motion (energy, actions, angular momenta) and chemistry ([Fe/H], [$\alpha$/Fe]) -- spaces in which co-eval stars are expected to cluster for timescales longer than the age of the universe. [Figures \ref{fig:feh_slices}, \ref{fig:feh_orbit_slices}, \ref{fig:confusogram_feh}]
    
    \item $92\%$ of our sample can be assigned to one of the following structures: Sagittarius, Aleph, the high-$\alpha$ disk + in-situ halo (the heated high-$\alpha$ disk), the Helmi Streams, Thamnos, Arjuna, Sequoia, I'itoi, Wukong, \textit{Gaia}-Sausage-Enceladus, and the metal-weak thick disk -- our key findings on each structure are distilled in \S\ref{subsec:structuresummary}. This leaves us with $8\%$ of the sample (``unclassified debris", $2\%$ disk-like, and $6\%$ halo-like) that can be largely accounted for as artifacts of our sharp selection boundaries. [\S\ref{subsec:structuresummary}, Table \ref{table:summary}]
    
    \item The high-$\alpha$ disk, the in-situ halo, GSE, and Sgr account for $\gtrsim75\%$ of all stars at all distances.  The high-$\alpha$ disk and in-situ halo are a major component at $\lesssim10$ kpc ($\approx50\%$), but their relative fraction rapidly declines to $\lesssim10\%$ beyond 15 kpc. GSE dominates between $\approx15-25$ kpc and Sgr forms the bulk of the halo beyond $30$ kpc.  The accreted halo within 50 kpc is therefore mainly built out of a small number of $10^{8}-10^{9}M_{\odot}$ galaxies (GSE, Sgr). That is, the mass function of accreted material is ``top-heavy". This explains the metallicity of the halo ([Fe/H]$\approx-1.2$, see also \citealt{Conroy19b}) that we find to be more metal-rich than several previous studies. [\S\ref{subsec:relfrac}, Figures \ref{fig:summary1}, \ref{fig:accretedinsitu}, \ref{fig:FeHaFe}]
    
    \item This inventory of substructure leaves very limited room for other proposed constituents of the halo including a spherical, retrograde, [Fe/H]$\sim-2.2$ ``dual halo" beyond 25 kpc, dissolved globular clusters, stars deposited by outflows, or stars born from smoothly accreted gas. [\S\ref{subsec:origin}, Figure \ref{fig:summary1}]
    
    \item There is no preference for retrograde orbits in the distant Galaxy as has been observed in the local halo. GSE shows net rotation consistent with zero ($V_{\rm{\phi}}=1.0\pm1.3\ \rm{km\,s^{-1}}$). In fact, setting the disk populations aside, prograde stars outnumber retrograde stars $\approx3:2$. [\S\ref{subsec:protretro}, Table \ref{table:summary}]
    
\end{itemize}

It has long been recognized that the distant halo is highly structured, and that this likely indicates an accretion-origin. Here, we have confirmed this picture, and further refined it by quantifying the exact proportions and extents of various in-situ and ex-situ components. In forthcoming work we will present detailed characterizations of the identified structures. With future \textit{Gaia} data releases we will extend this work even further into the halo using the $\sim1000$ H3 giants extending out to $100$ kpc that were excluded from this work due to uncertain proper motions. By resolving the stellar halo into its constituent pieces we are delivering on the promise of Galactic Archaeology as a powerful tool to determine the assembly history of our Galaxy.

\facilities{MMT (Hectochelle), \textit{Gaia}}

\software{
    \package{IPython} \citep{ipython},
    \package{matplotlib} \citep{matplotlib},
    \package{numpy} \citep{numpy},
    \package{scipy} \citep{scipy},
    \package{jupyter} \citep{jupyter},
    \package{dynesty} \citep{Speagle19},
    \package{gala} \citep{gala1, gala2},
    \package{GalPot}
    \citep{McMillan17, Dehnen98},
    \package{Astropy}
    \citep{astropy1, astropy2},
    }

\acknowledgments{It is a pleasure to acknowledge illuminating conversations with Marion Dierickx, Diederik Kruijssen, GyuChul Myeong, Kareem El-Badry, Helmer Koppelman, Peter Senchyna and Vasily Belokurov. RPN gratefully acknowledges an Ashford Fellowship and Peirce Fellowship granted by Harvard University. CC acknowledges funding from the Packard foundation. YST is supported by the NASA Hubble Fellowship grant HST-HF2-51425.001 awarded by the Space Telescope Science Institute. We thank the Hectochelle operators Chun Ly, ShiAnne Kattner, Perry Berlind, and Mike Calkins, and the CfA and U. Arizona TACs for their continued support of the H3 Survey. The computations in this paper were run on the FASRC Cannon cluster supported by the FAS Division of Science Research Computing Group at Harvard University.

This work has made use of data from the European Space Agency (ESA) mission
{\it Gaia} (\url{https://www.cosmos.esa.int/gaia}), processed by the {\it Gaia}
Data Processing and Analysis Consortium (DPAC,
\url{https://www.cosmos.esa.int/web/gaia/dpac/consortium}) \citep{dr2ack1, dr2ack2}. Funding for the DPAC
has been provided by national institutions, in particular the institutions
participating in the {\it Gaia} Multilateral Agreement.}
    
\bibliography{MasterBiblio}

\begin{thebibliography}{}
\expandafter\ifx\csname natexlab\endcsname\relax\def\natexlab#1{#1}\fi

\bibitem[{{Adibekyan} {et~al.}(2012){Adibekyan}, {Sousa}, {Santos}, {Delgado
  Mena}, {Gonz{\'a}lez Hern{\'a}ndez}, {Israelian}, {Mayor}, \&
  {Khachatryan}}]{Adibekyan12}
{Adibekyan}, V.~Z., {Sousa}, S.~G., {Santos}, N.~C., {et~al.} 2012, \aap, 545,
  A32

\bibitem[{{Alfaro-Cuello} {et~al.}(2019){Alfaro-Cuello}, {Kacharov},
  {Neumayer}, {L{\"u}tzgendorf}, {Seth}, {B{\"o}ker}, {Kamann}, {Leaman}, {van
  de Ven}, {Bianchini}, {Watkins}, \& {Lyubenova}}]{Alfaro-Cuello19}
{Alfaro-Cuello}, M., {Kacharov}, N., {Neumayer}, N., {et~al.} 2019, \apj, 886,
  57

\bibitem[{{Amarante} {et~al.}(2020){Amarante}, {Smith}, \&
  {Boeche}}]{Amarante20}
{Amarante}, J. A.~S., {Smith}, M.~C., \& {Boeche}, C. 2020, \mnras, 492, 3816

\bibitem[{{Amorisco}(2017)}]{Amorisco17}
{Amorisco}, N.~C. 2017, \mnras, 464, 2882

\bibitem[{{Astropy Collaboration} {et~al.}(2013){Astropy Collaboration},
  {Robitaille}, {Tollerud}, {Greenfield}, {Droettboom}, {Bray}, {Aldcroft},
  {Davis}, {Ginsburg}, {Price-Whelan}, {Kerzendorf}, {Conley}, {Crighton},
  {Barbary}, {Muna}, {Ferguson}, {Grollier}, {Parikh}, {Nair}, {Unther},
  {Deil}, {Woillez}, {Conseil}, {Kramer}, {Turner}, {Singer}, {Fox}, {Weaver},
  {Zabalza}, {Edwards}, {Azalee Bostroem}, {Burke}, {Casey}, {Crawford},
  {Dencheva}, {Ely}, {Jenness}, {Labrie}, {Lim}, {Pierfederici}, {Pontzen},
  {Ptak}, {Refsdal}, {Servillat}, \& {Streicher}}]{astropy1}
{Astropy Collaboration}, {Robitaille}, T.~P., {Tollerud}, E.~J., {et~al.} 2013,
  \aap, 558, A33

\bibitem[{{Astropy Collaboration} {et~al.}(2018){Astropy Collaboration},
  {Price-Whelan}, {Sip{\H o}cz}, {G{\"u}nther}, {Lim}, {Crawford}, {Conseil},
  {Shupe}, {Craig}, {Dencheva}, {Ginsburg}, {VanderPlas}, {Bradley},
  {P{\'e}rez-Su{\'a}rez}, {de Val-Borro}, {Aldcroft}, {Cruz}, {Robitaille},
  {Tollerud}, {Ardelean}, {Babej}, {Bach}, {Bachetti}, {Bakanov}, {Bamford},
  {Barentsen}, {Barmby}, {Baumbach}, {Berry}, {Biscani}, {Boquien}, {Bostroem},
  {Bouma}, {Brammer}, {Bray}, {Breytenbach}, {Buddelmeijer}, {Burke},
  {Calderone}, {Cano Rodr{\'{\i}}guez}, {Cara}, {Cardoso}, {Cheedella},
  {Copin}, {Corrales}, {Crichton}, {D'Avella}, {Deil}, {Depagne}, {Dietrich},
  {Donath}, {Droettboom}, {Earl}, {Erben}, {Fabbro}, {Ferreira}, {Finethy},
  {Fox}, {Garrison}, {Gibbons}, {Goldstein}, {Gommers}, {Greco}, {Greenfield},
  {Groener}, {Grollier}, {Hagen}, {Hirst}, {Homeier}, {Horton}, {Hosseinzadeh},
  {Hu}, {Hunkeler}, {Ivezi{\'c}}, {Jain}, {Jenness}, {Kanarek}, {Kendrew},
  {Kern}, {Kerzendorf}, {Khvalko}, {King}, {Kirkby}, {Kulkarni}, {Kumar},
  {Lee}, {Lenz}, {Littlefair}, {Ma}, {Macleod}, {Mastropietro}, {McCully},
  {Montagnac}, {Morris}, {Mueller}, {Mumford}, {Muna}, {Murphy}, {Nelson},
  {Nguyen}, {Ninan}, {N{\"o}the}, {Ogaz}, {Oh}, {Parejko}, {Parley}, {Pascual},
  {Patil}, {Patil}, {Plunkett}, {Prochaska}, {Rastogi}, {Reddy Janga},
  {Sabater}, {Sakurikar}, {Seifert}, {Sherbert}, {Sherwood-Taylor}, {Shih},
  {Sick}, {Silbiger}, {Singanamalla}, {Singer}, {Sladen}, {Sooley},
  {Sornarajah}, {Streicher}, {Teuben}, {Thomas}, {Tremblay}, {Turner},
  {Terr{\'o}n}, {van Kerkwijk}, {de la Vega}, {Watkins}, {Weaver}, {Whitmore},
  {Woillez}, {Zabalza}, \& {Astropy Contributors}}]{astropy2}
{Astropy Collaboration}, {Price-Whelan}, A.~M., {Sip{\H o}cz}, B.~M., {et~al.}
  2018, \aj, 156, 123

\bibitem[{{Barkana} \& {Loeb}(1999)}]{Barkana99}
{Barkana}, R., \& {Loeb}, A. 1999, \apj, 523, 54

\bibitem[{{Bastian} \& {Lardo}(2018)}]{Bastian18}
{Bastian}, N., \& {Lardo}, C. 2018, \araa, 56, 83

\bibitem[{{Baumgardt} {et~al.}(2019){Baumgardt}, {Hilker}, {Sollima}, \&
  {Bellini}}]{Baumgardt19}
{Baumgardt}, H., {Hilker}, M., {Sollima}, A., \& {Bellini}, A. 2019, \mnras,
  482, 5138

\bibitem[{{Beane} {et~al.}(2019){Beane}, {Sanderson}, {Ness}, {Johnston},
  {Grion Filho}, {Mac Low}, {Angl{\'e}s-Alc{\'a}zar}, {Hogg}, \&
  {Laporte}}]{Beane19}
{Beane}, A., {Sanderson}, R.~E., {Ness}, M.~K., {et~al.} 2019, \apj, 883, 103

\bibitem[{{Beers} {et~al.}(2012){Beers}, {Carollo}, {Ivezi{\'c}}, {An},
  {Chiba}, {Norris}, {Freeman}, {Lee}, {Munn}, {Re Fiorentin}, {Sivarani},
  {Wilhelm}, {Yanny}, \& {York}}]{Beers12}
{Beers}, T.~C., {Carollo}, D., {Ivezi{\'c}}, {\v{Z}}., {et~al.} 2012, \apj,
  746, 34

\bibitem[{{Bell} {et~al.}(2008){Bell}, {Zucker}, {Belokurov}, {Sharma},
  {Johnston}, {Bullock}, {Hogg}, {Jahnke}, {de Jong}, {Beers}, {Evans},
  {Grebel}, {Ivezi{\'c}}, {Koposov}, {Rix}, {Schneider}, {Steinmetz}, \&
  {Zolotov}}]{Bell08}
{Bell}, E.~F., {Zucker}, D.~B., {Belokurov}, V., {et~al.} 2008, \apj, 680, 295

\bibitem[{{Bellazzini} {et~al.}(2006){Bellazzini}, {Newberg}, {Correnti},
  {Ferraro}, \& {Monaco}}]{Bellazzini06}
{Bellazzini}, M., {Newberg}, H.~J., {Correnti}, M., {Ferraro}, F.~R., \&
  {Monaco}, L. 2006, \aap, 457, L21

\bibitem[{{Belokurov} {et~al.}(2018){Belokurov}, {Erkal}, {Evans}, {Koposov},
  \& {Deason}}]{Belokurov18}
{Belokurov}, V., {Erkal}, D., {Evans}, N.~W., {Koposov}, S.~E., \& {Deason},
  A.~J. 2018, \mnras, 478, 611

\bibitem[{{Belokurov} {et~al.}(2020){Belokurov}, {Sanders}, {Fattahi}, {Smith},
  {Deason}, {Evans}, \& {Grand }}]{Belokurov20}
{Belokurov}, V., {Sanders}, J.~L., {Fattahi}, A., {et~al.} 2020, \mnras, 494,
  3880

\bibitem[{{Belokurov} {et~al.}(2014){Belokurov}, {Koposov}, {Evans},
  {Pe{\~n}arrubia}, {Irwin}, {Smith}, {Lewis}, {Gieles}, {Wilkinson},
  {Gilmore}, {Olszewski}, \& {Niederste-Ostholt}}]{Belokurov14}
{Belokurov}, V., {Koposov}, S.~E., {Evans}, N.~W., {et~al.} 2014, \mnras, 437,
  116

\bibitem[{{Bennett} \& {Bovy}(2019)}]{BennettBovy19}
{Bennett}, M., \& {Bovy}, J. 2019, \mnras, 482, 1417

\bibitem[{{Bensby} {et~al.}(2003){Bensby}, {Feltzing}, \&
  {Lundstr{\"o}m}}]{Bensby03}
{Bensby}, T., {Feltzing}, S., \& {Lundstr{\"o}m}, I. 2003, \aap, 410, 527

\bibitem[{{Bergemann} {et~al.}(2018){Bergemann}, {Sesar}, {Cohen}, {Serenelli},
  {Sheffield}, {Li}, {Casagrande}, {Johnston}, {Laporte}, {Price-Whelan},
  {Sch{\"o}nrich}, \& {Gould}}]{Bergemann18}
{Bergemann}, M., {Sesar}, B., {Cohen}, J.~G., {et~al.} 2018, \nat, 555, 334

\bibitem[{{Bian} {et~al.}(2020){Bian}, {Kewley}, {Groves}, \&
  {Dopita}}]{Bian20}
{Bian}, F., {Kewley}, L.~J., {Groves}, B., \& {Dopita}, M.~A. 2020, \mnras,
  493, 580

\bibitem[{{Bignone} {et~al.}(2019){Bignone}, {Helmi}, \& {Tissera}}]{Bignone19}
{Bignone}, L.~A., {Helmi}, A., \& {Tissera}, P.~B. 2019, \apjl, 883, L5

\bibitem[{{Binney} \& {Tremaine}(2008)}]{BT}
{Binney}, J., \& {Tremaine}, S. 2008, {Galactic Dynamics: Second Edition}

\bibitem[{{Boberg} {et~al.}(2015){Boberg}, {Friel}, \& {Vesperini}}]{Boberg15}
{Boberg}, O.~M., {Friel}, E.~D., \& {Vesperini}, E. 2015, \apj, 804, 109

\bibitem[{{Boberg} {et~al.}(2016){Boberg}, {Friel}, \& {Vesperini}}]{Boberg16}
---. 2016, \apj, 824, 5

\bibitem[{{Bonaca} {et~al.}(2017){Bonaca}, {Conroy}, {Wetzel}, {Hopkins}, \&
  {Kere{\v{s}}}}]{Bonaca17}
{Bonaca}, A., {Conroy}, C., {Wetzel}, A., {Hopkins}, P.~F., \& {Kere{\v{s}}},
  D. 2017, \apj, 845, 101

\bibitem[{{Bonaca} {et~al.}(2020){Bonaca}, {Conroy}, {Cargile}, {Naidu},
  {Johnson}, {Zaritsky}, {Ting}, {Caldwell}, {Han}, \& {van Dokkum}}]{Bonaca20}
{Bonaca}, A., {Conroy}, C., {Cargile}, P.~A., {et~al.} 2020, arXiv e-prints,
  arXiv:2004.11384

\bibitem[{{Bovy}(2015)}]{Bovy15}
{Bovy}, J. 2015, \apjs, 216, 29

\bibitem[{{Bovy} {et~al.}(2014){Bovy}, {Nidever}, {Rix}, {Girardi}, {Zasowski},
  {Chojnowski}, {Holtzman}, {Epstein}, {Frinchaboy}, {Hayden}, {Rodrigues},
  {Majewski}, {Johnson}, {Pinsonneault}, {Stello}, {Allende Prieto}, {Andrews},
  {Basu}, {Beers}, {Bizyaev}, {Burton}, {Chaplin}, {Cunha}, {Elsworth},
  {Garc{\'\i}a}, {Garc{\'\i}a-Her{\'n}andez}, {Garc{\'\i}a P{\'e}rez},
  {Hearty}, {Hekker}, {Kallinger}, {Kinemuchi}, {Koesterke},
  {M{\'e}sz{\'a}ros}, {Mosser}, {O'Connell}, {Oravetz}, {Pan}, {Robin},
  {Schiavon}, {Schneider}, {Schultheis}, {Serenelli}, {Shetrone}, {Silva
  Aguirre}, {Simmons}, {Skrutskie}, {Smith}, {Stassun}, {Weinberg}, {Wilson},
  \& {Zamora}}]{Bovy14}
{Bovy}, J., {Nidever}, D.~L., {Rix}, H.-W., {et~al.} 2014, \apj, 790, 127

\bibitem[{{Boylan-Kolchin} {et~al.}(2016){Boylan-Kolchin}, {Weisz}, {Bullock},
  \& {Cooper}}]{Boylan-Kolchin16}
{Boylan-Kolchin}, M., {Weisz}, D.~R., {Bullock}, J.~S., \& {Cooper}, M.~C.
  2016, \mnras, 462, L51

\bibitem[{{Boylan-Kolchin} {et~al.}(2015){Boylan-Kolchin}, {Weisz}, {Johnson},
  {Bullock}, {Conroy}, \& {Fitts}}]{Boylan-Kolchin15}
{Boylan-Kolchin}, M., {Weisz}, D.~R., {Johnson}, B.~D., {et~al.} 2015, \mnras,
  453, 1503

\bibitem[{{Brown} {et~al.}(2005){Brown}, {Vel{\'a}zquez}, \&
  {Aguilar}}]{Brown05}
{Brown}, A. G.~A., {Vel{\'a}zquez}, H.~M., \& {Aguilar}, L.~A. 2005, \mnras,
  359, 1287

\bibitem[{{Bullock} \& {Johnston}(2005)}]{bj05_1}
{Bullock}, J.~S., \& {Johnston}, K.~V. 2005, \apj, 635, 931

\bibitem[{{Byrd} {et~al.}(1986){Byrd}, {Saarinen}, \& {Valtonen}}]{Byrd86}
{Byrd}, G.~G., {Saarinen}, S., \& {Valtonen}, M.~J. 1986, \mnras, 220, 619

\bibitem[{{Cargile} {et~al.}(2019){Cargile}, {Conroy}, {Johnson}, {Ting},
  {Bonaca}, \& {Dotter}}]{Cargile19}
{Cargile}, P.~A., {Conroy}, C., {Johnson}, B.~D., {et~al.} 2019, arXiv
  e-prints, arXiv:1907.07690

\bibitem[{{Carlin} {et~al.}(2012){Carlin}, {Majewski}, {Casetti-Dinescu},
  {Law}, {Girard}, \& {Patterson}}]{Carlin12}
{Carlin}, J.~L., {Majewski}, S.~R., {Casetti-Dinescu}, D.~I., {et~al.} 2012,
  \apj, 744, 25

\bibitem[{{Carnall} {et~al.}(2019){Carnall}, {Leja}, {Johnson}, {McLure},
  {Dunlop}, \& {Conroy}}]{Carnall19}
{Carnall}, A.~C., {Leja}, J., {Johnson}, B.~D., {et~al.} 2019, \apj, 873, 44

\bibitem[{{Carollo} {et~al.}(2007){Carollo}, {Scarlata}, {Stiavelli}, {Wyse},
  \& {Mayer}}]{Carollo07}
{Carollo}, C.~M., {Scarlata}, C., {Stiavelli}, M., {Wyse}, R.~F.~G., \&
  {Mayer}, L. 2007, \apj, 658, 960

\bibitem[{{Carollo} {et~al.}(2010){Carollo}, {Beers}, {Chiba}, {Norris},
  {Freeman}, {Lee}, {Ivezi{\'c}}, {Rockosi}, \& {Yanny}}]{Carollo10}
{Carollo}, D., {Beers}, T.~C., {Chiba}, M., {et~al.} 2010, \apj, 712, 692

\bibitem[{{Carollo} {et~al.}(2019){Carollo}, {Chiba}, {Ishigaki}, {Freeman},
  {Beers}, {Lee}, {Tissera}, {Battistini}, \& {Primas}}]{Carollo19}
{Carollo}, D., {Chiba}, M., {Ishigaki}, M., {et~al.} 2019, \apj, 887, 22

\bibitem[{{Carraro} {et~al.}(2007){Carraro}, {Zinn}, \& {Moni
  Bidin}}]{Carraro07}
{Carraro}, G., {Zinn}, R., \& {Moni Bidin}, C. 2007, \aap, 466, 181

\bibitem[{{Carretta}(2016)}]{Carretta16}
{Carretta}, E. 2016, in IAU Symposium, Vol. 317, The General Assembly of Galaxy
  Halos: Structure, Origin and Evolution, ed. A.~{Bragaglia}, M.~{Arnaboldi},
  M.~{Rejkuba}, \& D.~{Romano}, 97--103

\bibitem[{{Chambers} {et~al.}(2016){Chambers}, {Magnier}, {Metcalfe},
  {Flewelling}, {Huber}, {Waters}, {Denneau}, {Draper}, {Farrow}, {Finkbeiner},
  {Holmberg}, {Koppenhoefer}, {Price}, {Rest}, {Saglia}, {Schlafly}, {Smartt},
  {Sweeney}, {Wainscoat}, {Burgett}, {Chastel}, {Grav}, {Heasley}, {Hodapp},
  {Jedicke}, {Kaiser}, {Kudritzki}, {Luppino}, {Lupton}, {Monet}, {Morgan},
  {Onaka}, {Shiao}, {Stubbs}, {Tonry}, {White}, {Ba{\~n}ados}, {Bell},
  {Bender}, {Bernard}, {Boegner}, {Boffi}, {Botticella}, {Calamida},
  {Casertano}, {Chen}, {Chen}, {Cole}, {Deacon}, {Frenk}, {Fitzsimmons},
  {Gezari}, {Gibbs}, {Goessl}, {Goggia}, {Gourgue}, {Goldman}, {Grant},
  {Grebel}, {Hambly}, {Hasinger}, {Heavens}, {Heckman}, {Henderson}, {Henning},
  {Holman}, {Hopp}, {Ip}, {Isani}, {Jackson}, {Keyes}, {Koekemoer}, {Kotak},
  {Le}, {Liska}, {Long}, {Lucey}, {Liu}, {Martin}, {Masci}, {McLean}, {Mindel},
  {Misra}, {Morganson}, {Murphy}, {Obaika}, {Narayan}, {Nieto-Santisteban},
  {Norberg}, {Peacock}, {Pier}, {Postman}, {Primak}, {Rae}, {Rai}, {Riess},
  {Riffeser}, {Rix}, {R{\"o}ser}, {Russel}, {Rutz}, {Schilbach}, {Schultz},
  {Scolnic}, {Strolger}, {Szalay}, {Seitz}, {Small}, {Smith}, {Soderblom},
  {Taylor}, {Thomson}, {Taylor}, {Thakar}, {Thiel}, {Thilker}, {Unger},
  {Urata}, {Valenti}, {Wagner}, {Walder}, {Walter}, {Watters}, {Werner},
  {Wood-Vasey}, \& {Wyse}}]{psack1}
{Chambers}, K.~C., {Magnier}, E.~A., {Metcalfe}, N., {et~al.} 2016, arXiv
  e-prints, arXiv:1612.05560

\bibitem[{{Chen} {et~al.}(2000){Chen}, {Nissen}, {Zhao}, {Zhang}, \&
  {Benoni}}]{Chen00}
{Chen}, Y.~Q., {Nissen}, P.~E., {Zhao}, G., {Zhang}, H.~W., \& {Benoni}, T.
  2000, \aaps, 141, 491

\bibitem[{{Chiba} \& {Beers}(2000)}]{Chiba00}
{Chiba}, M., \& {Beers}, T.~C. 2000, \aj, 119, 2843

\bibitem[{{Choi} {et~al.}(2016){Choi}, {Dotter}, {Conroy}, {Cantiello},
  {Paxton}, \& {Johnson}}]{Choi16}
{Choi}, J., {Dotter}, A., {Conroy}, C., {et~al.} 2016, \apj, 823, 102

\bibitem[{{Chou} {et~al.}(2007){Chou}, {Majewski}, {Cunha}, {Smith},
  {Patterson}, {Mart{\'\i}nez-Delgado}, {Law}, {Crane}, {Mu{\~n}oz}, {Garcia
  L{\'o}pez}, {Geisler}, \& {Skrutskie}}]{Chou07}
{Chou}, M.-Y., {Majewski}, S.~R., {Cunha}, K., {et~al.} 2007, \apj, 670, 346

\bibitem[{{Cohen} {et~al.}(2017){Cohen}, {Sesar}, {Bahnolzer}, {He},
  {Kulkarni}, {Prince}, {Bellm}, \& {Laher}}]{Cohen17}
{Cohen}, J.~G., {Sesar}, B., {Bahnolzer}, S., {et~al.} 2017, \apj, 849, 150

\bibitem[{{Conroy} {et~al.}(2018){Conroy}, {Bonaca}, {Naidu}, {Eisenstein},
  {Johnson}, {Dotter}, \& {Finkbeiner}}]{Conroy18}
{Conroy}, C., {Bonaca}, A., {Naidu}, R.~P., {et~al.} 2018, \apjl, 861, L16

\bibitem[{{Conroy} {et~al.}(2019{\natexlab{a}}){Conroy}, {Naidu}, {Zaritsky},
  {Bonaca}, {Cargile}, {Johnson}, \& {Caldwell}}]{Conroy19b}
{Conroy}, C., {Naidu}, R.~P., {Zaritsky}, D., {et~al.} 2019{\natexlab{a}},
  \apj, 887, 237

\bibitem[{{Conroy} {et~al.}(2019{\natexlab{b}}){Conroy}, {Bonaca}, {Cargile},
  {Johnson}, {Caldwell}, {Naidu}, {Zaritsky}, {Fabricant}, {Moran}, {Rhee},
  {Szentgyorgyi}, {Berlind}, {Calkins}, {Kattner}, \& {Ly}}]{Conroy19}
{Conroy}, C., {Bonaca}, A., {Cargile}, P., {et~al.} 2019{\natexlab{b}}, \apj,
  883, 107

\bibitem[{{Cooper} {et~al.}(2015){Cooper}, {Parry}, {Lowing}, {Cole}, \&
  {Frenk}}]{Cooper15}
{Cooper}, A.~P., {Parry}, O.~H., {Lowing}, B., {Cole}, S., \& {Frenk}, C. 2015,
  \mnras, 454, 3185

\bibitem[{{Cui} {et~al.}(2012){Cui}, {Zhao}, {Chu}, {Li}, {Li}, {Zhang}, {Su},
  {Yao}, {Wang}, {Xing}, {Li}, {Zhu}, {Wang}, {Gu}, {Luo}, {Xu}, {Zhang},
  {Liu}, {Zhang}, {Yang}, {Cao}, {Chen}, {Chen}, {Chen}, {Chen}, {Chu}, {Feng},
  {Gong}, {Hou}, {Hu}, {Hu}, {Hu}, {Jia}, {Jiang}, {Jiang}, {Jiang}, {Jin},
  {Li}, {Li}, {Li}, {Liu}, {Liu}, {Lu}, {Mao}, {Men}, {Qi}, {Qi}, {Shi},
  {Tang}, {Tao}, {Wang}, {Wang}, {Wang}, {Wang}, {Wang}, {Wang}, {Wang},
  {Wang}, {Wang}, {Wang}, {Wang}, {Wang}, {Xu}, {Xu}, {Yang}, {Yu}, {Yuan},
  {Yuan}, {Zhai}, {Zhang}, {Zhang}, {Zhang}, {Zhao}, {Zhou}, {Zhou}, {Zhu}, \&
  {Zou}}]{LAMOST}
{Cui}, X.-Q., {Zhao}, Y.-H., {Chu}, Y.-Q., {et~al.} 2012, Research in Astronomy
  and Astrophysics, 12, 1197

\bibitem[{{Das} \& {Binney}(2016)}]{Das16b}
{Das}, P., \& {Binney}, J. 2016, \mnras, 460, 1725

\bibitem[{{Das} {et~al.}(2016){Das}, {Williams}, \& {Binney}}]{Das16}
{Das}, P., {Williams}, A., \& {Binney}, J. 2016, \mnras, 463, 3169

\bibitem[{{de Boer} {et~al.}(2015){de Boer}, {Belokurov}, \&
  {Koposov}}]{deboer15}
{de Boer}, T.~J.~L., {Belokurov}, V., \& {Koposov}, S. 2015, \mnras, 451, 3489

\bibitem[{{De Silva} {et~al.}(2015){De Silva}, {Freeman}, {Bland-Hawthorn},
  {Martell}, {de Boer}, {Asplund}, {Keller}, {Sharma}, {Zucker}, {Zwitter},
  {Anguiano}, {Bacigalupo}, {Bayliss}, {Beavis}, {Bergemann}, {Campbell},
  {Cannon}, {Carollo}, {Casagrande}, {Casey}, {Da Costa}, {D'Orazi}, {Dotter},
  {Duong}, {Heger}, {Ireland}, {Kafle}, {Kos}, {Lattanzio}, {Lewis}, {Lin},
  {Lind}, {Munari}, {Nataf}, {O'Toole}, {Parker}, {Reid}, {Schlesinger},
  {Sheinis}, {Simpson}, {Stello}, {Ting}, {Traven}, {Watson}, {Wittenmyer},
  {Yong}, \& {{\v{Z}}erjal}}]{GALAH}
{De Silva}, G.~M., {Freeman}, K.~C., {Bland-Hawthorn}, J., {et~al.} 2015,
  \mnras, 449, 2604

\bibitem[{{Deason} {et~al.}(2011){Deason}, {Belokurov}, \& {Evans}}]{Deason11}
{Deason}, A.~J., {Belokurov}, V., \& {Evans}, N.~W. 2011, \mnras, 416, 2903

\bibitem[{{Deason} {et~al.}(2013){Deason}, {Belokurov}, {Evans}, \&
  {Johnston}}]{Deason13}
{Deason}, A.~J., {Belokurov}, V., {Evans}, N.~W., \& {Johnston}, K.~V. 2013,
  \apj, 763, 113

\bibitem[{{Deason} {et~al.}(2018){Deason}, {Belokurov}, {Koposov}, \&
  {Lancaster}}]{Deason18}
{Deason}, A.~J., {Belokurov}, V., {Koposov}, S.~E., \& {Lancaster}, L. 2018,
  \apjl, 862, L1

\bibitem[{{Deason} {et~al.}(2014){Deason}, {Belokurov}, {Koposov}, \&
  {Rockosi}}]{Deason14}
{Deason}, A.~J., {Belokurov}, V., {Koposov}, S.~E., \& {Rockosi}, C.~M. 2014,
  \apj, 787, 30

\bibitem[{{Deason} {et~al.}(2019){Deason}, {Belokurov}, \&
  {Sanders}}]{Deason19}
{Deason}, A.~J., {Belokurov}, V., \& {Sanders}, J.~L. 2019, \mnras, 490, 3426

\bibitem[{{Deason} {et~al.}(2015){Deason}, {Belokurov}, \& {Weisz}}]{Deason15}
{Deason}, A.~J., {Belokurov}, V., \& {Weisz}, D.~R. 2015, \mnras, 448, L77

\bibitem[{{Deason} {et~al.}(2016){Deason}, {Mao}, \& {Wechsler}}]{Deason16}
{Deason}, A.~J., {Mao}, Y.-Y., \& {Wechsler}, R.~H. 2016, \apj, 821, 5

\bibitem[{{Dehnen} \& {Binney}(1998)}]{Dehnen98}
{Dehnen}, W., \& {Binney}, J.~J. 1998, \mnras, 298, 387

\bibitem[{{Di Matteo} {et~al.}(2019){Di Matteo}, {Haywood}, {Lehnert}, {Katz},
  {Khoperskov}, {Snaith}, {G{\'o}mez}, \& {Robichon}}]{DiMatteo19}
{Di Matteo}, P., {Haywood}, M., {Lehnert}, M.~D., {et~al.} 2019, \aap, 632, A4

\bibitem[{{Dierickx} \& {Loeb}(2017)}]{Dierickx17}
{Dierickx}, M. I.~P., \& {Loeb}, A. 2017, \apj, 836, 92

\bibitem[{{D'Onghia} {et~al.}(2010){D'Onghia}, {Springel}, {Hernquist}, \&
  {Keres}}]{donghia10}
{D'Onghia}, E., {Springel}, V., {Hernquist}, L., \& {Keres}, D. 2010, \apj,
  709, 1138

\bibitem[{{Dormand} \& {Prince}(1978)}]{DormandPrince}
{Dormand}, J.~R., \& {Prince}, P.~J. 1978, Celestial Mechanics, 18, 223

\bibitem[{{Drimmel} \& {Poggio}(2018)}]{Drimmel18}
{Drimmel}, R., \& {Poggio}, E. 2018, Research Notes of the American
  Astronomical Society, 2, 210

\bibitem[{{D'Souza} \& {Bell}(2018)}]{dsouza18}
{D'Souza}, R., \& {Bell}, E.~F. 2018, \mnras, 474, 5300

\bibitem[{{Edvardsson} {et~al.}(1993){Edvardsson}, {Andersen}, {Gustafsson},
  {Lambert}, {Nissen}, \& {Tomkin}}]{Edvardsson93}
{Edvardsson}, B., {Andersen}, J., {Gustafsson}, B., {et~al.} 1993, \aap, 500,
  391

\bibitem[{{Eggen} {et~al.}(1962){Eggen}, {Lynden-Bell}, \& {Sandage}}]{Eggen62}
{Eggen}, O.~J., {Lynden-Bell}, D., \& {Sandage}, A.~R. 1962, \apj, 136, 748

\bibitem[{{Elias} {et~al.}(2020){Elias}, {Sales}, {Helmi}, \&
  {Hernquist}}]{Elias20}
{Elias}, L.~M., {Sales}, L.~V., {Helmi}, A., \& {Hernquist}, L. 2020, arXiv
  e-prints, arXiv:2003.03381

\bibitem[{{Evans}(2020)}]{Evans20}
{Evans}, N.~W. 2020, arXiv e-prints, arXiv:2002.05740

\bibitem[{{Everall} \& {Das}(2020)}]{Everall20}
{Everall}, A., \& {Das}, P. 2020, \mnras, 493, 2042

\bibitem[{{Fardal} {et~al.}(2019){Fardal}, {van der Marel}, {Law}, {Sohn},
  {Sesar}, {Hernitschek}, \& {Rix}}]{Fardal19}
{Fardal}, M.~A., {van der Marel}, R.~P., {Law}, D.~R., {et~al.} 2019, \mnras,
  483, 4724

\bibitem[{{Fattahi} {et~al.}(2020){Fattahi}, {Deason}, {Frenk}, {Simpson},
  {Gomez}, {Grand }, {Monachesi}, {Marinacci}, \& {Pakmor}}]{Fattahi20}
{Fattahi}, A., {Deason}, A.~J., {Frenk}, C.~S., {et~al.} 2020, arXiv e-prints,
  arXiv:2002.12043

\bibitem[{{Fern{\'a}ndez-Alvar} {et~al.}(2018){Fern{\'a}ndez-Alvar}, {Carigi},
  {Schuster}, {Hayes}, {{\'A}vila-Vergara}, {Majewski}, {Allende Prieto},
  {Beers}, {S{\'a}nchez}, {Zamora}, {Garc{\'\i}a-Hern{\'a}ndez}, {Tang},
  {Fern{\'a}ndez-Trincado}, {Tissera}, {Geisler}, \&
  {Villanova}}]{Fernandez-Alvar18}
{Fern{\'a}ndez-Alvar}, E., {Carigi}, L., {Schuster}, W.~J., {et~al.} 2018,
  \apj, 852, 50

\bibitem[{{Feuillet} {et~al.}(2020){Feuillet}, {Feltzing}, {Sahlholdt}, \&
  {Casagrande}}]{Feuillet20}
{Feuillet}, D.~K., {Feltzing}, S., {Sahlholdt}, C., \& {Casagrande}, L. 2020,
  arXiv e-prints, arXiv:2003.11039

\bibitem[{{Flewelling} {et~al.}(2016){Flewelling}, {Magnier}, {Chambers},
  {Heasley}, {Holmberg}, {Huber}, {Sweeney}, {Waters}, {Calamida}, {Casertano},
  {Chen}, {Farrow}, {Hasinger}, {Henderson}, {Long}, {Metcalfe}, {Narayan},
  {Nieto-Santisteban}, {Norberg}, {Rest}, {Saglia}, {Szalay}, {Thakar},
  {Tonry}, {Valenti}, {Werner}, {White}, {Denneau}, {Draper}, {Hodapp},
  {Jedicke}, {Kaiser}, {Kudritzki}, {Price}, {Wainscoat}, {Builders},
  {Chastel}, {McLean}, {Postman}, \& {Shiao}}]{psack2}
{Flewelling}, H.~A., {Magnier}, E.~A., {Chambers}, K.~C., {et~al.} 2016, arXiv
  e-prints, arXiv:1612.05243

\bibitem[{{Font} {et~al.}(2006){Font}, {Johnston}, {Bullock}, \&
  {Robertson}}]{bj05_3}
{Font}, A.~S., {Johnston}, K.~V., {Bullock}, J.~S., \& {Robertson}, B.~E. 2006,
  \apj, 638, 585

\bibitem[{{Font} {et~al.}(2011){Font}, {McCarthy}, {Crain}, {Theuns}, {Schaye},
  {Wiersma}, \& {Dalla Vecchia}}]{Font11}
{Font}, A.~S., {McCarthy}, I.~G., {Crain}, R.~A., {et~al.} 2011, \mnras, 416,
  2802

\bibitem[{{Font} {et~al.}(2020){Font}, {McCarthy}, {Poole-Mckenzie},
  {Stafford}, {Brown}, {Schaye}, {Crain}, {Theuns}, \& {Schaller}}]{Font20}
{Font}, A.~S., {McCarthy}, I.~G., {Poole-Mckenzie}, R., {et~al.} 2020, arXiv
  e-prints, arXiv:2004.01914

\bibitem[{{Forbes}(2020)}]{Forbes20}
{Forbes}, D.~A. 2020, \mnras, 493, 847

\bibitem[{{Forbes} \& {Bridges}(2010)}]{Forbes10}
{Forbes}, D.~A., \& {Bridges}, T. 2010, \mnras, 404, 1203

\bibitem[{{Frebel} {et~al.}(2010){Frebel}, {Simon}, {Geha}, \&
  {Willman}}]{Frebel10}
{Frebel}, A., {Simon}, J.~D., {Geha}, M., \& {Willman}, B. 2010, \apj, 708, 560

\bibitem[{{Freeman} \& {Bland-Hawthorn}(2002)}]{Freeman02}
{Freeman}, K., \& {Bland-Hawthorn}, J. 2002, \araa, 40, 487

\bibitem[{{Fuhrmann}(1998)}]{Fuhrmann98}
{Fuhrmann}, K. 1998, \aap, 338, 161

\bibitem[{{Gaia Collaboration} {et~al.}(2016){Gaia Collaboration}, {Prusti},
  {de Bruijne}, {Brown}, {Vallenari}, {Babusiaux}, {Bailer-Jones}, {Bastian},
  {Biermann}, {Evans}, {Eyer}, {Jansen}, {Jordi}, {Klioner}, {Lammers},
  {Lindegren}, {Luri}, {Mignard}, {Milligan}, {Panem}, {Poinsignon},
  {Pourbaix}, {Randich}, {Sarri}, {Sartoretti}, {Siddiqui}, {Soubiran},
  {Valette}, {van Leeuwen}, {Walton}, {Aerts}, {Arenou}, {Cropper}, {Drimmel},
  {H{\o}g}, {Katz}, {Lattanzi}, {O'Mullane}, {Grebel}, {Holland}, {Huc},
  {Passot}, {Bramante}, {Cacciari}, {Casta{\~n}eda}, {Chaoul}, {Cheek}, {De
  Angeli}, {Fabricius}, {Guerra}, {Hern{\'a}ndez}, {Jean-Antoine-Piccolo},
  {Masana}, {Messineo}, {Mowlavi}, {Nienartowicz}, {Ord{\'o}{\~n}ez-Blanco},
  {Panuzzo}, {Portell}, {Richards}, {Riello}, {Seabroke}, {Tanga},
  {Th{\'e}venin}, {Torra}, {Els}, {Gracia-Abril}, {Comoretto},
  {Garcia-Reinaldos}, {Lock}, {Mercier}, {Altmann}, {Andrae}, {Astraatmadja},
  {Bellas-Velidis}, {Benson}, {Berthier}, {Blomme}, {Busso}, {Carry},
  {Cellino}, {Clementini}, {Cowell}, {Creevey}, {Cuypers}, {Davidson}, {De
  Ridder}, {de Torres}, {Delchambre}, {Dell'Oro}, {Ducourant}, {Fr{\'e}mat},
  {Garc{\'\i}a-Torres}, {Gosset}, {Halbwachs}, {Hambly}, {Harrison}, {Hauser},
  {Hestroffer}, {Hodgkin}, {Huckle}, {Hutton}, {Jasniewicz}, {Jordan},
  {Kontizas}, {Korn}, {Lanzafame}, {Manteiga}, {Moitinho}, {Muinonen},
  {Osinde}, {Pancino}, {Pauwels}, {Petit}, {Recio-Blanco}, {Robin}, {Sarro},
  {Siopis}, {Smith}, {Smith}, {Sozzetti}, {Thuillot}, {van Reeven}, {Viala},
  {Abbas}, {Abreu Aramburu}, {Accart}, {Aguado}, {Allan}, {Allasia},
  {Altavilla}, {{\'A}lvarez}, {Alves}, {Anderson}, {Andrei}, {Anglada Varela},
  {Antiche}, {Antoja}, {Ant{\'o}n}, {Arcay}, {Atzei}, {Ayache}, {Bach},
  {Baker}, {Balaguer-N{\'u}{\~n}ez}, {Barache}, {Barata}, {Barbier}, {Barblan},
  {Baroni}, {Barrado y Navascu{\'e}s}, {Barros}, {Barstow}, {Becciani},
  {Bellazzini}, {Bellei}, {Bello Garc{\'\i}a}, {Belokurov}, {Bendjoya},
  {Berihuete}, {Bianchi}, {Bienaym{\'e}}, {Billebaud}, {Blagorodnova},
  {Blanco-Cuaresma}, {Boch}, {Bombrun}, {Borrachero}, {Bouquillon}, {Bourda},
  {Bouy}, {Bragaglia}, {Breddels}, {Brouillet}, {Br{\"u}semeister},
  {Bucciarelli}, {Budnik}, {Burgess}, {Burgon}, {Burlacu}, {Busonero}, {Buzzi},
  {Caffau}, {Cambras}, {Campbell}, {Cancelliere}, {Cantat-Gaudin}, {Carlucci},
  {Carrasco}, {Castellani}, {Charlot}, {Charnas}, {Charvet}, {Chassat},
  {Chiavassa}, {Clotet}, {Cocozza}, {Collins}, {Collins}, {Costigan}, {Crifo},
  {Cross}, {Crosta}, {Crowley}, {Dafonte}, {Damerdji}, {Dapergolas}, {David},
  {David}, {De Cat}, {de Felice}, {de Laverny}, {De Luise}, {De March}, {de
  Martino}, {de Souza}, {Debosscher}, {del Pozo}, {Delbo}, {Delgado},
  {Delgado}, {di Marco}, {Di Matteo}, {Diakite}, {Distefano}, {Dolding}, {Dos
  Anjos}, {Drazinos}, {Dur{\'a}n}, {Dzigan}, {Ecale}, {Edvardsson}, {Enke},
  {Erdmann}, {Escolar}, {Espina}, {Evans}, {Eynard Bontemps}, {Fabre},
  {Fabrizio}, {Faigler}, {Falc{\~a}o}, {Farr{\`a}s Casas}, {Faye}, {Federici},
  {Fedorets}, {Fern{\'a}ndez-Hern{\'a}ndez}, {Fernique}, {Fienga}, {Figueras},
  {Filippi}, {Findeisen}, {Fonti}, {Fouesneau}, {Fraile}, {Fraser}, {Fuchs},
  {Furnell}, {Gai}, {Galleti}, {Galluccio}, {Garabato}, {Garc{\'\i}a-Sedano},
  {Gar{\'e}}, {Garofalo}, {Garralda}, {Gavras}, {Gerssen}, {Geyer}, {Gilmore},
  {Girona}, {Giuffrida}, {Gomes}, {Gonz{\'a}lez-Marcos},
  {Gonz{\'a}lez-N{\'u}{\~n}ez}, {Gonz{\'a}lez-Vidal}, {Granvik}, {Guerrier},
  {Guillout}, {Guiraud}, {G{\'u}rpide}, {Guti{\'e}rrez-S{\'a}nchez}, {Guy},
  {Haigron}, {Hatzidimitriou}, {Haywood}, {Heiter}, {Helmi}, {Hobbs},
  {Hofmann}, {Holl}, {Holland }, {Hunt}, {Hypki}, {Icardi}, {Irwin}, {Jevardat
  de Fombelle}, {Jofr{\'e}}, {Jonker}, {Jorissen}, {Julbe}, {Karampelas},
  {Kochoska}, {Kohley}, {Kolenberg}, {Kontizas}, {Koposov}, {Kordopatis},
  {Koubsky}, {Kowalczyk}, {Krone-Martins}, {Kudryashova}, {Kull}, {Bachchan},
  {Lacoste-Seris}, {Lanza}, {Lavigne}, {Le Poncin-Lafitte}, {Lebreton},
  {Lebzelter}, {Leccia}, {Leclerc}, {Lecoeur-Taibi}, {Lemaitre}, {Lenhardt},
  {Leroux}, {Liao}, {Licata}, {Lindstr{\o}m}, {Lister}, {Livanou}, {Lobel},
  {L{\"o}ffler}, {L{\'o}pez}, {Lopez-Lozano}, {Lorenz}, {Loureiro},
  {MacDonald}, {Magalh{\~a}es Fernandes}, {Managau}, {Mann}, {Mantelet},
  {Marchal}, {Marchant}, {Marconi}, {Marie}, {Marinoni}, {Marrese},
  {Marschalk{\'o}}, {Marshall}, {Mart{\'\i}n-Fleitas}, {Martino}, {Mary},
  {Matijevi{\v{c}}}, {Mazeh}, {McMillan}, {Messina}, {Mestre}, {Michalik},
  {Millar}, {Miranda}, {Molina}, {Molinaro}, {Molinaro}, {Moln{\'a}r},
  {Moniez}, {Montegriffo}, {Monteiro}, {Mor}, {Mora}, {Morbidelli}, {Morel},
  {Morgenthaler}, {Morley}, {Morris}, {Mulone}, {Muraveva}, {Musella},
  {Narbonne}, {Nelemans}, {Nicastro}, {Noval}, {Ord{\'e}novic},
  {Ordieres-Mer{\'e}}, {Osborne}, {Pagani}, {Pagano}, {Pailler}, {Palacin},
  {Palaversa}, {Parsons}, {Paulsen}, {Pecoraro}, {Pedrosa}, {Pentik{\"a}inen},
  {Pereira}, {Pichon}, {Piersimoni}, {Pineau}, {Plachy}, {Plum}, {Poujoulet},
  {Pr{\v{s}}a}, {Pulone}, {Ragaini}, {Rago}, {Rambaux}, {Ramos-Lerate},
  {Ranalli}, {Rauw}, {Read}, {Regibo}, {Renk}, {Reyl{\'e}}, {Ribeiro},
  {Rimoldini}, {Ripepi}, {Riva}, {Rixon}, {Roelens}, {Romero-G{\'o}mez},
  {Rowell}, {Royer}, {Rudolph}, {Ruiz-Dern}, {Sadowski}, {Sagrist{\`a}
  Sell{\'e}s}, {Sahlmann}, {Salgado}, {Salguero}, {Sarasso}, {Savietto},
  {Schnorhk}, {Schultheis}, {Sciacca}, {Segol}, {Segovia}, {Segransan},
  {Serpell}, {Shih}, {Smareglia}, {Smart}, {Smith}, {Solano}, {Solitro},
  {Sordo}, {Soria Nieto}, {Souchay}, {Spagna}, {Spoto}, {Stampa}, {Steele},
  {Steidelm{\"u}ller}, {Stephenson}, {Stoev}, {Suess}, {S{\"u}veges}, {Surdej},
  {Szabados}, {Szegedi-Elek}, {Tapiador}, {Taris}, {Tauran}, {Taylor},
  {Teixeira}, {Terrett}, {Tingley}, {Trager}, {Turon}, {Ulla}, {Utrilla},
  {Valentini}, {van Elteren}, {Van Hemelryck}, {van Leeuwen}, {Varadi},
  {Vecchiato}, {Veljanoski}, {Via}, {Vicente}, {Vogt}, {Voss}, {Votruba},
  {Voutsinas}, {Walmsley}, {Weiler}, {Weingrill}, {Werner}, {Wevers},
  {Whitehead}, {Wyrzykowski}, {Yoldas}, {{\v{Z}}erjal}, {Zucker}, {Zurbach},
  {Zwitter}, {Alecu}, {Allen}, {Allende Prieto}, {Amorim},
  {Anglada-Escud{\'e}}, {Arsenijevic}, {Azaz}, {Balm}, {Beck}, {Bernstein},
  {Bigot}, {Bijaoui}, {Blasco}, {Bonfigli}, {Bono}, {Boudreault}, {Bressan},
  {Brown}, {Brunet}, {Bunclark}, {Buonanno}, {Butkevich}, {Carret}, {Carrion},
  {Chemin}, {Ch{\'e}reau}, {Corcione}, {Darmigny}, {de Boer}, {de Teodoro}, {de
  Zeeuw}, {Delle Luche}, {Domingues}, {Dubath}, {Fodor}, {Fr{\'e}zouls},
  {Fries}, {Fustes}, {Fyfe}, {Gallardo}, {Gallegos}, {Gardiol}, {Gebran},
  {Gomboc}, {G{\'o}mez}, {Grux}, {Gueguen}, {Heyrovsky}, {Hoar}, {Iannicola},
  {Isasi Parache}, {Janotto}, {Joliet}, {Jonckheere}, {Keil}, {Kim},
  {Klagyivik}, {Klar}, {Knude}, {Kochukhov}, {Kolka}, {Kos}, {Kutka}, {Lainey},
  {LeBouquin}, {Liu}, {Loreggia}, {Makarov}, {Marseille}, {Martayan},
  {Martinez-Rubi}, {Massart}, {Meynadier}, {Mignot}, {Munari}, {Nguyen},
  {Nordlander}, {Ocvirk}, {O'Flaherty}, {Olias Sanz}, {Ortiz}, {Osorio},
  {Oszkiewicz}, {Ouzounis}, {Palmer}, {Park}, {Pasquato}, {Peltzer}, {Peralta},
  {P{\'e}turaud}, {Pieniluoma}, {Pigozzi}, {Poels}, {Prat}, {Prod'homme},
  {Raison}, {Rebordao}, {Risquez}, {Rocca-Volmerange}, {Rosen}, {Ruiz-Fuertes},
  {Russo}, {Sembay}, {Serraller Vizcaino}, {Short}, {Siebert}, {Silva},
  {Sinachopoulos}, {Slezak}, {Soffel}, {Sosnowska}, {Strai{\v{z}}ys}, {ter
  Linden}, {Terrell}, {Theil}, {Tiede}, {Troisi}, {Tsalmantza}, {Tur},
  {Vaccari}, {Vachier}, {Valles}, {Van Hamme}, {Veltz}, {Virtanen}, {Wallut},
  {Wichmann}, {Wilkinson}, {Ziaeepour}, \& {Zschocke}}]{dr2ack1}
{Gaia Collaboration}, {Prusti}, T., {de Bruijne}, J.~H.~J., {et~al.} 2016,
  \aap, 595, A1

\bibitem[{{Gaia Collaboration} {et~al.}(2018{\natexlab{a}}){Gaia
  Collaboration}, {Brown}, {Vallenari}, {Prusti}, {de Bruijne}, {Babusiaux},
  {Bailer-Jones}, {Biermann}, {Evans}, {Eyer}, {Jansen}, {Jordi}, {Klioner},
  {Lammers}, {Lindegren}, {Luri}, {Mignard}, {Panem}, {Pourbaix}, {Randich},
  {Sartoretti}, {Siddiqui}, {Soubiran}, {van Leeuwen}, {Walton}, {Arenou},
  {Bastian}, {Cropper}, {Drimmel}, {Katz}, {Lattanzi}, {Bakker}, {Cacciari},
  {Casta{\~n}eda}, {Chaoul}, {Cheek}, {De Angeli}, {Fabricius}, {Guerra},
  {Holl}, {Masana}, {Messineo}, {Mowlavi}, {Nienartowicz}, {Panuzzo},
  {Portell}, {Riello}, {Seabroke}, {Tanga}, {Th{\'e}venin}, {Gracia-Abril},
  {Comoretto}, {Garcia-Reinaldos}, {Teyssier}, {Altmann}, {Andrae}, {Audard},
  {Bellas-Velidis}, {Benson}, {Berthier}, {Blomme}, {Burgess}, {Busso},
  {Carry}, {Cellino}, {Clementini}, {Clotet}, {Creevey}, {Davidson}, {De
  Ridder}, {Delchambre}, {Dell'Oro}, {Ducourant},
  {Fern{\'a}ndez-Hern{\'a}ndez}, {Fouesneau}, {Fr{\'e}mat}, {Galluccio},
  {Garc{\'\i}a-Torres}, {Gonz{\'a}lez-N{\'u}{\~n}ez}, {Gonz{\'a}lez-Vidal},
  {Gosset}, {Guy}, {Halbwachs}, {Hambly}, {Harrison}, {Hern{\'a}ndez},
  {Hestroffer}, {Hodgkin}, {Hutton}, {Jasniewicz}, {Jean-Antoine-Piccolo},
  {Jordan}, {Korn}, {Krone-Martins}, {Lanzafame}, {Lebzelter}, {L{\"o}ffler},
  {Manteiga}, {Marrese}, {Mart{\'\i}n-Fleitas}, {Moitinho}, {Mora}, {Muinonen},
  {Osinde}, {Pancino}, {Pauwels}, {Petit}, {Recio-Blanco}, {Richards},
  {Rimoldini}, {Robin}, {Sarro}, {Siopis}, {Smith}, {Sozzetti}, {S{\"u}veges},
  {Torra}, {van Reeven}, {Abbas}, {Abreu Aramburu}, {Accart}, {Aerts},
  {Altavilla}, {{\'A}lvarez}, {Alvarez}, {Alves}, {Anderson}, {Andrei},
  {Anglada Varela}, {Antiche}, {Antoja}, {Arcay}, {Astraatmadja}, {Bach},
  {Baker}, {Balaguer-N{\'u}{\~n}ez}, {Balm}, {Barache}, {Barata}, {Barbato},
  {Barblan}, {Barklem}, {Barrado}, {Barros}, {Barstow}, {Bartholom{\'e}
  Mu{\~n}oz}, {Bassilana}, {Becciani}, {Bellazzini}, {Berihuete}, {Bertone},
  {Bianchi}, {Bienaym{\'e}}, {Blanco-Cuaresma}, {Boch}, {Boeche}, {Bombrun},
  {Borrachero}, {Bossini}, {Bouquillon}, {Bourda}, {Bragaglia}, {Bramante},
  {Breddels}, {Bressan}, {Brouillet}, {Br{\"u}semeister}, {Brugaletta},
  {Bucciarelli}, {Burlacu}, {Busonero}, {Butkevich}, {Buzzi}, {Caffau},
  {Cancelliere}, {Cannizzaro}, {Cantat-Gaudin}, {Carballo}, {Carlucci},
  {Carrasco}, {Casamiquela}, {Castellani}, {Castro-Ginard}, {Charlot},
  {Chemin}, {Chiavassa}, {Cocozza}, {Costigan}, {Cowell}, {Crifo}, {Crosta},
  {Crowley}, {Cuypers}, {Dafonte}, {Damerdji}, {Dapergolas}, {David}, {David},
  {de Laverny}, {De Luise}, {De March}, {de Martino}, {de Souza}, {de Torres},
  {Debosscher}, {del Pozo}, {Delbo}, {Delgado}, {Delgado}, {Di Matteo},
  {Diakite}, {Diener}, {Distefano}, {Dolding}, {Drazinos}, {Dur{\'a}n},
  {Edvardsson}, {Enke}, {Eriksson}, {Esquej}, {Eynard Bontemps}, {Fabre},
  {Fabrizio}, {Faigler}, {Falc{\~a}o}, {Farr{\`a}s Casas}, {Federici},
  {Fedorets}, {Fernique}, {Figueras}, {Filippi}, {Findeisen}, {Fonti},
  {Fraile}, {Fraser}, {Fr{\'e}zouls}, {Gai}, {Galleti}, {Garabato},
  {Garc{\'\i}a-Sedano}, {Garofalo}, {Garralda}, {Gavel}, {Gavras}, {Gerssen},
  {Geyer}, {Giacobbe}, {Gilmore}, {Girona}, {Giuffrida}, {Glass}, {Gomes},
  {Granvik}, {Gueguen}, {Guerrier}, {Guiraud}, {Guti{\'e}rrez-S{\'a}nchez},
  {Haigron}, {Hatzidimitriou}, {Hauser}, {Haywood}, {Heiter}, {Helmi}, {Heu},
  {Hilger}, {Hobbs}, {Hofmann}, {Holland}, {Huckle}, {Hypki}, {Icardi},
  {Jan{\ss}en}, {Jevardat de Fombelle}, {Jonker}, {Juh{\'a}sz}, {Julbe},
  {Karampelas}, {Kewley}, {Klar}, {Kochoska}, {Kohley}, {Kolenberg},
  {Kontizas}, {Kontizas}, {Koposov}, {Kordopatis}, {Kostrzewa-Rutkowska},
  {Koubsky}, {Lambert}, {Lanza}, {Lasne}, {Lavigne}, {Le Fustec}, {Le
  Poncin-Lafitte}, {Lebreton}, {Leccia}, {Leclerc}, {Lecoeur-Taibi},
  {Lenhardt}, {Leroux}, {Liao}, {Licata}, {Lindstr{\o}m}, {Lister}, {Livanou},
  {Lobel}, {L{\'o}pez}, {Managau}, {Mann}, {Mantelet}, {Marchal}, {Marchant},
  {Marconi}, {Marinoni}, {Marschalk{\'o}}, {Marshall}, {Martino}, {Marton},
  {Mary}, {Massari}, {Matijevi{\v{c}}}, {Mazeh}, {McMillan}, {Messina},
  {Michalik}, {Millar}, {Molina}, {Molinaro}, {Moln{\'a}r}, {Montegriffo},
  {Mor}, {Morbidelli}, {Morel}, {Morris}, {Mulone}, {Muraveva}, {Musella},
  {Nelemans}, {Nicastro}, {Noval}, {O'Mullane}, {Ord{\'e}novic},
  {Ord{\'o}{\~n}ez-Blanco}, {Osborne}, {Pagani}, {Pagano}, {Pailler},
  {Palacin}, {Palaversa}, {Panahi}, {Pawlak}, {Piersimoni}, {Pineau}, {Plachy},
  {Plum}, {Poggio}, {Poujoulet}, {Pr{\v{s}}a}, {Pulone}, {Racero}, {Ragaini},
  {Rambaux}, {Ramos-Lerate}, {Regibo}, {Reyl{\'e}}, {Riclet}, {Ripepi}, {Riva},
  {Rivard}, {Rixon}, {Roegiers}, {Roelens}, {Romero-G{\'o}mez}, {Rowell},
  {Royer}, {Ruiz-Dern}, {Sadowski}, {Sagrist{\`a} Sell{\'e}s}, {Sahlmann},
  {Salgado}, {Salguero}, {Sanna}, {Santana-Ros}, {Sarasso}, {Savietto},
  {Schultheis}, {Sciacca}, {Segol}, {Segovia}, {S{\'e}gransan}, {Shih},
  {Siltala}, {Silva}, {Smart}, {Smith}, {Solano}, {Solitro}, {Sordo}, {Soria
  Nieto}, {Souchay}, {Spagna}, {Spoto}, {Stampa}, {Steele},
  {Steidelm{\"u}ller}, {Stephenson}, {Stoev}, {Suess}, {Surdej}, {Szabados},
  {Szegedi-Elek}, {Tapiador}, {Taris}, {Tauran}, {Taylor}, {Teixeira},
  {Terrett}, {Teyssand ier}, {Thuillot}, {Titarenko}, {Torra Clotet}, {Turon},
  {Ulla}, {Utrilla}, {Uzzi}, {Vaillant}, {Valentini}, {Valette}, {van Elteren},
  {Van Hemelryck}, {van Leeuwen}, {Vaschetto}, {Vecchiato}, {Veljanoski},
  {Viala}, {Vicente}, {Vogt}, {von Essen}, {Voss}, {Votruba}, {Voutsinas},
  {Walmsley}, {Weiler}, {Wertz}, {Wevers}, {Wyrzykowski}, {Yoldas},
  {{\v{Z}}erjal}, {Ziaeepour}, {Zorec}, {Zschocke}, {Zucker}, {Zurbach}, \&
  {Zwitter}}]{gaia}
{Gaia Collaboration}, {Brown}, A.~G.~A., {Vallenari}, A., {et~al.}
  2018{\natexlab{a}}, \aap, 616, A1

\bibitem[{{Gaia Collaboration} {et~al.}(2018{\natexlab{b}}){Gaia
  Collaboration}, {Brown}, {Vallenari}, {Prusti}, {de Bruijne}, {Babusiaux},
  {Bailer-Jones}, {Biermann}, {Evans}, {Eyer}, {Jansen}, {Jordi}, {Klioner},
  {Lammers}, {Lindegren}, {Luri}, {Mignard}, {Panem}, {Pourbaix}, {Randich},
  {Sartoretti}, {Siddiqui}, {Soubiran}, {van Leeuwen}, {Walton}, {Arenou},
  {Bastian}, {Cropper}, {Drimmel}, {Katz}, {Lattanzi}, {Bakker}, {Cacciari},
  {Casta{\~n}eda}, {Chaoul}, {Cheek}, {De Angeli}, {Fabricius}, {Guerra},
  {Holl}, {Masana}, {Messineo}, {Mowlavi}, {Nienartowicz}, {Panuzzo},
  {Portell}, {Riello}, {Seabroke}, {Tanga}, {Th{\'e}venin}, {Gracia-Abril},
  {Comoretto}, {Garcia-Reinaldos}, {Teyssier}, {Altmann}, {Andrae}, {Audard},
  {Bellas-Velidis}, {Benson}, {Berthier}, {Blomme}, {Burgess}, {Busso},
  {Carry}, {Cellino}, {Clementini}, {Clotet}, {Creevey}, {Davidson}, {De
  Ridder}, {Delchambre}, {Dell'Oro}, {Ducourant},
  {Fern{\'a}ndez-Hern{\'a}ndez}, {Fouesneau}, {Fr{\'e}mat}, {Galluccio},
  {Garc{\'\i}a-Torres}, {Gonz{\'a}lez-N{\'u}{\~n}ez}, {Gonz{\'a}lez-Vidal},
  {Gosset}, {Guy}, {Halbwachs}, {Hambly}, {Harrison}, {Hern{\'a}ndez},
  {Hestroffer}, {Hodgkin}, {Hutton}, {Jasniewicz}, {Jean-Antoine-Piccolo},
  {Jordan}, {Korn}, {Krone-Martins}, {Lanzafame}, {Lebzelter}, {L{\"o}ffler},
  {Manteiga}, {Marrese}, {Mart{\'\i}n-Fleitas}, {Moitinho}, {Mora}, {Muinonen},
  {Osinde}, {Pancino}, {Pauwels}, {Petit}, {Recio-Blanco}, {Richards},
  {Rimoldini}, {Robin}, {Sarro}, {Siopis}, {Smith}, {Sozzetti}, {S{\"u}veges},
  {Torra}, {van Reeven}, {Abbas}, {Abreu Aramburu}, {Accart}, {Aerts},
  {Altavilla}, {{\'A}lvarez}, {Alvarez}, {Alves}, {Anderson}, {Andrei},
  {Anglada Varela}, {Antiche}, {Antoja}, {Arcay}, {Astraatmadja}, {Bach},
  {Baker}, {Balaguer-N{\'u}{\~n}ez}, {Balm}, {Barache}, {Barata}, {Barbato},
  {Barblan}, {Barklem}, {Barrado}, {Barros}, {Barstow}, {Bartholom{\'e}
  Mu{\~n}oz}, {Bassilana}, {Becciani}, {Bellazzini}, {Berihuete}, {Bertone},
  {Bianchi}, {Bienaym{\'e}}, {Blanco-Cuaresma}, {Boch}, {Boeche}, {Bombrun},
  {Borrachero}, {Bossini}, {Bouquillon}, {Bourda}, {Bragaglia}, {Bramante},
  {Breddels}, {Bressan}, {Brouillet}, {Br{\"u}semeister}, {Brugaletta},
  {Bucciarelli}, {Burlacu}, {Busonero}, {Butkevich}, {Buzzi}, {Caffau},
  {Cancelliere}, {Cannizzaro}, {Cantat-Gaudin}, {Carballo}, {Carlucci},
  {Carrasco}, {Casamiquela}, {Castellani}, {Castro-Ginard}, {Charlot},
  {Chemin}, {Chiavassa}, {Cocozza}, {Costigan}, {Cowell}, {Crifo}, {Crosta},
  {Crowley}, {Cuypers}, {Dafonte}, {Damerdji}, {Dapergolas}, {David}, {David},
  {de Laverny}, {De Luise}, {De March}, {de Martino}, {de Souza}, {de Torres},
  {Debosscher}, {del Pozo}, {Delbo}, {Delgado}, {Delgado}, {Di Matteo},
  {Diakite}, {Diener}, {Distefano}, {Dolding}, {Drazinos}, {Dur{\'a}n},
  {Edvardsson}, {Enke}, {Eriksson}, {Esquej}, {Eynard Bontemps}, {Fabre},
  {Fabrizio}, {Faigler}, {Falc{\~a}o}, {Farr{\`a}s Casas}, {Federici},
  {Fedorets}, {Fernique}, {Figueras}, {Filippi}, {Findeisen}, {Fonti},
  {Fraile}, {Fraser}, {Fr{\'e}zouls}, {Gai}, {Galleti}, {Garabato},
  {Garc{\'\i}a-Sedano}, {Garofalo}, {Garralda}, {Gavel}, {Gavras}, {Gerssen},
  {Geyer}, {Giacobbe}, {Gilmore}, {Girona}, {Giuffrida}, {Glass}, {Gomes},
  {Granvik}, {Gueguen}, {Guerrier}, {Guiraud}, {Guti{\'e}rrez-S{\'a}nchez},
  {Haigron}, {Hatzidimitriou}, {Hauser}, {Haywood}, {Heiter}, {Helmi}, {Heu},
  {Hilger}, {Hobbs}, {Hofmann}, {Holland}, {Huckle}, {Hypki}, {Icardi},
  {Jan{\ss}en}, {Jevardat de Fombelle}, {Jonker}, {Juh{\'a}sz}, {Julbe},
  {Karampelas}, {Kewley}, {Klar}, {Kochoska}, {Kohley}, {Kolenberg},
  {Kontizas}, {Kontizas}, {Koposov}, {Kordopatis}, {Kostrzewa-Rutkowska},
  {Koubsky}, {Lambert}, {Lanza}, {Lasne}, {Lavigne}, {Le Fustec}, {Le
  Poncin-Lafitte}, {Lebreton}, {Leccia}, {Leclerc}, {Lecoeur-Taibi},
  {Lenhardt}, {Leroux}, {Liao}, {Licata}, {Lindstr{\o}m}, {Lister}, {Livanou},
  {Lobel}, {L{\'o}pez}, {Managau}, {Mann}, {Mantelet}, {Marchal}, {Marchant},
  {Marconi}, {Marinoni}, {Marschalk{\'o}}, {Marshall}, {Martino}, {Marton},
  {Mary}, {Massari}, {Matijevi{\v{c}}}, {Mazeh}, {McMillan}, {Messina},
  {Michalik}, {Millar}, {Molina}, {Molinaro}, {Moln{\'a}r}, {Montegriffo},
  {Mor}, {Morbidelli}, {Morel}, {Morris}, {Mulone}, {Muraveva}, {Musella},
  {Nelemans}, {Nicastro}, {Noval}, {O'Mullane}, {Ord{\'e}novic},
  {Ord{\'o}{\~n}ez-Blanco}, {Osborne}, {Pagani}, {Pagano}, {Pailler},
  {Palacin}, {Palaversa}, {Panahi}, {Pawlak}, {Piersimoni}, {Pineau}, {Plachy},
  {Plum}, {Poggio}, {Poujoulet}, {Pr{\v{s}}a}, {Pulone}, {Racero}, {Ragaini},
  {Rambaux}, {Ramos-Lerate}, {Regibo}, {Reyl{\'e}}, {Riclet}, {Ripepi}, {Riva},
  {Rivard}, {Rixon}, {Roegiers}, {Roelens}, {Romero-G{\'o}mez}, {Rowell},
  {Royer}, {Ruiz-Dern}, {Sadowski}, {Sagrist{\`a} Sell{\'e}s}, {Sahlmann},
  {Salgado}, {Salguero}, {Sanna}, {Santana-Ros}, {Sarasso}, {Savietto},
  {Schultheis}, {Sciacca}, {Segol}, {Segovia}, {S{\'e}gransan}, {Shih},
  {Siltala}, {Silva}, {Smart}, {Smith}, {Solano}, {Solitro}, {Sordo}, {Soria
  Nieto}, {Souchay}, {Spagna}, {Spoto}, {Stampa}, {Steele},
  {Steidelm{\"u}ller}, {Stephenson}, {Stoev}, {Suess}, {Surdej}, {Szabados},
  {Szegedi-Elek}, {Tapiador}, {Taris}, {Tauran}, {Taylor}, {Teixeira},
  {Terrett}, {Teyssand ier}, {Thuillot}, {Titarenko}, {Torra Clotet}, {Turon},
  {Ulla}, {Utrilla}, {Uzzi}, {Vaillant}, {Valentini}, {Valette}, {van Elteren},
  {Van Hemelryck}, {van Leeuwen}, {Vaschetto}, {Vecchiato}, {Veljanoski},
  {Viala}, {Vicente}, {Vogt}, {von Essen}, {Voss}, {Votruba}, {Voutsinas},
  {Walmsley}, {Weiler}, {Wertz}, {Wevers}, {Wyrzykowski}, {Yoldas},
  {{\v{Z}}erjal}, {Ziaeepour}, {Zorec}, {Zschocke}, {Zucker}, {Zurbach}, \&
  {Zwitter}}]{dr2ack2}
---. 2018{\natexlab{b}}, \aap, 616, A1

\bibitem[{{Gallart} {et~al.}(2019){Gallart}, {Bernard}, {Brook}, {Ruiz-Lara},
  {Cassisi}, {Hill}, \& {Monelli}}]{Gallart19}
{Gallart}, C., {Bernard}, E.~J., {Brook}, C.~B., {et~al.} 2019, Nature
  Astronomy, 3, 932

\bibitem[{{Gibbons} {et~al.}(2017){Gibbons}, {Belokurov}, \&
  {Evans}}]{Gibbons17}
{Gibbons}, S.~L.~J., {Belokurov}, V., \& {Evans}, N.~W. 2017, \mnras, 464, 794

\bibitem[{{Gnedin} \& {Ostriker}(1997)}]{Gnedin97}
{Gnedin}, O.~Y., \& {Ostriker}, J.~P. 1997, \apj, 474, 223

\bibitem[{{G{\'o}mez} {et~al.}(2010){G{\'o}mez}, {Helmi}, {Brown}, \&
  {Li}}]{Gomez10}
{G{\'o}mez}, F.~A., {Helmi}, A., {Brown}, A. G.~A., \& {Li}, Y.-S. 2010,
  \mnras, 408, 935

\bibitem[{{G{\'o}mez} {et~al.}(2013){G{\'o}mez}, {Helmi}, {Cooper}, {Frenk},
  {Navarro}, \& {White}}]{Gomez13}
{G{\'o}mez}, F.~A., {Helmi}, A., {Cooper}, A.~P., {et~al.} 2013, \mnras, 436,
  3602

\bibitem[{{G{\'o}mez} {et~al.}(2017){G{\'o}mez}, {Grand}, {Monachesi}, {White},
  {Bustamante}, {Marinacci}, {Pakmor}, {Simpson}, {Springel}, \&
  {Frenk}}]{Gomez17}
{G{\'o}mez}, F.~A., {Grand}, R. J.~J., {Monachesi}, A., {et~al.} 2017, \mnras,
  472, 3722

\bibitem[{{Gravity Collaboration} {et~al.}(2019){Gravity Collaboration},
  {Abuter}, {Amorim}, {Baub{\"o}ck}, {Berger}, {Bonnet}, {Brand ner},
  {Cl{\'e}net}, {Coud{\'e} Du Foresto}, {de Zeeuw}, {Dexter}, {Duvert},
  {Eckart}, {Eisenhauer}, {F{\"o}rster Schreiber}, {Garcia}, {Gao}, {Gendron},
  {Genzel}, {Gerhard}, {Gillessen}, {Habibi}, {Haubois}, {Henning}, {Hippler},
  {Horrobin}, {Jim{\'e}nez-Rosales}, {Jocou}, {Kervella}, {Lacour},
  {Lapeyr{\`e}re}, {Le Bouquin}, {L{\'e}na}, {Ott}, {Paumard}, {Perraut},
  {Perrin}, {Pfuhl}, {Rabien}, {Rodriguez Coira}, {Rousset}, {Scheithauer},
  {Sternberg}, {Straub}, {Straubmeier}, {Sturm}, {Tacconi}, {Vincent}, {von
  Fellenberg}, {Waisberg}, {Widmann}, {Wieprecht}, {Wiezorrek}, {Woillez}, \&
  {Yazici}}]{Gravity19}
{Gravity Collaboration}, {Abuter}, R., {Amorim}, A., {et~al.} 2019, \aap, 625,
  L10

\bibitem[{{Grillmair} \& {Dionatos}(2006)}]{Grillmair06}
{Grillmair}, C.~J., \& {Dionatos}, O. 2006, \apjl, 643, L17

\bibitem[{{Hanke} {et~al.}(2020){Hanke}, {Koch}, {Prudil}, {Grebel}, \&
  {Bastian}}]{Hanke20}
{Hanke}, M., {Koch}, A., {Prudil}, Z., {Grebel}, E.~K., \& {Bastian}, U. 2020,
  \aap, 637, A98

\bibitem[{{Hayes} {et~al.}(2020){Hayes}, {Majewski}, {Hasselquist}, {Anguiano},
  {Shetrone}, {Law}, {Schiavon}, {Cunha}, {Smith}, {Beaton}, {Price-Whelan},
  {Allende Prieto}, {Battaglia}, {Bizyaev}, {Brownstein}, {Cohen},
  {Frinchaboy}, {Garc{\'\i}a-Hern{\'a}ndez}, {Lacerna}, {Lane},
  {M{\'e}sz{\'a}ros}, {Bidin}, {M{\~{u}}noz}, {Nidever}, {Oravetz}, {Oravetz},
  {Pan}, {Roman-Lopes}, {Sobeck}, \& {Stringfellow}}]{Hayes20}
{Hayes}, C.~R., {Majewski}, S.~R., {Hasselquist}, S., {et~al.} 2020, \apj, 889,
  63

\bibitem[{{Haywood} {et~al.}(2018){Haywood}, {Di Matteo}, {Lehnert}, {Snaith},
  {Khoperskov}, \& {G{\'o}mez}}]{Haywood18}
{Haywood}, M., {Di Matteo}, P., {Lehnert}, M.~D., {et~al.} 2018, \apj, 863, 113

\bibitem[{{Helmi}(2020)}]{Helmi20}
{Helmi}, A. 2020, arXiv e-prints, arXiv:2002.04340

\bibitem[{{Helmi} {et~al.}(2018){Helmi}, {Babusiaux}, {Koppelman}, {Massari},
  {Veljanoski}, \& {Brown}}]{Helmi18}
{Helmi}, A., {Babusiaux}, C., {Koppelman}, H.~H., {et~al.} 2018, \nat, 563, 85

\bibitem[{{Helmi} \& {de Zeeuw}(2000)}]{Helmi00}
{Helmi}, A., \& {de Zeeuw}, P.~T. 2000, \mnras, 319, 657

\bibitem[{{Helmi} {et~al.}(2017){Helmi}, {Veljanoski}, {Breddels}, {Tian}, \&
  {Sales}}]{Helmi17}
{Helmi}, A., {Veljanoski}, J., {Breddels}, M.~A., {Tian}, H., \& {Sales}, L.~V.
  2017, \aap, 598, A58

\bibitem[{{Helmi} {et~al.}(1999){Helmi}, {White}, {de Zeeuw}, \&
  {Zhao}}]{Helmi99}
{Helmi}, A., {White}, S. D.~M., {de Zeeuw}, P.~T., \& {Zhao}, H. 1999, \nat,
  402, 53

\bibitem[{{Hernitschek} {et~al.}(2017){Hernitschek}, {Sesar}, {Rix},
  {Belokurov}, {Martinez-Delgado}, {Martin}, {Kaiser}, {Hodapp}, {Chambers},
  {Wainscoat}, {Magnier}, {Kudritzki}, {Metcalfe}, \& {Draper}}]{Hernitschek17}
{Hernitschek}, N., {Sesar}, B., {Rix}, H.-W., {et~al.} 2017, \apj, 850, 96

\bibitem[{{Hernquist}(1990)}]{Hernquist90}
{Hernquist}, L. 1990, \apj, 356, 359

\bibitem[{Hunter(2007)}]{matplotlib}
Hunter, J.~D. 2007, Computing In Science \& Engineering, 9, 90

\bibitem[{{Ibata} {et~al.}(1994){Ibata}, {Gilmore}, \& {Irwin}}]{Ibata94}
{Ibata}, R.~A., {Gilmore}, G., \& {Irwin}, M.~J. 1994, \nat, 370, 194

\bibitem[{{Iorio} \& {Belokurov}(2019)}]{Iorio19}
{Iorio}, G., \& {Belokurov}, V. 2019, \mnras, 482, 3868

\bibitem[{{Ivezi{\'c}} {et~al.}(2008){Ivezi{\'c}}, {Sesar}, {Juri{\'c}},
  {Bond}, {Dalcanton}, {Rockosi}, {Yanny}, {Newberg}, {Beers}, {Allende
  Prieto}, {Wilhelm}, {Lee}, {Sivarani}, {Norris}, {Bailer-Jones}, {Re
  Fiorentin}, {Schlegel}, {Uomoto}, {Lupton}, {Knapp}, {Gunn}, {Covey}, {Allyn
  Smith}, {Miknaitis}, {Doi}, {Tanaka}, {Fukugita}, {Kent}, {Finkbeiner},
  {Munn}, {Pier}, {Quinn}, {Hawley}, {Anderson}, {Kiuchi}, {Chen}, {Bushong},
  {Sohi}, {Haggard}, {Kimball}, {Barentine}, {Brewington}, {Harvanek},
  {Kleinman}, {Krzesinski}, {Long}, {Nitta}, {Snedden}, {Lee}, {Harris},
  {Brinkmann}, {Schneider}, \& {York}}]{Ivezic08}
{Ivezi{\'c}}, {\v{Z}}., {Sesar}, B., {Juri{\'c}}, M., {et~al.} 2008, \apj, 684,
  287

\bibitem[{{Janesh} {et~al.}(2016){Janesh}, {Morrison}, {Ma}, {Rockosi},
  {Starkenburg}, {Xue}, {Rix}, {Harding}, {Beers}, {Johnson}, {Lee}, \&
  {Schneider}}]{Janesh16}
{Janesh}, W., {Morrison}, H.~L., {Ma}, Z., {et~al.} 2016, \apj, 816, 80

\bibitem[{{Jean-Baptiste} {et~al.}(2017){Jean-Baptiste}, {Di Matteo},
  {Haywood}, {G{\'o}mez}, {Montuori}, {Combes}, \& {Semelin}}]{Jean-Baptiste17}
{Jean-Baptiste}, I., {Di Matteo}, P., {Haywood}, M., {et~al.} 2017, \aap, 604,
  A106

\bibitem[{{Johnson} {et~al.}(2020){Johnson}, {Conroy}, {Naidu}, {Bonaca},
  {Zaritsky}, {Ting}, {Cargile}, \& {Han}}]{Johnson20}
{Johnson}, B.~D., {Conroy}, C., {Naidu}, R.~P., {et~al.} 2020, submitted to ApJ

\bibitem[{{Johnston} {et~al.}(1995){Johnston}, {Spergel}, \&
  {Hernquist}}]{Johnston95}
{Johnston}, K.~V., {Spergel}, D.~N., \& {Hernquist}, L. 1995, \apj, 451, 598

\bibitem[{{Kafle} {et~al.}(2012){Kafle}, {Sharma}, {Lewis}, \& {Bland
  -Hawthorn}}]{Kafle12}
{Kafle}, P.~R., {Sharma}, S., {Lewis}, G.~F., \& {Bland -Hawthorn}, J. 2012,
  \apj, 761, 98

\bibitem[{{Kirby} {et~al.}(2013){Kirby}, {Cohen}, {Guhathakurta}, {Cheng},
  {Bullock}, \& {Gallazzi}}]{Kirby13}
{Kirby}, E.~N., {Cohen}, J.~G., {Guhathakurta}, P., {et~al.} 2013, \apj, 779,
  102

\bibitem[{{Kirby} {et~al.}(2011){Kirby}, {Lanfranchi}, {Simon}, {Cohen}, \&
  {Guhathakurta}}]{Kirby11}
{Kirby}, E.~N., {Lanfranchi}, G.~A., {Simon}, J.~D., {Cohen}, J.~G., \&
  {Guhathakurta}, P. 2011, \apj, 727, 78

\bibitem[{Kluyver {et~al.}(2016)Kluyver, Ragan-Kelley, P{\'e}rez, Granger,
  Bussonnier, Frederic, Kelley, Hamrick, Grout, Corlay, Ivanov, Avila, Abdalla,
  \& Willing}]{jupyter}
Kluyver, T., Ragan-Kelley, B., P{\'e}rez, F., {et~al.} 2016, in Positioning and
  Power in Academic Publishing: Players, Agents and Agendas, ed. F.~Loizides \&
  B.~Schmidt, IOS Press, 87 -- 90

\bibitem[{{Koch} {et~al.}(2019){Koch}, {Grebel}, \& {Martell}}]{Koch19}
{Koch}, A., {Grebel}, E.~K., \& {Martell}, S.~L. 2019, \aap, 625, A75

\bibitem[{{Koposov} {et~al.}(2007){Koposov}, {de Jong}, {Belokurov}, {Rix},
  {Zucker}, {Evans}, {Gilmore}, {Irwin}, \& {Bell}}]{Koposov07}
{Koposov}, S., {de Jong}, J.~T.~A., {Belokurov}, V., {et~al.} 2007, \apj, 669,
  337

\bibitem[{{Koppelman} {et~al.}(2018){Koppelman}, {Helmi}, \&
  {Veljanoski}}]{Koppelman18}
{Koppelman}, H., {Helmi}, A., \& {Veljanoski}, J. 2018, \apjl, 860, L11

\bibitem[{{Koppelman} \& {Helmi}(2020)}]{Koppelman20}
{Koppelman}, H.~H., \& {Helmi}, A. 2020, arXiv e-prints, arXiv:2004.07328

\bibitem[{{Koppelman} {et~al.}(2019{\natexlab{a}}){Koppelman}, {Helmi},
  {Massari}, {Price-Whelan}, \& {Starkenburg}}]{Koppelman19}
{Koppelman}, H.~H., {Helmi}, A., {Massari}, D., {Price-Whelan}, A.~M., \&
  {Starkenburg}, T.~K. 2019{\natexlab{a}}, \aap, 631, L9

\bibitem[{{Koppelman} {et~al.}(2019{\natexlab{b}}){Koppelman}, {Helmi},
  {Massari}, {Roelenga}, \& {Bastian}}]{Koppelman19HS}
{Koppelman}, H.~H., {Helmi}, A., {Massari}, D., {Roelenga}, S., \& {Bastian},
  U. 2019{\natexlab{b}}, \aap, 625, A5

\bibitem[{{Kroupa}(2001)}]{Kroupa01}
{Kroupa}, P. 2001, \mnras, 322, 231

\bibitem[{{Kruijssen} {et~al.}(2019){Kruijssen}, {Pfeffer}, {Reina-Campos},
  {Crain}, \& {Bastian}}]{Kruijssen19}
{Kruijssen}, J.~M.~D., {Pfeffer}, J.~L., {Reina-Campos}, M., {Crain}, R.~A., \&
  {Bastian}, N. 2019, \mnras, 486, 3180

\bibitem[{{Kruijssen} {et~al.}(2020){Kruijssen}, {Pfeffer}, {Chevance},
  {Bonaca}, {Trujillo-Gomez}, {Bastian}, {Reina-Campos}, {Crain}, \&
  {Hughes}}]{Kruijssen20}
{Kruijssen}, J.~M.~D., {Pfeffer}, J.~L., {Chevance}, M., {et~al.} 2020, arXiv
  e-prints, arXiv:2003.01119

\bibitem[{{Lancaster} {et~al.}(2019){Lancaster}, {Koposov}, {Belokurov},
  {Evans}, \& {Deason}}]{Lancaster19}
{Lancaster}, L., {Koposov}, S.~E., {Belokurov}, V., {Evans}, N.~W., \&
  {Deason}, A.~J. 2019, \mnras, 486, 378

\bibitem[{{Laporte} {et~al.}(2018){Laporte}, {Johnston}, {G{\'o}mez},
  {Garavito-Camargo}, \& {Besla}}]{Laporte18}
{Laporte}, C. F.~P., {Johnston}, K.~V., {G{\'o}mez}, F.~A., {Garavito-Camargo},
  N., \& {Besla}, G. 2018, \mnras, 481, 286

\bibitem[{{Law} {et~al.}(2005){Law}, {Johnston}, \& {Majewski}}]{LM05}
{Law}, D.~R., {Johnston}, K.~V., \& {Majewski}, S.~R. 2005, \apj, 619, 807

\bibitem[{{Law} \& {Majewski}(2010)}]{LM10}
{Law}, D.~R., \& {Majewski}, S.~R. 2010, \apj, 714, 229

\bibitem[{{Lee} {et~al.}(2015){Lee}, {Johnston}, {Sen}, \& {Jessop}}]{Lee15}
{Lee}, D.~M., {Johnston}, K.~V., {Sen}, B., \& {Jessop}, W. 2015, \apj, 802, 48

\bibitem[{{Leja} {et~al.}(2019){Leja}, {Carnall}, {Johnson}, {Conroy}, \&
  {Speagle}}]{Leja19}
{Leja}, J., {Carnall}, A.~C., {Johnson}, B.~D., {Conroy}, C., \& {Speagle},
  J.~S. 2019, \apj, 876, 3

\bibitem[{{Li} {et~al.}(2019){Li}, {FELLOW}, {Liu}, {Xue}, {Zhong}, {Weiss},
  {Carlin}, {Tian}, \& {FELLOW}}]{Li19}
{Li}, J., {FELLOW}, L., {Liu}, C., {et~al.} 2019, \apj, 874, 138

\bibitem[{{Li} {et~al.}(2017){Li}, {Sheffield}, {Johnston}, {Marshall},
  {Majewski}, {Price-Whelan}, {Damke}, {Beaton}, {Bernard}, {Richardson},
  {Sharma}, \& {Sesar}}]{Li17}
{Li}, T.~S., {Sheffield}, A.~A., {Johnston}, K.~V., {et~al.} 2017, \apj, 844,
  74

\bibitem[{{Lian} {et~al.}(2020){Lian}, {Thomas}, {Maraston}, {Zamora}, {Tayar},
  {Pan}, {Tissera}, {Fern{\'a}ndez-Trincado}, \& {Anibal
  Garcia-Hernandez}}]{Lian20}
{Lian}, J., {Thomas}, D., {Maraston}, C., {et~al.} 2020, arXiv e-prints,
  arXiv:2003.11549

\bibitem[{{Lilleengen} {et~al.}(2020){Lilleengen}, {Trick}, \& {van de
  Ven}}]{Lilleengen20}
{Lilleengen}, S., {Trick}, W., \& {van de Ven}, G. 2020, in IAU Symposium, Vol.
  353, IAU Symposium, ed. M.~{Valluri} \& J.~A. {Sellwood}, 266--270

\bibitem[{{Liu} {et~al.}(2018){Liu}, {Du}, {Newberg}, {Chen}, {Wu}, {Ma},
  {Zhou}, {Cao}, {Hou}, {Wang}, \& {Zhang}}]{Liu18}
{Liu}, S., {Du}, C., {Newberg}, H.~J., {et~al.} 2018, \apj, 862, 163

\bibitem[{{Lynden-Bell}(1975)}]{Lynden-Bell75}
{Lynden-Bell}, D. 1975, Vistas in Astronomy, 19, 299

\bibitem[{{Ma} {et~al.}(2016{\natexlab{a}}){Ma}, {Hopkins},
  {Faucher-Gigu{\`e}re}, {Zolman}, {Muratov}, {Kere{\v{s}}}, \&
  {Quataert}}]{Ma16MZR}
{Ma}, X., {Hopkins}, P.~F., {Faucher-Gigu{\`e}re}, C.-A., {et~al.}
  2016{\natexlab{a}}, \mnras, 456, 2140

\bibitem[{{Ma} {et~al.}(2016{\natexlab{b}}){Ma}, {Hopkins}, {Kasen},
  {Quataert}, {Faucher-Gigu{\`e}re}, {Kere{\v s}}, {Murray}, \& {Strom}}]{Ma16}
{Ma}, X., {Hopkins}, P.~F., {Kasen}, D., {et~al.} 2016{\natexlab{b}}, \mnras,
  459, 3614

\bibitem[{{Mackereth} \& {Bovy}(2020)}]{Mackereth20}
{Mackereth}, J.~T., \& {Bovy}, J. 2020, \mnras, 492, 3631

\bibitem[{{Mackereth} {et~al.}(2017){Mackereth}, {Bovy}, {Schiavon},
  {Zasowski}, {Cunha}, {Frinchaboy}, {Garc{\'\i}a Perez}, {Hayden}, {Holtzman},
  {Majewski}, {M{\'e}sz{\'a}ros}, {Nidever}, {Pinsonneault}, \&
  {Shetrone}}]{Mackereth17}
{Mackereth}, J.~T., {Bovy}, J., {Schiavon}, R.~P., {et~al.} 2017, \mnras, 471,
  3057

\bibitem[{{Mackereth} {et~al.}(2019){Mackereth}, {Schiavon}, {Pfeffer},
  {Hayes}, {Bovy}, {Anguiano}, {Allende Prieto}, {Hasselquist}, {Holtzman},
  {Johnson}, {Majewski}, {O'Connell}, {Shetrone}, {Tissera}, \&
  {Fern{\'a}ndez-Trincado}}]{Mackereth19}
{Mackereth}, J.~T., {Schiavon}, R.~P., {Pfeffer}, J., {et~al.} 2019, \mnras,
  482, 3426

\bibitem[{{Maiolino} \& {Mannucci}(2019)}]{Maiolino19}
{Maiolino}, R., \& {Mannucci}, F. 2019, \aapr, 27, 3

\bibitem[{{Majewski} {et~al.}(2003){Majewski}, {Skrutskie}, {Weinberg}, \&
  {Ostheimer}}]{Majewski03}
{Majewski}, S.~R., {Skrutskie}, M.~F., {Weinberg}, M.~D., \& {Ostheimer}, J.~C.
  2003, \apj, 599, 1082

\bibitem[{{Majewski} {et~al.}(2017){Majewski}, {Schiavon}, {Frinchaboy},
  {Allende Prieto}, {Barkhouser}, {Bizyaev}, {Blank}, {Brunner}, {Burton},
  {Carrera}, {Chojnowski}, {Cunha}, {Epstein}, {Fitzgerald}, {Garc{\'\i}a
  P{\'e}rez}, {Hearty}, {Henderson}, {Holtzman}, {Johnson}, {Lam}, {Lawler},
  {Maseman}, {M{\'e}sz{\'a}ros}, {Nelson}, {Nguyen}, {Nidever}, {Pinsonneault},
  {Shetrone}, {Smee}, {Smith}, {Stolberg}, {Skrutskie}, {Walker}, {Wilson},
  {Zasowski}, {Anders}, {Basu}, {Beland}, {Blanton}, {Bovy}, {Brownstein},
  {Carlberg}, {Chaplin}, {Chiappini}, {Eisenstein}, {Elsworth}, {Feuillet},
  {Fleming}, {Galbraith-Frew}, {Garc{\'\i}a}, {Garc{\'\i}a-Hern{\'a}ndez},
  {Gillespie}, {Girardi}, {Gunn}, {Hasselquist}, {Hayden}, {Hekker}, {Ivans},
  {Kinemuchi}, {Klaene}, {Mahadevan}, {Mathur}, {Mosser}, {Muna}, {Munn},
  {Nichol}, {O'Connell}, {Parejko}, {Robin}, {Rocha-Pinto}, {Schultheis},
  {Serenelli}, {Shane}, {Silva Aguirre}, {Sobeck}, {Thompson}, {Troup},
  {Weinberg}, \& {Zamora}}]{APOGEE}
{Majewski}, S.~R., {Schiavon}, R.~P., {Frinchaboy}, P.~M., {et~al.} 2017, \aj,
  154, 94

\bibitem[{{Malhan} {et~al.}(2018){Malhan}, {Ibata}, \& {Martin}}]{Malhan18}
{Malhan}, K., {Ibata}, R.~A., \& {Martin}, N.~F. 2018, \mnras, 481, 3442

\bibitem[{{Martell} {et~al.}(2011){Martell}, {Smolinski}, {Beers}, \&
  {Grebel}}]{Martell11}
{Martell}, S.~L., {Smolinski}, J.~P., {Beers}, T.~C., \& {Grebel}, E.~K. 2011,
  \aap, 534, A136

\bibitem[{{Martell} {et~al.}(2016){Martell}, {Shetrone}, {Lucatello},
  {Schiavon}, {M{\'e}sz{\'a}ros}, {Allende Prieto},
  {Garc{\'\i}a-Hern{\'a}ndez}, {Beers}, \& {Nidever}}]{Martell16}
{Martell}, S.~L., {Shetrone}, M.~D., {Lucatello}, S., {et~al.} 2016, \apj, 825,
  146

\bibitem[{{Massari} {et~al.}(2019){Massari}, {Koppelman}, \&
  {Helmi}}]{Massari19}
{Massari}, D., {Koppelman}, H.~H., \& {Helmi}, A. 2019, \aap, 630, L4

\bibitem[{{Masseron} \& {Hawkins}(2017)}]{Masseron17}
{Masseron}, T., \& {Hawkins}, K. 2017, \aap, 597, L3

\bibitem[{{Matsuno} {et~al.}(2019){Matsuno}, {Aoki}, \& {Suda}}]{Matsuno19}
{Matsuno}, T., {Aoki}, W., \& {Suda}, T. 2019, \apjl, 874, L35

\bibitem[{{Matteucci} \& {Greggio}(1986)}]{Matteucci86}
{Matteucci}, F., \& {Greggio}, L. 1986, \aap, 154, 279

\bibitem[{{McMillan}(2017)}]{McMillan17}
{McMillan}, P.~J. 2017, \mnras, 465, 76

\bibitem[{{Miyamoto} \& {Nagai}(1975)}]{MiyamotoNagai75}
{Miyamoto}, M., \& {Nagai}, R. 1975, \pasj, 27, 533

\bibitem[{{Monachesi} {et~al.}(2019){Monachesi}, {G{\'o}mez}, {Grand },
  {Simpson}, {Kauffmann}, {Bustamante}, {Marinacci}, {Pakmor}, {Springel},
  {Frenk}, {White}, \& {Tissera}}]{Monachesi19}
{Monachesi}, A., {G{\'o}mez}, F.~A., {Grand }, R. J.~J., {et~al.} 2019, \mnras,
  485, 2589

\bibitem[{{Monaco} {et~al.}(2007){Monaco}, {Bellazzini}, {Bonifacio},
  {Buzzoni}, {Ferraro}, {Marconi}, {Sbordone}, \& {Zaggia}}]{Moncao07}
{Monaco}, L., {Bellazzini}, M., {Bonifacio}, P., {et~al.} 2007, \aap, 464, 201

\bibitem[{{Monty} {et~al.}(2019){Monty}, {Venn}, {Lane}, {Lokhorst}, \&
  {Yong}}]{Monty19}
{Monty}, S., {Venn}, K.~A., {Lane}, J. M.~M., {Lokhorst}, D., \& {Yong}, D.
  2019, arXiv e-prints, arXiv:1909.11969

\bibitem[{{Myeong} {et~al.}(2018{\natexlab{a}}){Myeong}, {Evans}, {Belokurov},
  {Amorisco}, \& {Koposov}}]{Myeong18d}
{Myeong}, G.~C., {Evans}, N.~W., {Belokurov}, V., {Amorisco}, N.~C., \&
  {Koposov}, S.~E. 2018{\natexlab{a}}, \mnras, 475, 1537

\bibitem[{{Myeong} {et~al.}(2018{\natexlab{b}}){Myeong}, {Evans}, {Belokurov},
  {Sand ers}, \& {Koposov}}]{Myeong18b}
{Myeong}, G.~C., {Evans}, N.~W., {Belokurov}, V., {Sand ers}, J.~L., \&
  {Koposov}, S.~E. 2018{\natexlab{b}}, \mnras, 478, 5449

\bibitem[{{Myeong} {et~al.}(2018{\natexlab{c}}){Myeong}, {Evans}, {Belokurov},
  {Sand ers}, \& {Koposov}}]{Myeong18c}
---. 2018{\natexlab{c}}, \apjl, 856, L26

\bibitem[{{Myeong} {et~al.}(2018{\natexlab{d}}){Myeong}, {Evans}, {Belokurov},
  {Sand ers}, \& {Koposov}}]{Myeong18}
---. 2018{\natexlab{d}}, \apjl, 863, L28

\bibitem[{{Myeong} {et~al.}(2017){Myeong}, {Jerjen}, {Mackey}, \& {Da
  Costa}}]{Myeong17}
{Myeong}, G.~C., {Jerjen}, H., {Mackey}, D., \& {Da Costa}, G.~S. 2017, \apjl,
  840, L25

\bibitem[{{Myeong} {et~al.}(2019){Myeong}, {Vasiliev}, {Iorio}, {Evans}, \&
  {Belokurov}}]{Myeong19}
{Myeong}, G.~C., {Vasiliev}, E., {Iorio}, G., {Evans}, N.~W., \& {Belokurov},
  V. 2019, \mnras, 488, 1235

\bibitem[{{Naidu} {et~al.}(2020){Naidu}, {Tacchella}, {Mason}, {Bose}, {Oesch},
  \& {Conroy}}]{Naidu20}
{Naidu}, R.~P., {Tacchella}, S., {Mason}, C.~A., {et~al.} 2020, \apj, 892, 109

\bibitem[{{Navarro} {et~al.}(1997){Navarro}, {Frenk}, \& {White}}]{Navarro97}
{Navarro}, J.~F., {Frenk}, C.~S., \& {White}, S. D.~M. 1997, \apj, 490, 493

\bibitem[{{Newberg} {et~al.}(2002){Newberg}, {Yanny}, {Rockosi}, {Grebel},
  {Rix}, {Brinkmann}, {Csabai}, {Hennessy}, {Hindsley}, {Ibata}, {Ivezi{\'c}},
  {Lamb}, {Nash}, {Odenkirchen}, {Rave}, {Schneider}, {Smith}, {Stolte}, \&
  {York}}]{Newberg02}
{Newberg}, H.~J., {Yanny}, B., {Rockosi}, C., {et~al.} 2002, \apj, 569, 245

\bibitem[{{Newberg} {et~al.}(2003){Newberg}, {Yanny}, {Grebel}, {Hennessy},
  {Ivezi{\'c}}, {Martinez-Delgado}, {Odenkirchen}, {Rix}, {Brinkmann}, {Lamb},
  {Schneider}, \& {York}}]{Newberg03}
{Newberg}, H.~J., {Yanny}, B., {Grebel}, E.~K., {et~al.} 2003, \apjl, 596, L191

\bibitem[{{Niederste-Ostholt} {et~al.}(2010){Niederste-Ostholt}, {Belokurov},
  {Evans}, {Koposov}, {Gieles}, \& {Irwin}}]{Niederste-Ostholt10}
{Niederste-Ostholt}, M., {Belokurov}, V., {Evans}, N.~W., {et~al.} 2010,
  \mnras, 408, L66

\bibitem[{{Norris} \& {Ryan}(1989)}]{Norris89}
{Norris}, J.~E., \& {Ryan}, S.~G. 1989, \apjl, 336, L17

\bibitem[{{Oke} \& {Gunn}(1983)}]{Oke83}
{Oke}, J.~B., \& {Gunn}, J.~E. 1983, \apj, 266, 713

\bibitem[{Oliphant(2006--)}]{numpy}
Oliphant, T. 2006--, {NumPy}: A guide to {NumPy}, USA: Trelgol Publishing, , ,
  [Online; accessed <today>]

\bibitem[{P\'erez \& Granger(2007)}]{ipython}
P\'erez, F., \& Granger, B.~E. 2007, Computing in Science and Engineering, 9,
  21

\bibitem[{{Pfeffer} {et~al.}(2020){Pfeffer}, {Trujillo-Gomez}, {Kruijssen},
  {Crain}, {Hughes}, {Reina-Campos}, \& {Bastian}}]{Pfeffer20}
{Pfeffer}, J.~L., {Trujillo-Gomez}, S., {Kruijssen}, J.~M.~D., {et~al.} 2020,
  arXiv e-prints, arXiv:2003.00076

\bibitem[{{Pillepich} {et~al.}(2015){Pillepich}, {Madau}, \&
  {Mayer}}]{Pillepich15}
{Pillepich}, A., {Madau}, P., \& {Mayer}, L. 2015, \apj, 799, 184

\bibitem[{{Pillepich} {et~al.}(2018){Pillepich}, {Nelson}, {Hernquist},
  {Springel}, {Pakmor}, {Torrey}, {Weinberger}, {Genel}, {Naiman}, {Marinacci},
  \& {Vogelsberger}}]{Pillepich18}
{Pillepich}, A., {Nelson}, D., {Hernquist}, L., {et~al.} 2018, \mnras, 475, 648

\bibitem[{Price-Whelan {et~al.}(2017)Price-Whelan, Sipocz, Major, \&
  Oh}]{gala2}
Price-Whelan, A., Sipocz, B., Major, S., \& Oh, S. 2017, adrn/gala: v0.2.1, , ,
  doi:10.5281/zenodo.833339

\bibitem[{Price-Whelan(2017)}]{gala1}
Price-Whelan, A.~M. 2017, The Journal of Open Source Software, 2,
  doi:10.21105/joss.00388

\bibitem[{{Price-Whelan} {et~al.}(2015){Price-Whelan}, {Johnston}, {Sheffield},
  {Laporte}, \& {Sesar}}]{Price-Whelan15}
{Price-Whelan}, A.~M., {Johnston}, K.~V., {Sheffield}, A.~A., {Laporte}, C.
  F.~P., \& {Sesar}, B. 2015, \mnras, 452, 676

\bibitem[{{Purcell} {et~al.}(2010){Purcell}, {Bullock}, \&
  {Kazantzidis}}]{Purcell10}
{Purcell}, C.~W., {Bullock}, J.~S., \& {Kazantzidis}, S. 2010, \mnras, 404,
  1711

\bibitem[{{Quinn} \& {Goodman}(1986)}]{Quinn86}
{Quinn}, P.~J., \& {Goodman}, J. 1986, \apj, 309, 472

\bibitem[{{Reina-Campos} {et~al.}(2020){Reina-Campos}, {Hughes}, {Kruijssen},
  {Pfeffer}, {Bastian}, {Crain}, {Koch}, \& {Grebel}}]{Reina-Campos20}
{Reina-Campos}, M., {Hughes}, M.~E., {Kruijssen}, J.~M.~D., {et~al.} 2020,
  \mnras, 493, 3422

\bibitem[{{Robertson} {et~al.}(2005){Robertson}, {Bullock}, {Font}, {Johnston},
  \& {Hernquist}}]{bj05_2}
{Robertson}, B., {Bullock}, J.~S., {Font}, A.~S., {Johnston}, K.~V., \&
  {Hernquist}, L. 2005, \apj, 632, 872

\bibitem[{{Rosenberg} {et~al.}(1998{\natexlab{a}}){Rosenberg}, {Piotto},
  {Saviane}, {Aparicio}, \& {Gratton}}]{Rosenberg98}
{Rosenberg}, A., {Piotto}, G., {Saviane}, I., {Aparicio}, A., \& {Gratton}, R.
  1998{\natexlab{a}}, \aj, 115, 658

\bibitem[{{Rosenberg} {et~al.}(1998{\natexlab{b}}){Rosenberg}, {Piotto},
  {Saviane}, {Aparicio}, \& {Gratton}}]{Rosenberg98b}
---. 1998{\natexlab{b}}, \aj, 115, 658

\bibitem[{{Ruiz-Lara} {et~al.}(2020){Ruiz-Lara}, {Gallart}, {Bernard}, \&
  {Cassisi}}]{Ruiz-Lara20}
{Ruiz-Lara}, T., {Gallart}, C., {Bernard}, E.~J., \& {Cassisi}, S. 2020, Nature
  Astronomy, arXiv:2003.12577

\bibitem[{{Sahlholdt} {et~al.}(2019){Sahlholdt}, {Casagrande}, \&
  {Feltzing}}]{Sahlholdt19}
{Sahlholdt}, C.~L., {Casagrande}, L., \& {Feltzing}, S. 2019, \apjl, 881, L10

\bibitem[{{Sakari} {et~al.}(2011){Sakari}, {Venn}, {Irwin}, {Aoki}, {Arimoto},
  \& {Dotter}}]{Sakari11}
{Sakari}, C.~M., {Venn}, K.~A., {Irwin}, M., {et~al.} 2011, \apj, 740, 106

\bibitem[{{Sanders} \& {Binney}(2016)}]{Sanders16}
{Sanders}, J.~L., \& {Binney}, J. 2016, \mnras, 457, 2107

\bibitem[{{Sanders} \& {Das}(2018)}]{Sanders18}
{Sanders}, J.~L., \& {Das}, P. 2018, \mnras, 481, 4093

\bibitem[{{Sanders} {et~al.}(2015){Sanders}, {Shapley}, {Kriek}, {Reddy},
  {Freeman}, {Coil}, {Siana}, {Mobasher}, {Shivaei}, {Price}, \& {de
  Groot}}]{Sanders15mosdef}
{Sanders}, R.~L., {Shapley}, A.~E., {Kriek}, M., {et~al.} 2015, \apj, 799, 138

\bibitem[{{Sanderson} {et~al.}(2015){Sanderson}, {Helmi}, \&
  {Hogg}}]{Sanderson15}
{Sanderson}, R.~E., {Helmi}, A., \& {Hogg}, D.~W. 2015, \apj, 801, 98

\bibitem[{{Santistevan} {et~al.}(2020){Santistevan}, {Wetzel}, {El-Badry},
  {Bland-Hawthorn}, {Boylan-Kolchin}, {Bailin}, {Faucher-Giguere}, \&
  {Benincasa}}]{Santistevan20}
{Santistevan}, I.~B., {Wetzel}, A., {El-Badry}, K., {et~al.} 2020, arXiv
  e-prints, arXiv:2001.03178

\bibitem[{{Sarajedini} {et~al.}(2011){Sarajedini}, {Koo}, {Klesman}, {Laird},
  {Perez Gonzalez}, \& {Mozena}}]{Sarajedini11}
{Sarajedini}, V.~L., {Koo}, D.~C., {Klesman}, A.~J., {et~al.} 2011, \apj, 731,
  97

\bibitem[{{Schaerer} \& {Charbonnel}(2011)}]{Schaerer11}
{Schaerer}, D., \& {Charbonnel}, C. 2011, \mnras, 413, 2297

\bibitem[{{Schiavon} {et~al.}(2017){Schiavon}, {Zamora}, {Carrera},
  {Lucatello}, {Robin}, {Ness}, {Martell}, {Smith},
  {Garc{\'\i}a-Hern{\'a}ndez}, {Manchado}, {Sch{\"o}nrich}, {Bastian},
  {Chiappini}, {Shetrone}, {Mackereth}, {Williams}, {M{\'e}sz{\'a}ros},
  {Allende Prieto}, {Anders}, {Bizyaev}, {Beers}, {Chojnowski}, {Cunha},
  {Epstein}, {Frinchaboy}, {Garc{\'\i}a P{\'e}rez}, {Hearty}, {Holtzman},
  {Johnson}, {Kinemuchi}, {Majewski}, {Muna}, {Nidever}, {Nguyen}, {O'Connell},
  {Oravetz}, {Pan}, {Pinsonneault}, {Schneider}, {Schultheis}, {Simmons},
  {Skrutskie}, {Sobeck}, {Wilson}, \& {Zasowski}}]{Schiavon17}
{Schiavon}, R.~P., {Zamora}, O., {Carrera}, R., {et~al.} 2017, \mnras, 465, 501

\bibitem[{{Schlaufman} {et~al.}(2012){Schlaufman}, {Rockosi}, {Lee}, {Beers},
  {Allende Prieto}, {Rashkov}, {Madau}, \& {Bizyaev}}]{Schlaufman12}
{Schlaufman}, K.~C., {Rockosi}, C.~M., {Lee}, Y.~S., {et~al.} 2012, \apj, 749,
  77

\bibitem[{{Sch{\"o}nrich} {et~al.}(2011){Sch{\"o}nrich}, {Asplund}, \&
  {Casagrande}}]{Schonrich11}
{Sch{\"o}nrich}, R., {Asplund}, M., \& {Casagrande}, L. 2011, \mnras, 415, 3807

\bibitem[{{Searle} \& {Zinn}(1978)}]{Searle78}
{Searle}, L., \& {Zinn}, R. 1978, \apj, 225, 357

\bibitem[{{Sesar} {et~al.}(2017{\natexlab{a}}){Sesar}, {Hernitschek},
  {Dierickx}, {Fardal}, \& {Rix}}]{Sesar17}
{Sesar}, B., {Hernitschek}, N., {Dierickx}, M. I.~P., {Fardal}, M.~A., \&
  {Rix}, H.-W. 2017{\natexlab{a}}, \apjl, 844, L4

\bibitem[{{Sesar} {et~al.}(2011){Sesar}, {Juri{\'c}}, \&
  {Ivezi{\'c}}}]{Sesar11}
{Sesar}, B., {Juri{\'c}}, M., \& {Ivezi{\'c}}, {\v{Z}}. 2011, \apj, 731, 4

\bibitem[{{Sesar} {et~al.}(2017{\natexlab{b}}){Sesar}, {Hernitschek},
  {Mitrovi{\'c}}, {Ivezi{\'c}}, {Rix}, {Cohen}, {Bernard}, {Grebel}, {Martin},
  {Schlafly}, {Burgett}, {Draper}, {Flewelling}, {Kaiser}, {Kudritzki},
  {Magnier}, {Metcalfe}, {Tonry}, \& {Waters}}]{Sesar17rrl}
{Sesar}, B., {Hernitschek}, N., {Mitrovi{\'c}}, S., {et~al.}
  2017{\natexlab{b}}, \aj, 153, 204

\bibitem[{{Sharma} {et~al.}(2011){Sharma}, {Bland-Hawthorn}, {Johnston}, \&
  {Binney}}]{Sharma11}
{Sharma}, S., {Bland-Hawthorn}, J., {Johnston}, K.~V., \& {Binney}, J. 2011,
  \apj, 730, 3

\bibitem[{{Shipp} {et~al.}(2018){Shipp}, {Drlica-Wagner}, {Balbinot},
  {Ferguson}, {Erkal}, {Li}, {Bechtol}, {Belokurov}, {Buncher}, {Carollo},
  {Carrasco Kind}, {Kuehn}, {Marshall}, {Pace}, {Rykoff}, {Sevilla-Noarbe},
  {Sheldon}, {Strigari}, {Vivas}, {Yanny}, {Zenteno}, {Abbott}, {Abdalla},
  {Allam}, {Avila}, {Bertin}, {Brooks}, {Burke}, {Carretero}, {Castander},
  {Cawthon}, {Crocce}, {Cunha}, {D'Andrea}, {da Costa}, {Davis}, {De Vicente},
  {Desai}, {Diehl}, {Doel}, {Evrard}, {Flaugher}, {Fosalba}, {Frieman},
  {Garc{\'\i}a-Bellido}, {Gaztanaga}, {Gerdes}, {Gruen}, {Gruendl}, {Gschwend},
  {Gutierrez}, {Hartley}, {Honscheid}, {Hoyle}, {James}, {Johnson}, {Krause},
  {Kuropatkin}, {Lahav}, {Lin}, {Maia}, {March}, {Martini}, {Menanteau},
  {Miller}, {Miquel}, {Nichol}, {Plazas}, {Romer}, {Sako}, {Sanchez},
  {Santiago}, {Scarpine}, {Schindler}, {Schubnell}, {Smith}, {Smith},
  {Sobreira}, {Suchyta}, {Swanson}, {Tarle}, {Thomas}, {Tucker}, {Walker},
  {Wechsler}, \& {DES Collaboration}}]{Shipp18}
{Shipp}, N., {Drlica-Wagner}, A., {Balbinot}, E., {et~al.} 2018, \apj, 862, 114

\bibitem[{{Siegel} {et~al.}(2007){Siegel}, {Dotter}, {Majewski}, {Sarajedini},
  {Chaboyer}, {Nidever}, {Anderson}, {Mar{\'\i}n-Franch}, {Rosenberg}, {Bedin},
  {Aparicio}, {King}, {Piotto}, \& {Reid}}]{Siegel07}
{Siegel}, M.~H., {Dotter}, A., {Majewski}, S.~R., {et~al.} 2007, \apjl, 667,
  L57

\bibitem[{{Simpson} {et~al.}(2019){Simpson}, {Gargiulo}, {G{\'o}mez}, {Grand},
  {Maffione}, {Cooper}, {Deason}, {Frenk}, {Helly}, {Marinacci}, \&
  {Pakmor}}]{Simpson19}
{Simpson}, C.~M., {Gargiulo}, I., {G{\'o}mez}, F.~A., {et~al.} 2019, \mnras,
  490, L32

\bibitem[{{Simpson} \& {Martell}(2019)}]{Simpson19b}
{Simpson}, J.~D., \& {Martell}, S.~L. 2019, \mnras, 490, 741

\bibitem[{{Speagle}(2019)}]{Speagle19}
{Speagle}, J.~S. 2019, arXiv e-prints, arXiv:1904.02180

\bibitem[{{Starkenburg} {et~al.}(2009){Starkenburg}, {Helmi}, {Morrison},
  {Harding}, {van Woerden}, {Mateo}, {Olszewski}, {Sivarani}, {Norris},
  {Freeman}, {Shectman}, {Dohm-Palmer}, {Frey}, \& {Oravetz}}]{Starkenburg09}
{Starkenburg}, E., {Helmi}, A., {Morrison}, H.~L., {et~al.} 2009, \apj, 698,
  567

\bibitem[{{Steidel} {et~al.}(2016){Steidel}, {Strom}, {Pettini}, {Rudie},
  {Reddy}, \& {Trainor}}]{Steidel16}
{Steidel}, C.~C., {Strom}, A.~L., {Pettini}, M., {et~al.} 2016, \apj, 826, 159

\bibitem[{{Steidel} {et~al.}(2014){Steidel}, {Rudie}, {Strom}, {Pettini},
  {Reddy}, {Shapley}, {Trainor}, {Erb}, {Turner}, {Konidaris}, {Kulas}, {Mace},
  {Matthews}, \& {McLean}}]{Steidel14}
{Steidel}, C.~C., {Rudie}, G.~C., {Strom}, A.~L., {et~al.} 2014, \apj, 795, 165

\bibitem[{{Steinmetz} {et~al.}(2006){Steinmetz}, {Zwitter}, {Siebert},
  {Watson}, {Freeman}, {Munari}, {Campbell}, {Williams}, {Seabroke}, {Wyse},
  {Parker}, {Bienaym{\'e}}, {Roeser}, {Gibson}, {Gilmore}, {Grebel}, {Helmi},
  {Navarro}, {Burton}, {Cass}, {Dawe}, {Fiegert}, {Hartley}, {Russell},
  {Saunders}, {Enke}, {Bailin}, {Binney}, {Bland -Hawthorn}, {Boeche},
  {Dehnen}, {Eisenstein}, {Evans}, {Fiorucci}, {Fulbright}, {Gerhard},
  {Jauregi}, {Kelz}, {Mijovi{\'c}}, {Minchev}, {Parmentier}, {Pe{\~n}arrubia},
  {Quillen}, {Read}, {Ruchti}, {Scholz}, {Siviero}, {Smith}, {Sordo}, {Veltz},
  {Vidrih}, {von Berlepsch}, {Boyle}, \& {Schilbach}}]{RAVE}
{Steinmetz}, M., {Zwitter}, T., {Siebert}, A., {et~al.} 2006, \aj, 132, 1645

\bibitem[{{Stonkut{\.{e}}} {et~al.}(2016){Stonkut{\.{e}}}, {Koposov}, {Howes},
  {Feltzing}, {Worley}, {Gilmore}, {Ruchti}, {Kordopatis}, {Randich},
  {Zwitter}, {Bensby}, {Bragaglia}, {Smiljanic}, {Costado},
  {Tautvai{\v{s}}ien{\.{e}}}, {Casey}, {Korn}, {Lanzafame}, {Pancino},
  {Franciosini}, {Hourihane}, {Jofr{\'e}}, {Lardo}, {Lewis}, {Magrini},
  {Monaco}, {Morbidelli}, {Sacco}, \& {Sbordone}}]{Stonkute16}
{Stonkut{\.{e}}}, E., {Koposov}, S.~E., {Howes}, L.~M., {et~al.} 2016, \mnras,
  460, 1131

\bibitem[{{Ting} \& {Rix}(2019)}]{Ting19}
{Ting}, Y.-S., \& {Rix}, H.-W. 2019, \apj, 878, 21

\bibitem[{{Tinsley}(1980)}]{Tinsley80}
{Tinsley}, B.~M. 1980, \fcp, 5, 287

\bibitem[{{Tissera} {et~al.}(2014){Tissera}, {Beers}, {Carollo}, \&
  {Scannapieco}}]{Tissera14}
{Tissera}, P.~B., {Beers}, T.~C., {Carollo}, D., \& {Scannapieco}, C. 2014,
  \mnras, 439, 3128

\bibitem[{{Torrey} {et~al.}(2019){Torrey}, {Vogelsberger}, {Marinacci},
  {Pakmor}, {Springel}, {Nelson}, {Naiman}, {Pillepich}, {Genel}, {Weinberger},
  \& {Hernquist}}]{Torrey19}
{Torrey}, P., {Vogelsberger}, M., {Marinacci}, F., {et~al.} 2019, \mnras, 484,
  5587

\bibitem[{{van den Bergh} \& {Mackey}(2004)}]{vanderbergh04}
{van den Bergh}, S., \& {Mackey}, A.~D. 2004, \mnras, 354, 713

\bibitem[{{Vasiliev}(2019)}]{Vasiliev19}
{Vasiliev}, E. 2019, \mnras, 484, 2832

\bibitem[{{Venn} {et~al.}(2004){Venn}, {Irwin}, {Shetrone}, {Tout}, {Hill}, \&
  {Tolstoy}}]{Venn04}
{Venn}, K.~A., {Irwin}, M., {Shetrone}, M.~D., {et~al.} 2004, \aj, 128, 1177

\bibitem[{{Vickers} \& {Smith}(2018)}]{Vickers18}
{Vickers}, J.~J., \& {Smith}, M.~C. 2018, \apj, 860, 91

\bibitem[{{Vincenzo} {et~al.}(2019){Vincenzo}, {Spitoni}, {Calura},
  {Matteucci}, {Silva Aguirre}, {Miglio}, \& {Cescutti}}]{Vincenzo19}
{Vincenzo}, F., {Spitoni}, E., {Calura}, F., {et~al.} 2019, \mnras, 487, L47

\bibitem[{{Virtanen} {et~al.}(2020){Virtanen}, {Gommers}, {Oliphant},
  {Haberland}, {Reddy}, {Cournapeau}, {Burovski}, {Peterson}, {Weckesser},
  {Bright}, {van der Walt}, {Brett}, {Wilson}, {Jarrod Millman}, {Mayorov},
  {Nelson}, {Jones}, {Kern}, {Larson}, {Carey}, {Polat}, {Feng}, {Moore}, {Vand
  erPlas}, {Laxalde}, {Perktold}, {Cimrman}, {Henriksen}, {Quintero}, {Harris},
  {Archibald}, {Ribeiro}, {Pedregosa}, {van Mulbregt}, \&
  {Contributors}}]{scipy}
{Virtanen}, P., {Gommers}, R., {Oliphant}, T.~E., {et~al.} 2020, Nature
  Methods, 17, 261

\bibitem[{{Weisz} {et~al.}(2014){Weisz}, {Dolphin}, {Skillman}, {Holtzman},
  {Gilbert}, {Dalcanton}, \& {Williams}}]{Weisz14}
{Weisz}, D.~R., {Dolphin}, A.~E., {Skillman}, E.~D., {et~al.} 2014, \apj, 789,
  148

\bibitem[{{Woolley}(1957)}]{Woolley57}
{Woolley}, R.~V.~D.~R. 1957, \mnras, 117, 198

\bibitem[{{Xue} {et~al.}(2015){Xue}, {Rix}, {Ma}, {Morrison}, {Bovy}, {Sesar},
  \& {Janesh}}]{Xue15}
{Xue}, X.-X., {Rix}, H.-W., {Ma}, Z., {et~al.} 2015, \apj, 809, 144

\bibitem[{{Xue} {et~al.}(2011){Xue}, {Luo}, {Brandt}, {Bauer}, {Lehmer},
  {Broos}, {Schneider}, {Alexander}, {Brusa}, {Comastri}, {Fabian}, {Gilli},
  {Hasinger}, {Hornschemeier}, {Koekemoer}, {Liu}, {Mainieri}, {Paolillo},
  {Rafferty}, {Rosati}, {Shemmer}, {Silverman}, {Smail}, {Tozzi}, \&
  {Vignali}}]{Xue11}
{Xue}, Y.~Q., {Luo}, B., {Brandt}, W.~N., {et~al.} 2011, \apjs, 195, 10

\bibitem[{{Yang} {et~al.}(2019){Yang}, {Xue}, {Li}, {Liu}, {Zhang}, {Rix},
  {Zhang}, {Zhao}, {Tian}, {Zhong}, {Xing}, {Wu}, {Li}, {Carlin}, \&
  {Chang}}]{Yang19}
{Yang}, C., {Xue}, X.-X., {Li}, J., {et~al.} 2019, \apj, 886, 154

\bibitem[{{Yanny} {et~al.}(2009){Yanny}, {Rockosi}, {Newberg}, {Knapp},
  {Adelman-McCarthy}, {Alcorn}, {Allam}, {Allende Prieto}, {An}, {Anderson},
  {Anderson}, {Bailer-Jones}, {Bastian}, {Beers}, {Bell}, {Belokurov},
  {Bizyaev}, {Blythe}, {Bochanski}, {Boroski}, {Brinchmann}, {Brinkmann},
  {Brewington}, {Carey}, {Cudworth}, {Evans}, {Evans}, {Gates}, {G{\"a}nsicke},
  {Gillespie}, {Gilmore}, {Nebot Gomez-Moran}, {Grebel}, {Greenwell}, {Gunn},
  {Jordan}, {Jordan}, {Harding}, {Harris}, {Hendry}, {Holder}, {Ivans},
  {Ivezi{\v{c}}}, {Jester}, {Johnson}, {Kent}, {Kleinman}, {Kniazev},
  {Krzesinski}, {Kron}, {Kuropatkin}, {Lebedeva}, {Lee}, {French Leger},
  {L{\'e}pine}, {Levine}, {Lin}, {Long}, {Loomis}, {Lupton}, {Malanushenko},
  {Malanushenko}, {Margon}, {Martinez-Delgado}, {McGehee}, {Monet}, {Morrison},
  {Munn}, {Neilsen}, {Nitta}, {Norris}, {Oravetz}, {Owen}, {Padmanabhan},
  {Pan}, {Peterson}, {Pier}, {Platson}, {Re Fiorentin}, {Richards}, {Rix},
  {Schlegel}, {Schneider}, {Schreiber}, {Schwope}, {Sibley}, {Simmons},
  {Snedden}, {Allyn Smith}, {Stark}, {Stauffer}, {Steinmetz}, {Stoughton},
  {SubbaRao}, {Szalay}, {Szkody}, {Thakar}, {Sivarani}, {Tucker}, {Uomoto},
  {Vanden Berk}, {Vidrih}, {Wadadekar}, {Watters}, {Wilhelm}, {Wyse}, {Yarger},
  \& {Zucker}}]{SEGUE}
{Yanny}, B., {Rockosi}, C., {Newberg}, H.~J., {et~al.} 2009, \aj, 137, 4377

\bibitem[{{Yu} {et~al.}(2020){Yu}, {Bullock}, {Wetzel}, {Sand erson}, {Graus},
  {Boylan-Kolchin}, {Nierenberg}, {Grudi{\'c}}, {Hopkins}, {Kere{\v{s}}}, \&
  {Faucher-Gigu{\`e}re}}]{Yu20}
{Yu}, S., {Bullock}, J.~S., {Wetzel}, A., {et~al.} 2020, \mnras,
  arXiv:1912.03316

\bibitem[{{Yuan} {et~al.}(2018){Yuan}, {Chang}, {Banerjee}, {Han}, {Kang}, \&
  {Smith}}]{Yuan18}
{Yuan}, Z., {Chang}, J., {Banerjee}, P., {et~al.} 2018, \apj, 863, 26

\bibitem[{{Yuan} {et~al.}(2020){Yuan}, {Myeong}, {Beers}, {Evans}, {Lee},
  {Banerjee}, {Gudin}, {Hattori}, {Li}, {Matsuno}, {Placco}, {Smith},
  {Whitten}, \& {Zhao}}]{Yuan20}
{Yuan}, Z., {Myeong}, G.~C., {Beers}, T.~C., {et~al.} 2020, \apj, 891, 39

\bibitem[{{Zahid} {et~al.}(2014){Zahid}, {Dima}, {Kudritzki}, {Kewley},
  {Geller}, {Hwang}, {Silverman}, \& {Kashino}}]{Zahid14}
{Zahid}, H.~J., {Dima}, G.~I., {Kudritzki}, R.-P., {et~al.} 2014, \apj, 791,
  130

\bibitem[{{Zaritsky} {et~al.}(2020){Zaritsky}, {Conroy}, {Zhang}, {Naidu},
  {Bonaca}, {Caldwell}, {Cargile}, \& {Johnson}}]{Zaritsky20}
{Zaritsky}, D., {Conroy}, C., {Zhang}, H., {et~al.} 2020, \apj, 888, 114

\bibitem[{{Zolotov} {et~al.}(2009){Zolotov}, {Willman}, {Brooks}, {Governato},
  {Brook}, {Hogg}, {Quinn}, \& {Stinson}}]{Zolotov09}
{Zolotov}, A., {Willman}, B., {Brooks}, A.~M., {et~al.} 2009, \apj, 702, 1058

\bibitem[{{Zuo} {et~al.}(2017){Zuo}, {Du}, {Jing}, {Gu}, {Newberg}, {Wu}, {Ma},
  \& {Zhou}}]{Zuo17}
{Zuo}, W., {Du}, C., {Jing}, Y., {et~al.} 2017, \apj, 841, 59

\end{thebibliography}
\bibliographystyle{apj}

\appendix
\section{Error Propagation in Phase Space}
\label{appendix:errorvec}

\begin{figure}[h]
\centering
\includegraphics[width=0.8\linewidth]{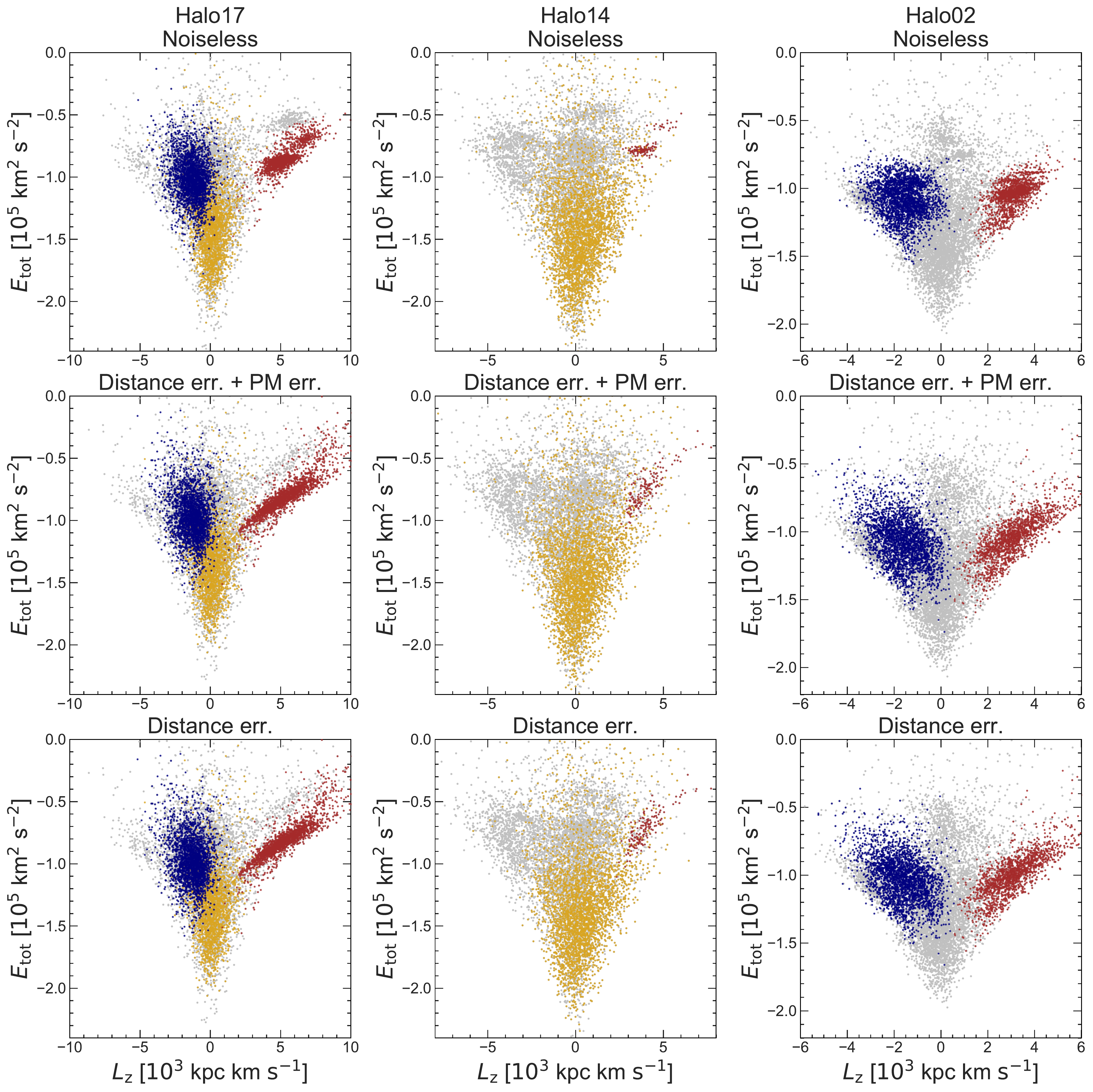}
\caption{Comparison of mocks from three different halos in the \citet{bj05_1} suite with no errors \textbf{(top)},  with distance and PM errors \textbf{(center)}, and with only distance errors \textbf{(bottom)}. Sgr-like (navy blue), high-energy retrograde halo-like (brown), and GSE-like (golden) progenitors are highlighted here. While the Sgr-like and GSE-like locii retain their morphology to first order across all three rows, the retrograde progenitors with high $L_{\rm{z}}$ are dispersed dramatically along a diagonal track in \elz. This is likely why disambiguating the various components of the high-energy retrograde halo (Arjuna+Sequoia+I'itoi) purely in phase-space without relying on chemistry is challenging. The center and bottom panels are virtually indistinguishable, emphasizing that the $10\%$ distance error is the dominant piece of the error budget.}
\label{fig:bj05_02}
\end{figure}

Here we explore how measurement errors -- the $\lesssim10\%$ distance uncertainty, and the error on \textit{Gaia} PMs -- distort substructure in phase-space. We use stellar halos built through hierarchical accretion from the \citet{bj05_1, bj05_2, bj05_3} simulations. These halos feature a realistic, evolving potential, including a disk component. Using the default settings of \texttt{Galaxia}, a code to generate synthetic surveys of the Milky Way from analytical and N-body models \citep{Sharma11}, we generate an H3-like survey -- with a $10\%$ error on distances, and errors on PMs as per the \textit{Gaia} DR2 error model. Potential-dependent phase-space quantities are computed using the $z=0$ potential described in \citet{bj05_1}. 
Figure \ref{fig:bj05_02} shows \elzs diagrams for three halos, both for the noiseless mock catalogs (top panels) and noisy mocks (center and bottom panels).  We also highlight GSE-like, Sgr-like, and high-energy retrograde halo-like progenitors. While GSE-like and Sagittarius-like progenitors, which have comparatively lower $L_{\mathrm{z}}$, retain their general morphology, the strongly retrograde progenitors are significantly smeared out along diagonal tracks in \elz. This is likely why we find it difficult to differentiate between Arjuna, Sequoia, and I'itoi in phase-space even though they are chemically distinct. By comparing the center panels (noisy PMs and noisy distances) and bottom panels (perfect PMs and noisy distances) we see the distance errors are the most significant component of the error budget.

\section{Comparison with alternate potential}
\label{appendix:pot}
\begin{figure}[h]
\centering
\includegraphics[width=0.9\linewidth]{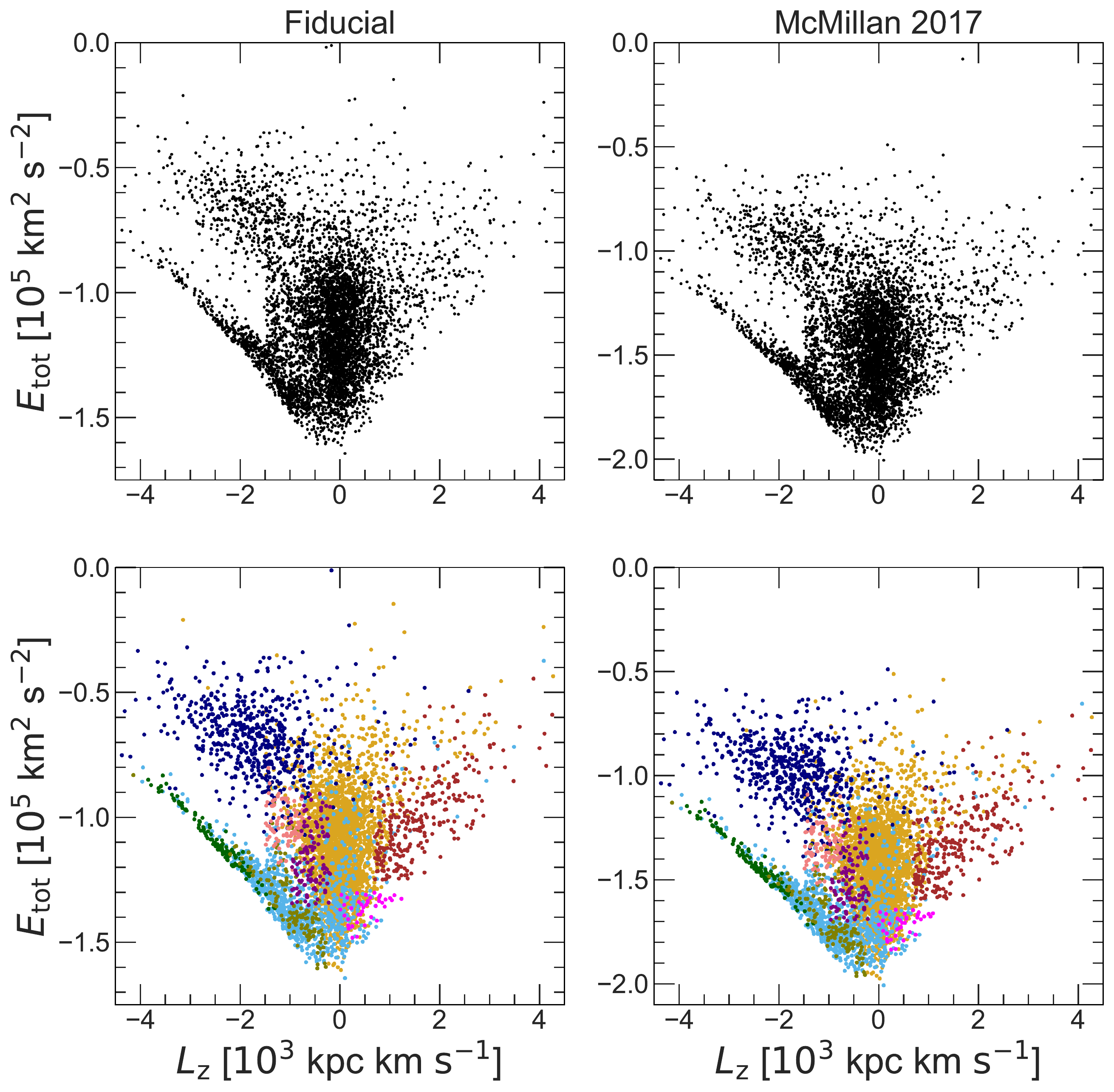}
\caption{\elzs diagrams in the fiducial potential (\textbf{left}) compared against the \citet{McMillan17} potential (\textbf{right}). In the top row we show the data as is, and in the bottom row colored as per the substructure we have assigned these stars to (same as Figure \ref{fig:summary2}). \elzs appears clumpier in the \citet[][]{McMillan17} potential due to the larger virial mass \citep[e.g.,][]{Sanderson15}. Importantly, all the proposed structures (e.g., Wukong, Thamnos, the retrograde shards) correspond to clear clumps and overdensities in this potential.}
\label{fig:M17}
\end{figure}

Here we provide a comparison against the \citet[][]{McMillan17} potential which is widely employed in the halo literature \citep[e.g.,][]{Myeong19,Koppelman19} and features a more massive Milky Way than in the adopted fiducial potential ($1.3\times10^{12}M_{\rm{\odot}}$ versus $9.9\times10^{11}M_{\rm{\odot}}$ within 200 kpc), with several differences in how the potential is parametrized (thick and thin disks, gas disks, a different form for the bulge). Figure \ref{fig:M17} allows for a straightforward visual conversion between the locations of various substructures across these two potentials. $L_{\rm{z}}$ is independent of the potential and so is the same in the left and right panels. It is encouraging that the structures identified in our fiducial potential remain coherent and well-defined in the alternative potential.

\end{CJK*}
\end{document}